\documentclass[superscriptaddress,onecolumn]{revtex4}
\usepackage{graphicx,epsfig}
\usepackage{amsmath}
\usepackage{amssymb}
\usepackage{float}

\usepackage{tikz}
\usetikzlibrary{trees,arrows}

\makeindex

\begin{document}

\title{Dynamical system analysis and observational constraints of cosmological models in mimetic gravity}
\author{Alberto Fritis}
\email{albertofripu@gmail.com}
\affiliation{Departamento de F\'{\i}sica, Facultad de Ciencias, Universidad de La Serena,\\ 
              Avenida Cisternas 1200, La Serena, Chile.}

\author{Daniel Villalobos-Silva}
\email{daniel.villalobos@userena.cl}
\affiliation{Departamento de F\'{\i}sica, Facultad de Ciencias, Universidad de La Serena,\\ 
              Avenida Cisternas 1200, La Serena, Chile.}

\author{Yerko V\'asquez}
\email{yvasquez@userena.cl}
\affiliation{Departamento de F\'{\i}sica, Facultad de Ciencias, Universidad de La Serena,\\ 
              Avenida Cisternas 1200, La Serena, Chile.}

\author{Carlos H. L\'opez-Caraballo}
\email{clopez@iac.es}
\affiliation{Instituto de Astrof\'{\i}sica de Canarias, E-38200 La Laguna, Tenerife, Spain.}
\affiliation{Departamento de Astrof\'{\i}sica, Universidad de La Laguna, E-38206 La Laguna, Tenerife, Spain.}

\author{Juan Carlos Helo}
\email{jchelo@userena.cl}
\affiliation{Departamento de F\'{\i}sica, Facultad de Ciencias, Universidad de La Serena,\\ 
              Avenida Cisternas 1200, La Serena, Chile.}
\affiliation{Millennium Institute for Subatomic Physics at the High Energy Frontier (SAPHIR), Fern\'andez Concha 700, Santiago, Chile.}
\date{\today}

\begin{abstract}

We study the dynamics of homogeneous and isotropic 
Friedmann-Lema\^itre-Robertson-Walker cosmological models with 
positive spatial curvature 
within the context of
mimetic gravity theory
by
employing dynamical system techniques. Our analysis yields phase-space 
trajectories that describe physically relevant solutions, 
capturing various stages of 
cosmic evolution. 
We 
also
employ Bayesian statistical analysis to constrain
the cosmological parameters of the models, utilizing data from 
Type Ia supernovae
and 
Hubble parameter data sets.
The observational data sets provide support for the viability 
of mimetic gravity models, which 
can 
effectively
describe the late-time accelerated expansion of the universe.

\end{abstract}

\maketitle

\section{Introduction}
\label{sec:introduction}

Observational data from various surveys and  experiments, 
such as 
Type Ia  Supernovae (SNIa) 
\cite{SupernovaSearchTeam:1998fmf, SupernovaCosmologyProject:1998vns}, 
the cosmic microwave background (CMB) \cite{WMAP:2003elm, Planck:2013pxb}, 
and measurements of the Hubble parameter and 
baryonic
acoustic oscillations (BAO) \cite{SDSS:2005xqv}, among others, 
indicate that the universe is currently undergoing accelerated 
expansion. This suggests a scenario 
in which
the matter--energy content 
of the universe comprises two extra components that only interact 
gravitationally, and their nature remains unknown. One component, 
referred to as dark energy, plays a crucial role in driving late-time 
cosmological acceleration. The other component, known as dark matter, 
behaves as a pressureless fluid and is essential for explaining 
observed galactic rotation curves.
\newline

Gravity theories based on modifications to General Relativity (GR) 
provide mechanics to explain late-time cosmological acceleration 
in a more interesting form than introducing an exotic 
component
of matter. Various proposals for modified gravity have been explored 
in the literature, including $f(R)$-gravity, where the action of GR 
is generalized to an arbitrary function of the curvature scalar. For 
a comprehensive review, 
see 
\cite{Sotiriou:2008rp}.
%
Gravity models arising from the inclusion of additional 
curvature invariants in the 
Lagrangian 
also
have intriguing cosmological 
implications. Dark energy, for instance, could be elucidated 
in
terms relevant at late time. Moreover, during the early stages of the universe,
 higher curvature corrections to GR, such as the Starobinsky model 
of gravity \cite{Starobinsky:1980te},
which
posits the form
$f(R) = R + \alpha R^2$, should be important for the early time inflation. 
Therefore, modified gravity serves as a natural scenario for a 
theory that unifies and explains both
the inflationary paradigm 
and the dark energy problem \cite{Nojiri:2017ncd}. 
Ref.\ \cite{Nojiri:2006be} 
reviewed various modified 
gravities as gravitational alternatives for explaining dark energy. 
A particularly interesting modification 
of
the gravitational theory 
is mimetic gravity \cite{Chamseddine:2013kea} that has 
drawn
considerable attention 
to
providing a geometric description of 
dark matter. Theories of this type were introduced previously in 
\cite{Lim:2010yk}, which can be obtained through conformal degenerate 
transformations of the metric \cite{Jirousek:2022kli}. In this 
theory, the conformal degree of freedom of the gravitational field 
is isolated by parametrizing the physical metric in terms of an 
auxiliary metric and a scalar field. This conformal degree of freedom, 
being dynamical, can serve as a source of cold dark matter. For a 
comprehensive review of mimetic gravity, 
see
Ref. \cite{Sebastiani:2016ras}.
In Ref.\ \cite{Chamseddine:2014vna} a potential for the mimetic 
field was introduced into the action 
and 
demonstrated
that the model 
can support both late-time accelerating and inflationary solutions. 
Several extensions of mimetic gravity have been considered by 
incorporating higher-order curvature invariants 
into 
the action,  
\cite[see][]{Nojiri:2014zqa, Myrzakulov:2015qaa, Astashenok:2015haa, Nojiri:2016vhu, Nojiri:2017ygt, Kaczmarek:2021psy},
to address the cosmological inflation, or to provide explanations 
for
dark energy. Some models also have the ability to unify and 
address both the inflationary paradigm and dark energy. Also, 
Ho\v{r}ava-like and mimetic Horndeski gravity were considered 
by
\cite{Cognola:2016gjy, Casalino:2018wnc, Casalino:2018tcd}. 
Yet
other theories featuring a scalar degree of 
freedom include quintessence \cite{Tsujikawa:2013fta}, scalar-tensor 
theories \cite{Faraoni:2004pi} and Horndeski's theory 
\cite{Kobayashi:2019hrl}, among others. 
\newline
 
The cosmological equations 
may
be analyzed using nonlinear 
dynamics techniques. The use of dynamical systems techniques 
in cosmology is a powerful tool for studying the 
overall
dynamics of a given cosmological model, if the suitable dynamical 
variables are identified, 
and thus
provide
a very good 
path to
qualitative understanding.
%
The phase space and stability test allows us to circumvent 
the nonlinearities of the cosmological equations and obtain 
a description of the global dynamics 
independently 
of the initial 
conditions of the universe. This connects critical points 
to epochs of evolutionary history that are of special relevance.
Typically, a late expansion period corresponds to an attractor, 
while epochs dominated by radiation and matter often correspond 
to saddle points. Previous 
studies
have applied dynamical system 
techniques to 
the analysis of
cosmological models, including those 
involving $f(R)$-gravity 
\cite{Starobinsky:1980te, Capozziello:1993xn, Odintsov:2017tbc}, 
canonical and phantom scalar fields 
\cite{Copeland:1997et, Roy:2014yta, Hao:2003ww}, and models 
featuring a mimetic field \cite{Odintsov:2015wwp, Dutta:2017fjw, Leon:2014yua} 
(see also \cite{Hrycyna:2013hla, Boehmer:2022wln, Carloni:2017ucm} 
and references therein). The dynamical system perspective has 
also been applied to 
the
study 
of
cosmological models with positive 
spatial curvature \cite{Goliath:1998na, Yilmaz:2021rsc}. For a 
comprehensive exploration of dynamical systems in cosmology, 
with 
particular emphasis on the late-time 
behavior 
of the 
universe, please refer to the following references 
\cite{Boehmer:2014vea, Bahamonde:2017ize}.
\newline

In this 
paper
we perform a dynamical system analysis of the field 
equations of mimetic gravity
by
incorporating a potential for the 
mimetic field, and also study the observational constraints for models which consider exponential scalar field potentials and inverse quartic potentials. In Ref. \cite{Dutta:2017fjw} a dynamical system study of cosmological models in mimetic gravity was performed considering the flat Friedmann-Lema\^itre-Robertson-Walker (FLRW) metric. Although recent observations generally support the spatially flat $\Lambda$CDM model, notable tensions suggest potential deviations from flatness, sometimes on the order of a few percent \cite{DiValentino:2019qzk, Handley:2019tkm, DiValentino:2020hov}. Analyses of the Planck 2018 results, as well as the lensing amplitude in the CMB power spectra, reveal an observed enhancement in the lensing amplitude compared to the predictions of the standard $\Lambda$CDM model. This discrepancy might be explained by a closed universe, which could provide a physical rationale for the observed effect and indicate a preference for a closed universe with more than 99\% confidence. This anomaly might be attributed to new physics, an unresolved systematic error, or merely a statistical fluctuation \cite{DiValentino:2019qzk}.
Our investigation into dynamical systems is conducted within a cosmological framework described by the FLRW metric, incorporating positive spatial curvature as motivated by these results. We derive phase 
space trajectories that depict physically relevant solutions
representing distinct stages of cosmic evolution.
Additionally, we employ a Bayesian statistical analysis to 
constrain
the free parameters of the models, utilizing 
observational data from the cosmic chronometers 
(the 
Hubble database) 
and the Pantheon database from 
Type Ia Supernovae.
\newline

The paper is structured as follows. In section \ref{general} we 
review 
mimetic gravity theory and write the field 
equations for the FLRW metric. In section \ref{dyn} we write the 
motion equations as an autonomous dynamical system, identify
critical points, and 
investigate
the stability of each point 
for a generic mimetic field potential. Specific potentials for 
the mimetic field are explored in detail, and the phase space 
of each model is analyzed. In section \ref{sec:potential} we 
numerically solve the field equations for two mimetic field 
potentials
and compare
the results with the non-flat $\Lambda$CDM 
solutions. Then, in section \ref{sec:description_model_Hz} we 
constrain
the free parameters of the models using Hubble parameter
 and SNIa measurements, 
and in section \ref{sec:conclusion} we 
present our conclusions.

\section{Mimetic Gravity}
\label{general}

In this section, we provide a concise overview of the mimetic 
gravity theory \cite{Chamseddine:2013kea} and write the field 
equations for the FLRW metric. In mimetic gravity the conformal 
degree of freedom of the gravitational field is isolated by 
parametrizing the physical metric $g_{\mu \nu}$ in terms of an 
auxiliary metric $\hat{g}_{\mu \nu}$ and a scalar field $\phi$, 
referred to as the mimetic field:
\begin{equation}
g_{\mu \nu} =- \hat{g}_{\mu \nu} \hat{g}^{\alpha \beta} \partial_{\alpha} \phi \partial_{\beta} \phi  \,.
\end{equation}
The gravitational action can be varied with respect to the 
auxiliary metric and the scalar field, rather than the physical 
metric, and the following constraint is obtained as a consistency 
condition:
\begin{equation}
g^{\mu \nu}(\hat{g}_{\mu \nu}, \phi) \partial_{\mu} \phi \partial_{\nu} \phi +1 =0 \,.
\end{equation}

The field equations obtained from varying the action written 
in terms of the physical metric and the imposition of the mimetic 
constraint are completely equivalent to the field equations 
derived from the action written in terms of the auxiliary metric. 
Therefore, the constraint can be 
implemented 
at the level of the 
action by introducing a Lagrange multiplier, 
$\lambda$. Thus, the 
action describing mimetic gravity can be expressed as:
\begin{equation}
S = \int d^4x \sqrt{-g} \left(  \frac{R}{2 \kappa^2} - \frac{\lambda}{2} (\partial_{\mu} \phi \partial^{\mu} \phi +1) - V(\phi) \right) +S_m \,,
\end{equation}
where $\kappa^2 = 8 \pi G$ is the gravitational constant, 
$\phi$ is the mimetic field, $\lambda$ is a Lagrange multiplier 
field, $V(\phi)$ is a potential for the mimetic field and $S_m$ 
is a generic matter action that we shall consider as a perfect 
fluid. The inclusion of a potential for the mimetic field was 
first considered in Ref. \cite{Chamseddine:2014vna}. 
\newline

The variation of the action with respect to the physical metric yields:
\begin{equation}
R_{\mu \nu}-\frac{1}{2} g_{\mu \nu} R = \kappa^2  \left(  T_{\mu \nu} + \lambda \partial_{\mu} \phi \partial_{\nu} \phi - g_{\mu \nu} V(\phi) \right)\,,
\end{equation}
where $T^{\mu \nu}$ is the energy-momentum tensor of 
the
matter fields. The variation of the action with respect to the Lagrangian 
multiplier yields the constraint
%
\begin{equation} \label{phi}
g^{\mu \nu} \partial_{\mu} \phi \partial_{\nu} \phi = -1 \,,
\end{equation}
and the variation with respect to the mimetic field produces
%
\begin{equation} \label{mim}
\nabla^{\mu} (\lambda \partial_{\mu} \phi)- \frac{dV}{d \phi} = 0 \,.
\end{equation}

We 
shall
consider a cosmological setting where the spacetime is 
described by the FLRW metric
%
\begin{equation} \label{metricaFLRW}
    ds^2 = -dt^2 + a(t)^2 \left( \frac{dr^2}{1-k r^2} + r^2 d\theta^2 + r^2 \sin^2 \theta  d\phi^2 \right) \,,
\end{equation}
where $k=1, 0, -1$ for a closed, flat and an open universe, respectively.
\newline

From the gravitational equations, considering an energy-momentum 
tensor of a perfect fluid given by 
$T^{\mu}_{\,\,\, \nu} = diag (-\rho, p,p,p)$ in the co-moving system, 
the following Friedmann equations are obtained:
\begin{eqnarray}
\label{friedmann} 3 H^2 &=& \kappa^2 \left(\rho +\lambda+V  \right) - \frac{3 k}{a^2} \,,  \\
 3 H^2 + 2 \dot{H} &=& - \kappa^2 \left( p- V \right) - \frac{k}{a^2} \,,
\end{eqnarray}
where $H = \frac{\dot{a}}{a}$ is the Hubble parameter, and the 
equation for the mimetic field (\ref{mim}) reads:
\begin{equation}
\dot{\lambda} + 3 H \lambda + \frac{dV}{d \phi} =  0 \,.
\end{equation}
Note that $\lambda$ represents a non-relativistic dark matter 
component. Also, using the above equation,
together
with the Friedmann equations, the conservation equation for the matter content is obtained
\begin{equation} \label{eqcon}
    \dot{\rho} + 3 H(\rho + p) =0 \,.
\end{equation}

The constraint equation (\ref{phi}) is straightforward to solve
and leads
to the solution $\phi(t) = t$.
\newline

In what follows, we 
assume a barotropic equation of state 
for the perfect fluid, i.e., $p = \omega \rho$, where the equation 
of state parameter is $\omega = 0$ for dust matter and 
$\omega = 1/3$ for radiation. 
\newline

It is useful to introduce 
dimensionless density parameters, 
which constitute the physical observables, defined by:
\begin{equation}
\Omega_{\rho} = \frac{\kappa^2 \rho}{3 H^2}\,, \,\,\, \Omega_{\lambda}
 = \frac{\kappa^2 \lambda}{3 H^2}\,, \,\,\, \Omega_{\phi} 
= \frac{\kappa^2 V}{3 H^2} \,, \,\,\, \Omega_{k}=- \frac{k}{H^2 a^2} \,.
\end{equation}
In terms of 
these parameters,
the Friedmann equation (\ref{friedmann}) reads:
\begin{equation} \label{density}
 \Omega_{\rho}+\Omega_{\lambda}+ \Omega_{\phi}+ \Omega_{k} =1 \,.
\end{equation}
It is also useful to introduce the deceleration parameter, 
defined as $q= - a(t) \ddot{a}(t)/\dot{a}(t)^2$, which quantifies 
how the expansion rate of the universe changes over time. 
A value $q<0$ ($q>0$) indicates that the expansion of the 
universe is accelerating (decelerating). The deceleration 
parameter is related to the density parameters by 
$q=\frac{1}{2}+\frac{3 }{2}\omega \Omega_{\rho}-\frac{3}{2} \Omega_{\phi}-\frac{1}{2}\Omega_{k}$, 
which will be very useful later. All the parameters 
defined above constitute the physical observables we are 
interested in. In the next section we will study the 
cosmological equations as an autonomous dynamical system.
\newline

\section{Dynamical system approach}
\label{dyn}

In this section we 
analyze the 
behavior 
of the 
cosmological models
with positive spatial curvature 
$k=1$
described in the previous section from a dynamical 
system perspective. To perform the analysis it is convenient 
to define the following dimensionless dynamical variables:
\begin{equation}
Q = \frac{H}{D} \,, \,\,\,\,\, \tilde{\Omega}_{\rho} 
= \frac{\kappa^2 \rho}{3 D^2} \,, \,\,\,\,\, \tilde{\Omega}_{\lambda} 
= \frac{\kappa^2 \lambda}{3 D^2} \,, \,\,\,\,\, \beta 
= \frac{\kappa \sqrt{V}}{\sqrt{3} D} \,, \,\,\,\,\, \Gamma 
= - \frac{k}{a^2 D^2} \,, \,\,\,\,\, \zeta 
= - \frac{1}{\kappa V^{3/2}} \frac{d V}{d \phi} \,,
\end{equation}
where $D \equiv \sqrt{H^2 + \frac{k}{a^2}}$. Here 
$Q$ is a compact variable that can take values only 
in the interval $[-1,1]$, where $Q>0$ corresponds to 
an expanding epoch, and $Q<0$ corresponds to a contracting epoch.
\newline

The Friedmann equation (\ref{friedmann}) and the 
definition of $D$ yield the constraints:
\begin{eqnarray} \label{rest}
 \tilde{\Omega}_{\rho} + \tilde{\Omega}_{\lambda}+\beta^2 &=& 1 \,, \\
Q^2 - \Gamma &=& 1 \,.
\end{eqnarray}
Next, using these constraints and introducing the new 
time variable $d\tau = D dt$, the cosmological equations 
can be written as an autonomous dynamical system:
\begin{eqnarray}
\label{Q} Q' &=& \left(1-Q^2 \right) \left( \frac{3}{2} \left( (1+\omega) (\beta^2-1)+ \omega \tilde{\Omega}_{\lambda} \right)+1 \right) \,,  \\ \label{OME}
\tilde{\Omega}_{\lambda}' &=& - 3 Q \tilde{\Omega}_{\lambda} + \sqrt{3} \beta^{3} \zeta - 3 Q \tilde{\Omega}_{\lambda} \left( (1+\omega) (\beta^2-1)+ \omega \tilde{\Omega}_{\lambda} \right) \,,  \\  \label{beta}
\beta' &=& -\frac{\sqrt{3}}{2} \beta^2 \zeta -\frac{3}{2} \beta Q \left( (1+\omega) (\beta^2-1)+ \omega \tilde{\Omega}_{\lambda} \right) \,, \\
\label{z} \zeta' &=& - \sqrt{3} \zeta^2 \beta \left(\mu(\zeta) - \frac{3}{2}  \right) \,,
\end{eqnarray}
where we have defined $\mu = \ddot{V} V / \dot{V}^2$ 
and the prime denotes a derivative with respect to $\tau$. 
The dimensionless dynamical variables constitute the basis
 of the phase space, and the first order differential 
equations (\ref{Q})-(\ref{z}) are completely equivalent 
to the original field equations and give the phase trajectories. 
Additionally, the weak energy condition 
$\tilde{\Omega}_{\rho} \geq 0$,
 in conjunction with Eq. (\ref{rest}),
 produces $-\infty < \tilde{\Omega}_{\lambda} \leq 1- \beta^2$. 
Therefore, the physical phase space is contained in the 
region $\beta > 0$, $\zeta \in \mathbb{R}$, 
$Q \in [-1,1]$ and $-\infty < \tilde{\Omega}_{\lambda} \leq 1- \beta^2$.
\newline

Furthermore, it can be found that 
$\tilde{\Omega}_{\rho}' = - 3 Q \tilde{\Omega}_{\rho} \left( (1+\omega) \beta^2 
+ \omega \tilde{\Omega}_{\lambda}   \right)$, 
which shows that $\tilde{\Omega}_{\rho}=0$ is an 
invariant submanifold. The invariant submanifolds 
of the reduced dynamical system are the following:
\newline

\begin{itemize}
\item $\tilde{\Omega}_{\rho}=0$,
which
corresponds to the $\rho=0$ boundary.
\item $Q=\pm 1$ are the flat submanifolds ($k=0$).
\item The $\beta=0$ submanifold.
\item The $\zeta=0$ submanifold, with $\zeta^2 \mu(\zeta) =0$ for $\zeta=0$.
\end{itemize}

The dynamical variables defined to describe the physical 
system are not physical observables
but 
are related to the density parameters by 
$\Omega_{\rho} = \frac{\tilde{\Omega}_{\rho}}{Q^2}$, $\Omega_{\lambda} 
= \frac{\tilde{\Omega}_{\lambda}}{Q^2}$, $\Omega_{\phi} 
= \frac{\beta^2}{Q^2}$, $\Omega_{k} = 1- \frac{1}{Q^2}$.
\newline

\subsection{Critical Points}

Applying the standard procedure of cosmological dynamical 
systems, we start by looking at the critical points of the 
system (\ref{Q})--(\ref{z}) for an arbitrary scalar field potential. 
\newline

The critical points correspond to the points where the system 
is in equilibrium  and are defined by the vanishing of the 
right-hand sides of Eqs.\ (\ref{Q})--(\ref{z}).
In Table \ref{tabla1} we show the critical points of the 
system and the values of some physical quantities associated 
with each of them. Note
that, with the exception of points 
$A_1$, $A_2$ and $D$, all other points lie on the flat 
submanifolds.
\newline

We now
examine the linear stability of the critical points 
by linearizing the evolution equations in the vicinity of 
these critical points. The Jacobian matrix of the system is 
given by:

{\tiny
\begin{equation}
   \begin{bmatrix}
 \notag   -2Q\left(\frac{3}{2}\left((\beta^2-1)(\omega+1)+\tilde{\Omega}_{\lambda}\omega\right)+1\right) & \frac{3}{2} \omega (1-Q^2) & 3 \beta (1-Q^2) (1+\omega) & 0 \\    -3\tilde{\Omega}_{\lambda}\left(\beta^2 (1+\omega)+\omega (\tilde{\Omega}_{\lambda}-1) \right) & -3Q\left(\beta^2 (1+\omega)+\omega (2\tilde{\Omega}_{\lambda}-1) \right) & 3\beta \left( \sqrt{3}\zeta\beta -2 Q \tilde{\Omega}_{\lambda}(\omega+1) \right)  & \sqrt{3} \beta^3 \\
    -\frac{3}{2}\beta \left((\beta^2 -1)(\omega+1)+\omega \tilde{\Omega}_{\lambda}\right)  & -\frac{3}{2}Q\beta \tilde{\Omega}_{\lambda} & -\sqrt{3} \beta \zeta + \frac{3Q}{2} \left( (1-3 \beta^2)(1+\omega)-\omega \tilde{\Omega}_{\lambda} \right) & \frac{-\sqrt{3}}{2} \beta^2 \\
    0 & 0 & -\sqrt{3}\zeta^2 \left( \mu(\zeta)-\frac{3}{2}\right) & -\sqrt{3}\zeta\beta \left( \zeta \frac{d \mu (\zeta)}{d \zeta} +2 \mu (\zeta) -3 \right)
    \end{bmatrix}
\end{equation}
}
\bigskip

The eigenvalues of the Jacobian matrix, computed at a 
critical point, enable the analysis of the linear stability of 
that 
point. In Table \ref{tabla2} we present the eigenvalues 
corresponding to each critical point. However, linear 
stability theory can only be applied to examine the stability 
of points whose eigenvalues have a non-null real part
(referred to as hyperbolic critical points). To assess the 
stability of non-hyperbolic critical points it is necessary 
to employ methods beyond linear stability theory, such as the 
Lyapunov
method or centre manifold theory. By employing the 
centre manifold, we 
shall
analyze certain cases that cannot 
be categorized as hyperbolically stable or unstable 
because of
 the presence of null eigenvalues in the variation matrix. 
The cases that always satisfy this condition are analyzed, 
as well as those that could meet this condition 
in
specific circumstances.
\newline
\begin{table}[H]
\begin{center}
\caption{Fixed points of the dynamical system (\ref{Q})-(\ref{z}) 
and some physical quantities for an arbitrary mimetic field 
potential. $\zeta_*$ is the solution of $\mu (\zeta) -3/2=0$}
\label{tabla1}
\resizebox{\columnwidth}{!}{
\begin{tabular}{|c|c|c c c|c|c c c c|c|}
\hline
Label & $\tilde{\Omega}_{\lambda}$ & $\beta$ & $\zeta$ & $Q$ & Existence & $\Omega_{\rho}$ & $\Omega_{\lambda}$ & $\Omega_{\phi}$ & $\Omega_{k}$ & $q$ \\
\hline
 $A_1$ & $-\frac{2}{3 \omega} + (1-\beta^2) \left(1+\frac{1}{\omega}\right)$ & $\beta \geq  \frac{1}{\sqrt{3}}$ & 0 & 0 & $\omega \neq 0$, $\zeta^2 \mu =0$ for $\zeta=0$ & $\infty$ & $\infty$ for $\tilde{\Omega_{\lambda}}>0$, $-\infty$ for $\tilde{\Omega_{\lambda}}<0$ & $\infty$ & $-\infty$ & $\infty$ \\
 $A_2$ & $\tilde{\Omega}_{\lambda} \leq \frac{2}{3}$ & $\frac{1}{\sqrt{3}}$ & 0 & 0 & $\omega = 0$, $\zeta^2 \mu =0$ for $\zeta=0$ & $\infty$ & $\infty$ for $\tilde{\Omega_{\lambda}}>0$, $-\infty$ for $\tilde{\Omega_{\lambda}}<0$ & $\infty$ & $-\infty$ & $\infty$ \\
 $B_+$ & 0 & 1 & 0 & 1 & $\zeta^2 \mu =0$ for $\zeta=0$ & 0 & 0 & 1 & 0 & -1 \\
 $B_-$ & 0 & 1 & 0 & -1 & $\zeta^2 \mu =0$ for $\zeta=0$ & 0 & 0 & 1 & 0 & -1 \\  
 $C_{1+}$ & $-\frac{3(1+\omega)^3}{\omega \zeta_{\ast}^2}$ & $\frac{(1+\omega) \sqrt{3}}{\zeta_{\ast}}$ & $\zeta_{\ast}$ & 1 & $\omega \neq 0$, $\zeta_{\ast} > 0$ &  $\frac{3+\omega (\zeta_*^2 +6 +3 \omega)}{\zeta_*^2 \omega}$  & $-\frac{3(1+\omega)^3}{\omega \zeta_*^2}$ & $\frac{3(1+\omega)^2}{\zeta_*^2}$ & 0 & $\frac{1}{2}(1+3\omega)$ \\
  $C_{1-}$ &  $-\frac{3(1+\omega)^3}{\omega \zeta_{\ast}^2}$  &  $-\frac{(1+\omega) \sqrt{3}}{\zeta_{\ast}}$   & $\zeta_{\ast}$ &  -1 &  $\omega \neq 0$, $\zeta_{\ast} < 0$  & $\frac{3+\omega (\zeta_*^2 +6 +3 \omega)}{\zeta_*^2 \omega}$  &  $-\frac{3(1+\omega)^3}{\omega \zeta_*^2}$  &  $\frac{3(1+\omega)^2}{\zeta_*^2}$  & 0 &  $\frac{1}{2}(1+3\omega)$  \\
 $C_{2+}$  & $\frac{1}{6} \zeta_{\ast} ( \sqrt{12+\zeta_{\ast}^2} -\zeta_{\ast})$ & $\frac{\sqrt{12+\zeta_{\ast}^2}-\zeta_{\ast}}{ 2 \sqrt{3}}$ & $\zeta_{\ast}$ & 1 &  & 0 & $\frac{1}{6} \zeta_{\ast} ( \sqrt{12+\zeta_{\ast}^2} -\zeta_{\ast})$ &  $\frac{1}{12} (\sqrt{\zeta_*^2 +12}-\zeta_*)^2$  & 0 & $\frac{1}{4} \zeta_* (\sqrt{\zeta_*^2+12}- \zeta_*) -1$ \\
 $C_{2-}$ & $-\frac{1}{6} \zeta_*(\sqrt{12+\zeta_{\ast}^2}+\zeta_*)$ & $\frac{\sqrt{12+\zeta_*^2}+\zeta_*}{ 2 \sqrt{3}}$ &$\zeta_{\ast}$  & -1 &  & 0 & $-\frac{1}{6} \zeta_* (\sqrt{12+\zeta_*^2}+\zeta_*)$ & $\frac{1}{12} (\sqrt{\zeta_*^2+12}+\zeta_*)^2$ & 0 & $-\frac{1}{4} \zeta_* (\sqrt{\zeta_*^2+12}+\zeta_*)-1$ \\
 $D$ & $\frac{2}{3}$ & $\frac{1}{\sqrt{3}}$ & $\zeta_{\ast}$ & $\frac{\zeta_{\ast}}{2}$ & $\zeta_{\ast} \neq 0$, $Q \neq \pm 1$ & 0 & $\frac{8}{3 \zeta_*^2}$ & $\frac{4}{3 \zeta_*^2}$ &  $1-\frac{4}{\zeta_*^2}$  & 0 \\   
 $L_{1+}$ & 0 & 0 & $\zeta$ & 1 & critical line for all $\zeta$ for $\omega \neq 0$& 1 & 0 & 0 & 0 & $\frac{1}{2}(1+3\omega)$ \\
 $L_{2+}$ & 1 & 0 & $\zeta$ & 1 & critical line for all $\zeta$ for $\omega \neq 0$ & 0 & 1 & 0 & 0 & $\frac{1}{2}$ \\
 $L_{1-}$ & 0 & 0 & $\zeta$ & -1 & critical line for all $\zeta$ for $\omega \neq 0$ & 1 & 0  & 0 & 0 & $\frac{1}{2}(1+3 \omega)$ \\
 $L_{2-}$ & 1 & 0 & $\zeta$ & -1 & critical line for all $\zeta$ for $\omega \neq 0$ & 0 & 1  & 0 & 0 & $\frac{1}{2}$ \\
$P_+$ & $\tilde{\Omega_{\lambda}} \leq 1$ & 0 & $\zeta$ & 1 & critical plane for $\omega=0$& $1-\Omega_{\lambda}$ & $\Omega_{\lambda} \leq 1$  & 0 & 0 & $\frac{1}{2}$ \\
$P_-$ & $\tilde{\Omega_{\lambda}} \leq 1$ & 0 & $\zeta$ & -1 & critical plane for $\omega=0$& $1-\Omega_{\lambda}$ & $\Omega_{\lambda} \leq 1$  & 0 & 0 & $\frac{1}{2}$ \\
\hline
\end{tabular}
}
\end{center}
\end{table}
\begin{table}[ht]
\begin{center}
\caption{Eigenvalues and stability of fixed points of the 
dynamical system (\ref{Q})-(\ref{z}). In the formulas below 
we 
have
defined 
$\xi_{\pm}(\zeta_*) = \zeta_*^2 \pm \zeta_* \sqrt{\zeta_*^2+12} +12 \omega$. 
$\zeta_*$ is the solution of $\mu (\zeta) -3/2=0$} \label{tabla2}
\resizebox{\columnwidth}{!}{
\begin{tabular}{|c|c|c|c|c|c|} 
\hline
Label & $\lambda_1$ & $\lambda_2$ & $\lambda_3$ & $\lambda_4$ & Dynamical Character \\
\hline
$A_1$ & $-\frac{\sqrt{-(1+3\omega)+ 9 \beta^2 (1+\omega)}}{\sqrt{2}}$ & $\frac{\sqrt{-(1+3\omega)+ 9 \beta^2 (1+\omega)}}{\sqrt{2}}$ &  $-\sqrt{3} \beta \frac{d}{d\zeta}(\zeta^2 \mu(\zeta))|_{\zeta=0}$ & 0 & non-hyperbolic, behaves as saddle \\ 
 $A_2$ & -1 & 1 & $-\frac{d}{d\zeta}(\zeta^2 \mu(\zeta))|_{\zeta=0}$ & 0 & non-hyperbolic, behaves as saddle \\
$B_+$  & -2 & $-3(1+\omega)$ &   $-\sqrt{3} \frac{d}{d\zeta}(\zeta^2 \mu(\zeta))|_{\zeta=0}$ & -3 & mostly attractive, potentially non-hyperbolic   \\
  $B_-$  & 2 & $3(1+\omega)$ & $-\sqrt{3} \frac{d}{d\zeta}(\zeta^2 \mu(\zeta))|_{\zeta=0}$ & 3 & mostly repulsive, potentially non-hyperbolic \\ 
$C_{1+}$ & $-\frac{3 \left((1-\omega)\zeta_*+\sqrt{24(1+\omega)^3+(1+3 \omega)^2 \zeta_*^2}\right)}{4\zeta_*}$ &  $-\frac{3 \left((1-\omega)\zeta_*-\sqrt{24(1+\omega)^3+(1+3 \omega)^2 \zeta_*^2}\right)}{4\zeta_*}$   &  $-3 (1+\omega) \zeta_* \mu'(\zeta_*)$  & $1+3 \omega$ &  saddle   \\ 
$C_{1-}$ & $\frac{3 \left((1-\omega)\zeta_*+\sqrt{24(1+\omega)^3+(1+3 \omega)^2 \zeta_*^2}\right)}{4\zeta_*}$ &  $\frac{3 \left((1-\omega)\zeta_*-\sqrt{24(1+\omega)^3+(1+3 \omega)^2 \zeta_*^2}\right)}{4\zeta_*}$   &  $3 (1+\omega) \zeta_* \mu'(\zeta_*)$  & $-(1+3 \omega)$ &  saddle   \\
 $C_{2+}$ & -$\frac{3}{8} \left( \xi_-(\zeta_*) + 8 (1- \omega) +\frac{1}{3} |\xi_-(\zeta_*)| \right) $ &  -$\frac{3}{8} \left( \xi_-(\zeta_*) + 8 (1- \omega) -  \frac{1}{3} |\xi_-(\zeta_*)| \right)$  &  $\frac{1}{2} \zeta_*^2 (\zeta_* - \sqrt{\zeta_*^2 +12}) \mu'(\zeta_*)$  & $ -2 + \frac{1}{2} \zeta_* (-\zeta_* + \sqrt{\zeta_*^2 +12}) $ & see Appendix \ref{appendixa} \\ 
 $C_{2-}$ &  $\frac{3}{8} \left( \xi_+ (\zeta_*) + 8 (1- \omega) - \frac{1}{3} |\xi_+(\zeta_*)| \right) $  & $\frac{3}{8} \left( \xi_+(\zeta_*) + 8 (1- \omega) + \frac{1}{3} |\xi_+(\zeta_*)| \right) $ & $-\frac{1}{2} \zeta_*^2 (\zeta_* + \sqrt{\zeta_*^2 +12}) \mu'(\zeta_*)$ &  $ 2 + \frac{1}{2} \zeta_* (\zeta_* + \sqrt{\zeta_*^2 +12}) $  &  see Appendix \ref{appendixa} \\ 
 $D$ & $-1-\frac{\zeta_*}{2}$ & $-\frac{1}{2} \zeta_* (1+3\omega)$ & $-\zeta_*^2 \mu'(\zeta_*)$ & $1-\frac{\zeta_*}{2}$ & see analysis \\ 
 $L_{1+}$ & $3 \omega$ & $\frac{3 (1+\omega)}{2}$ & $1+3\omega$ & 0 & non-hyperbolic, unstable \\ 
 $L_{2+}$ & 1  & 0 & $-3 \omega$ & $\frac{3}{2}$ &  saddle \\ 
 $L_{1-}$ & $-3 \omega$ & $-\frac{3}{2}(1+\omega)$ & $-(1+ 3\omega)$ & 0 & non-hyperbolic \\ 
 $L_{2-}$ & -1 & 0 & $3 \omega$ & $-\frac{3}{2}$ & saddle  \\ 
$P_+$ & 0 & 0 & $\frac{3}{2}$ &  1  & non-hyperbolic, unstable  \\ 
$P_-$ & 0 & 0 & -$\frac{3}{2}$ &  -1  & non-hyperbolic    \\
\hline
\end{tabular}
}
\end{center}
\end{table}

The information presented in Tables \ref{tabla1} and 
\ref{tabla2} enables the description of the behavior of 
the dynamical system in the neighborhood of each critical 
point. In what follows $\zeta_*$ corresponds to the 
solution of $\mu (\zeta) -3/2=0$.

\begin{itemize}
\item $A_1$ and $A_2$ correspond to critical curves where 
$Q$ vanishes and the deceleration parameter diverges, 
thus
resembling the behavior of the Einstein static solution 
where $H=0$ and $\dot{H}=0$. $A_1$ exists for 
$\omega \neq 0$ and $\zeta^2 \mu =0$ for $\zeta=0$, 
whereas $A_2$ exists for $\omega = 0$ and $\zeta^2 \mu =0$ 
for $\zeta=0$. Moreover, these curves are non-hyperbolic 
and act as saddles, as the 
non-zero 
eigenvalues have opposite signs. 
\item $B_+$ exists if $\zeta^2 \mu = 0$ for $\zeta=0$, 
representing an expanding de Sitter accelerated solution 
dominated by the mimetic field, $\Omega_{\phi} =1$. The 
stability analysis involves examining the sign of 
$-\sqrt{3}\frac{d}{d\zeta}\left(\zeta^2 \mu (\zeta) \right)$ 
at $\zeta=0$. If it is negative, the critical point is 
an attractor;
if it is positive, it is a saddle 
point. If 
$-\sqrt{3}\frac{d}{d\zeta}\left(\zeta^2 \mu (\zeta) \right)$ 
at $\zeta=0$ equals $0$, the point is non-hyperbolic, and 
its stability is analyzed using the centre manifold method, 
as detailed in appendix A. In this case, several possibilities 
arise:
    \begin{itemize}
        \item If $-\sqrt{3}\frac{d^2}{d\zeta^2}\left(\zeta^2 \mu (\zeta) \right)|_{\zeta=0}\neq 0$, 
the critical point will manifest unstable behavior in a 
specific direction (it is easy to see that the direction 
is locally $\zeta$)
        \item Otherwise, with 
$\sigma=-\sqrt{3}\frac{d^3}{d\zeta^3}\left(\zeta^2 \mu (\zeta) \right)|_{\zeta=0}$, 
if $\sigma <0$ it has an asymptotically stable behavior. 
If $\sigma > 0$ it displays unstable behavior in this 
direction. If $\sigma=0$,
 higher order terms must be considered in the series.
    \end{itemize}
\item $B_-$ is the contracting analogue of $B_+$ and is unstable.
For $-\sqrt{3}\frac{d}{d\zeta}\left(\zeta^2 \mu (\zeta) \right)|_{\zeta=0}=0$, 
it is non-hyperbolic, leading to the following possibilities:
\begin{itemize}
        \item If $-\sqrt{3}\frac{d^2}{d\zeta^2}\left(\zeta^2 \mu (\zeta) \right)|_{\zeta=0}\neq 0$, 
the critical point will manifest stable behavior in one direction 
only (locally $\zeta$)
        \item Otherwise, with 
$\sigma=-\sqrt{3}\frac{d^3}{d\zeta^3}\left(\zeta^2 \mu (\zeta) \right)|_{\zeta=0}$, 
if $\sigma <0$, the point shows stable behavior in one direction. 
If $\sigma > 0$, 
its
behavior 
is unstable
in all directions. If 
$\sigma=0$ 
higher-order terms must be considered in the series. 
    \end{itemize}
\item $C_{1+}$ and $C_{1-}$ behave as scaling solutions, which 
can alleviate the cosmic coincidence problem. In these scenarios, both 
a
scalar field and matter densities coexist, meaning 
that
the universe undergoes evolution 
influenced 
by both matter 
(baryonic 
plus mimetic matter) and the scalar field. However, the universe 
expands as if it were dominated by matter. Here, the deceleration 
parameter is given by $q= \frac{1}{2} (1+ 3\omega)>0$ for matter
 and radiation, indicating non-accelerating solutions. The stability 
analysis reveals that these 
are saddle points. 
\item  The point $C_{2+}$ stands for an expanding cosmological 
solution where the mimetic matter and the mimetic field dominate. 
When $\zeta_* < 2$, accelerating solutions are possible. In the 
limit as $\zeta_{\ast} \rightarrow 0$, the universe is predominantly 
governed by the mimetic field, resulting in an expanding de 
Sitter solution. Note that point $B_+$ is a special case of 
$C_{2+}$ with $\zeta_*=0$. Conversely, $C_{2-}$ is the contracting 
analogue of $C_{2+}$. 
\item Critical point D corresponds to a uniformly 
expanding/contracting solution ($q=0$) and exists for $\zeta_* \neq 0$ 
and $Q \neq \pm 1$.
    \begin{itemize}
        \item If $-2<\zeta_*<2$, it behaves as saddle
        \item If $\zeta_*=-2$, and for $\mu'(\zeta)>0$, 
it behaves as 
a 
saddle. We are interested in the case $\mu'(\zeta)<0$. 
$\mu'(\zeta)=0$,
however,
 is a more particular case 
that we 
shall 
rule out. There is also the possibility of choosing 
this value to coincide with another eigenvalue, and it is possible 
that the system may not be diagonalizable. If this is the case, 
the variation matrix must be expressed in its Jordan form to 
separate the spaces associated with each eigenvalue. We do not 
intend to analyze these special cases here.
        \begin{itemize}
            \item For $\mu'(\zeta)<0$ only one direction 
(locally $Q$) shows  attractive behavior.
        \end{itemize}
        Note that for this value of $\zeta_*$ , $D$ does not 
match $Q \neq -1$; however, for this value it matches $C_{2-}$, 
which is covered in depth in appendix \ref{appendixa}
        \item If  $\zeta_*=2$, and for $\mu'(\zeta)<0$ it
 behaves as 
a 
saddle. Therefore, we are interested in the case 
$\mu'(\zeta)>0$. Besides, there may be exceptions equivalent 
to the previous case that we 
shall 
not analyze.
        \begin{itemize}
            \item For $\mu'(\zeta)>0$ only one direction 
(locally $Q$) shows  repulsive behavior.
        \end{itemize}
          Note that for this value of $\zeta_*$ , $D$ does 
not match $Q \neq 1$; however, for this value it 
does match
$C_{2+}$,
which
is covered in depth in appendix \ref{appendixa}.
    \end{itemize}
\item $L_{1+}$ is a critical line that corresponds to a 
non-accelerating radiation-dominated universe, it never 
represents a stable line (note that, as shown in Table I, the matter component is dominant, and in the evaluated cases, only the case with $\omega=1/3$ is possible.). 
\item $L_{2+}$ is a critical line that corresponds to a 
universe dominated by mimetic dark matter, and its behavior 
is similar to 
that of 
$L_{1+}$. 
\item $L_{1-}$ and $L_{2-}$ are the contracting analogs of 
lines $L_{1+}$ and $L_{2+}$, respectively.
\item $P_+$ is a non-hyperbolic unstable critical plane that 
describes a non-accelerating universe dominated by baryonic 
plus mimetic dark matter, and $P_-$ is its contracting analog.
\end{itemize}

\subsection{Dynamics in invariant manifolds}
\label{sub}

Given the existence of principal invariant submanifolds and 
under the good behavior of the system variables in them, it 
becomes possible to gain some level of understanding of the 
neighborhoods of the submanifolds by analyzing them. For this 
reason, the submanifolds $\beta =0$ , $\zeta=0$ and 
$\tilde{\Omega}_{\rho}=0$ are good indicators.
The first two do not depend on the shape of the potential, 
therefore, 
their dynamics 
is independent 
of the mimetic field potential. 
In contrast,
$\tilde{\Omega}_{\rho}$ 
and $Q=\pm 1$ could only be analyzed given the potential. 
However, the submanifolds $Q=\pm 1$ do not represent their 
neighborhoods well. Instead, they constitute inaccessible 
exceptions, and they will not be 
the subject of 
further analysis.

\subsubsection{Submanifold $\zeta=0$}
\begin{figure}[H] 
\begin{center}
\includegraphics[width=0.5\textwidth]{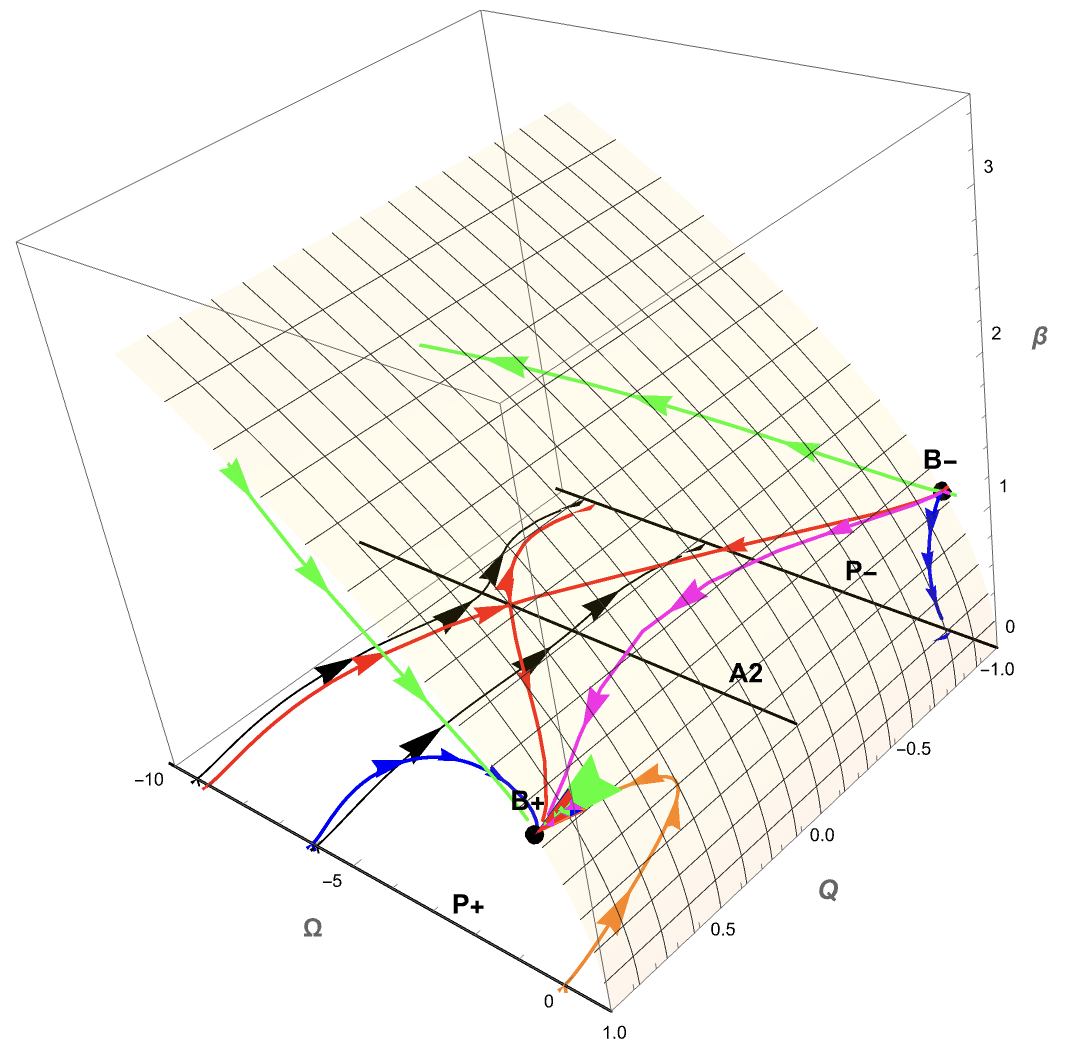}
\caption{Phase portrait of the invariant submanifold $\zeta=0$ 
for $\omega=0$, here \( \Omega \) is the dynamical variable \( \tilde{\Omega}_{\lambda} \). Some trajectories starting from the 
matter-dominated 
critical line $P_+$ as $\tau \rightarrow - \infty$ 
converge to the dark energy dominated point $B_+$ as 
$\tau \rightarrow \infty$, while others after some expansion 
begin contracting and collapse to a big crunch at $P_-$. Some 
trajectories which are past asymptotic to $B_-$ evolve to the 
future attractor $B_+$, while others evolve to $P_-$.}
\label{Fig1}
\end{center}
\end{figure}

\begin{figure}[H] 
\begin{center}
\includegraphics[width=0.5\textwidth]{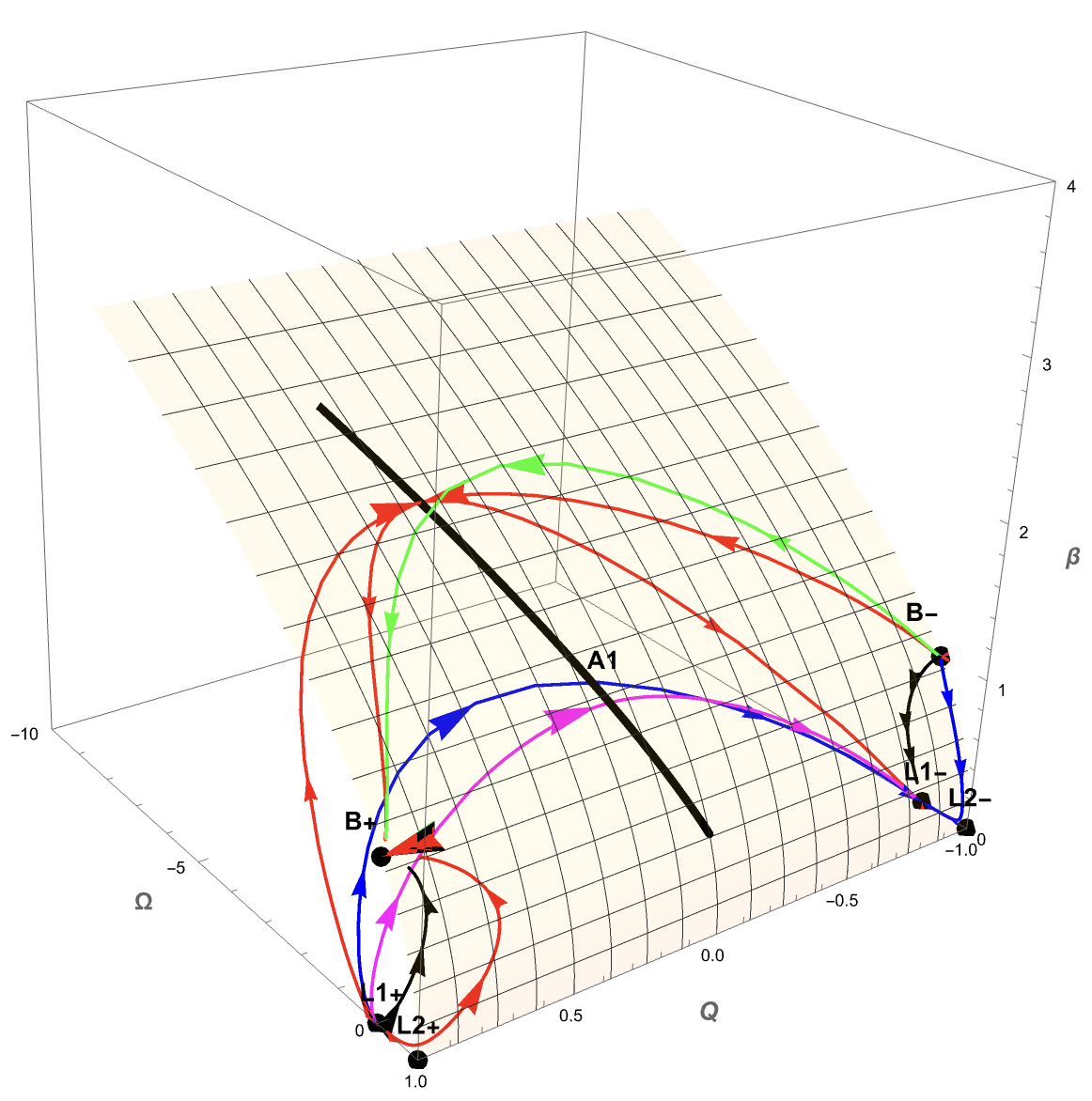}
\caption{Phase portrait of the invariant submanifold $\zeta=0$ 
for $\omega=1/3$, here \( \Omega \) is the dynamical variable \( \tilde{\Omega}_{\lambda} \). Some trajectories starting from the ordinary 
matter-dominated critical line $L_{1+}$ and from the dark matter 
dominated critical line $L_{2+}$ as $\tau \rightarrow -\infty$ 
converge to the dark energy dominated point $B_+$ as 
$\tau \rightarrow \infty$, while others after some expansion 
begin contracting and collapse to a big crunch at $P_-$. Some 
trajectories which are past asymptotic to $B_-$ evolve to the 
future attractor $B_+$, while others evolve to $L_{1-}$ or $L_{2-}$.}
\label{Fig2}
\end{center}
\end{figure}

The critical points and their respective dynamics in the 
submanifold are detailed in Figs. \ref{Fig1} and \ref{Fig2}:
\begin{itemize}
    \item For $\omega =0$: $A_2$ exhibits saddle behavior, 
$P_+$ displays repulsive behavior in directions other than 
the critical axis and $P_-$ is the contracting analogue of $P_+$.
    \item For $\omega= 1/3$: $A_1$ shows saddle behavior, 
$L_{1+}$ exhibits repulsive behavior, $L_{2+}$ exhibits 
saddle behavior, $L_{1-}$ shows attractor behavior and 
$L_{2-}$ displays saddle behavior.
    \item For both values of $\omega$: $B_+$ shows attractor 
behavior and $B_-$ exhibits repulsive behavior.
\end{itemize}

In Figs.~\ref{Fig1}~and~\ref{Fig2} we depict the phase 
portraits of the invariant submanifold $\zeta=0$ for $\omega=0$ 
and $\omega=1/3$, respectively. The phase space is divided 
into two halves, one corresponding to a contracting epoch 
($Q<0$) and the other to an expanding epoch ($Q>0$).
For $\omega=0$ we can see that certain trajectories 
that 
are past asymptotic to the matter-dominated critical line 
$P_+$, with a sufficiently small value of $\Omega_{\phi}$ 
after some expansion enter the contracting phase space and 
collapse to a big crunch at $P_-$. 
There are also
models starting from $P_+$ with a 
high  
enough value of 
$\Omega_{\phi}$ that evolve to the future attractor $B_+$, 
which represents an expanding de Sitter accelerated solution 
dominated by the mimetic field.
Some trajectories which are past asymptotic to $B_-$ evolve 
to point $B_+$, while others evolve to $P_{-}$. There are  
also trajectories which are past asymptotic as well as future 
asymptotic to the Einstein static solution, line $A_2$. 
Similar behavior is found for the case $\omega=1/3$, with 
the matter-dominated critical line $P_{+}$/$P_-$ replaced by 
the points $L_{1+}$/$L_{1-}$ (ordinary matter dominated) 
and $L_{2+}$/$L_{2-}$ (mimetic matter dominated) and $A_2$ 
by $A_1$.


The behavior of the variable $\zeta$ for some points close 
to the submanifold depends almost exclusively on the potential. 
By explicitly analyzing the behavior of $\zeta$ as a function
 of $t$, one can gain insights into the curves and expect a 
relationship between the behaviors with cosmological time and 
transformed time.
However, for more precision, it is advisable to examine the 
dynamics around the points belonging to the submanifold by
using 
either
the eigenvalues of the variation matrix, if applicable, 
or the center manifold, if necessary.
Lastly, 
whether the submanifold is reachable from outside
will depend on the potential. 

\subsubsection{Submanifold $\beta=0$}

\begin{figure}[H]
 \centering{
    \includegraphics[width=0.4\textwidth]{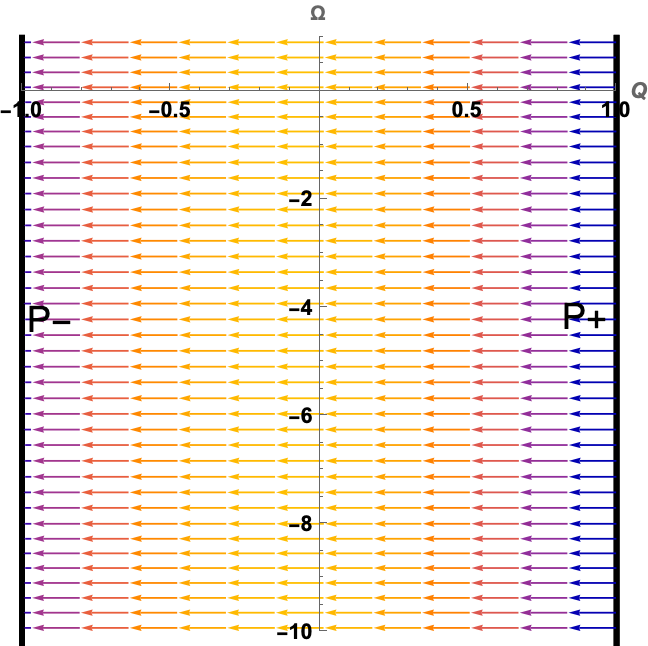}}{
    \includegraphics[width=0.4\textwidth]{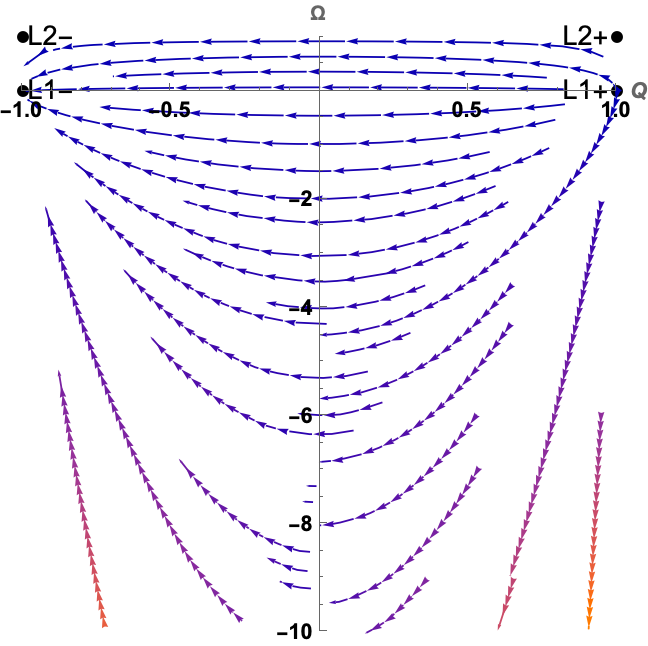}}
 \caption{Phase portraits of the invariant submanifold $\beta=0$ 
for $\omega=0$ (left panel) and $\omega= 1/3 $ (right panel), \( \Omega \) is the dynamical variable \( \tilde{\Omega}_{\lambda} \). 
Trajectories starting from the matter-dominated line $P_+$ as 
$\tau \rightarrow -\infty$ end at $P_-$ as $\tau \rightarrow \infty$. 
Analogously, trajectories starting from the ordinary matter-dominated 
line $L_{1+}$ or from the dark matter dominated line $L_{2+}$ end at 
$L_{1-}$ or $L_{2-}$.}
 \label{Fig3}
\end{figure}

The critical points and their corresponding dynamics in the 
submanifold are illustrated in Fig. \ref{Fig3}:
\begin{itemize}
    \item For $\omega = 0$ : $P_+$ exhibits repulsive behavior 
in the noncritical axes, while $P_-$ displays attractive behavior 
in the noncritical axes.
\item For $\omega = 1/3$ : $L_{1+}$ exhibits repulsive behavior, 
$L_{1-}$ has attractive behavior, and both $L_{2+}$ and  $L_{2-}$ 
behave as saddles.
\end{itemize}

Note that for $\beta=0$, the equation for $\zeta$ vanishes 
(assuming that $\zeta^2 (\mu(\zeta)-3/2)$ does not diverge for 
the conditions in which $\beta$ vanishes); 
the contribution of $\zeta$ to the remaining system is 
also 
null; therefore, it is 
enough to graph the coordinates $(Q,\tilde{\Omega}_{\lambda})$.
In Fig. \ref{Fig3} we plot the phase portrait of the invariant 
submanifold $\beta = 0$ for $\omega=0$ and $\omega = 1/3$. The 
trajectories observed are past asymptotic to a matter-dominated 
critical line which after some expansion enter the contracting 
phase space and collapse to a big crunch.

Analysis of the behavior on the $\beta$ axis, which is 
important 
for
determining
whether 
a trajectory stays in the neighborhood 
of the submanifold, is difficult 
owing
to the dependence of $\beta$ on $V$ and $D$.
It is possible to obtain this information by analyzing the 
behavior in the neighborhood  of the critical points belonging 
to the submanifold. For instance, for $P_-$ the behavior is 
attractive, while for $P_+$ it is repulsive along this axis.
Again, regarding whether the submanifold can be accessed from an 
external point depends on the potential.

\subsection{Specific potentials}
\label{particular}

In this section we explore the dynamics of the system for two 
specific potentials of the mimetic field: the inverse square 
potential and the exponential potential.

\subsubsection{Inverse square potential}
\label{section}

Here, we consider the potential $V(\phi) = \gamma / \phi^2$, 
where $\gamma$ is a coupling constant.  In this case we find 
that $\mu(\zeta) = 3/2$ and $\zeta = \frac{2}{\kappa \gamma^{1/2}}$ 
become constants; therefore, the dynamical system reduces to a 
three-dimensional autonomous dynamical system given by 
Eqs. (\ref{Q})-(\ref{beta}) with $\zeta$ constant.
\newline

The critical points of the system are shown in Table \ref{tabla3} 
and the eigenvalues of the Jacobian matrix 
that
characterize their 
dynamical behavior are shown in Table \ref{tabla4}. Note that the 
Einstein static solutions $A_1$ and $A_2$,
and also the solutions $B_+$ and $B_-$, disappear for this 
potential.
\newline

\begin{table}[H]
\begin{center}
\caption{Fixed points of the dynamical system (\ref{Q})-(\ref{z}) 
and some physical quantities for the inverse square potential. 
In the formulas below $\zeta = \frac{2}{\kappa \gamma^{1/2}}$.} \label{tabla3}
\resizebox{\columnwidth}{!}{
\begin{tabular}{|c|c|c|c|c|c|c|c|c|c|} 
\hline
Label & $\tilde{\Omega}_{\lambda}$ & $\beta$ & $Q$ & Existence & $\Omega_{\rho}$ & $\Omega_{\lambda}$ & $\Omega_{\phi}$ & $\Omega_{k}$ & $q$ \\
\hline
 $C_{1+}$ & $-\frac{3(1+\omega)^3}{\omega \zeta^2}$ & $\frac{(1+\omega) \sqrt{3}}{\zeta}$ & 1 & $\omega \neq 0$, $\zeta > 0$ &  $\frac{3+\omega (\zeta^2 +6 +3 \omega)}{\zeta^2 \omega}$  & $-\frac{3(1+\omega)^3}{\omega \zeta^2}$ & $\frac{3(1+\omega)^2}{\zeta^2}$ & 0 & $\frac{1}{2}(1+3\omega)$ \\
  $C_{1-}$ &  $-\frac{3(1+\omega)^3}{\omega \zeta^2}$  &  $-\frac{(1+\omega) \sqrt{3}}{\zeta}$   &  -1 &  $\omega \neq 0$, $\zeta < 0$  & $\frac{3+\omega (\zeta^2 +6 +3 \omega)}{\zeta^2 \omega}$  &  $-\frac{3(1+\omega)^3}{\omega \zeta^2}$  &  $\frac{3(1+\omega)^2}{\zeta^2}$  & 0 &  $\frac{1}{2}(1+3\omega)$  \\
 $C_{2+}$  & $\frac{1}{6} \zeta ( \sqrt{12+\zeta^2} -\zeta)$ & $\frac{\sqrt{12+\zeta^2}-\zeta}{ 2 \sqrt{3}}$ & 1 &  & 0 & $\frac{1}{6} \zeta ( \sqrt{12+\zeta^2} -\zeta)$ &  $\frac{1}{12} (\sqrt{\zeta^2 +12}-\zeta)^2$  & 0 & $\frac{1}{4} \zeta (\sqrt{\zeta^2+12}- \zeta) -1$ \\
 $C_{2-}$ & $-\frac{1}{6} \zeta(\sqrt{12+\zeta^2}+\zeta)$ & $\frac{\sqrt{12+\zeta^2}+\zeta}{ 2 \sqrt{3}}$ & -1 &  & 0 & $-\frac{1}{6} \zeta (\sqrt{12+\zeta^2}+\zeta)$ & $\frac{1}{12} (\sqrt{\zeta^2+12}+\zeta)^2$ & 0 & $-\frac{1}{4} \zeta(\sqrt{\zeta^2+12}+\zeta)-1$ \\  
 $D$ & $\frac{2}{3}$ & $\frac{1}{\sqrt{3}}$ & $\frac{\zeta}{2}$ & $\zeta \neq 0$, $Q \neq \pm 1$ & 0 & $\frac{8}{3 \zeta^2}$ & $\frac{4}{3 \zeta^2}$ &  $1-\frac{4}{\zeta^2}$  & 0 \\   
 $L_{1+}$ & 0 & 0 & 1 & & 1 & 0 & 0 & 0 & $\frac{1}{2}(1+3\omega)$ \\
 $L_{2+}$ & 1 & 0 & 1 & & 0 & 1 & 0 & 0 & $\frac{1}{2}$ \\
 $L_{1-}$ & 0 & 0 & -1 & & 1 & 0  & 0 & 0 & $\frac{1}{2}(1+3 \omega)$ \\
 $L_{2-}$ & 1 & 0 & -1 & & 0 & 1  & 0 & 0 & $\frac{1}{2}$ \\ 
  $P_+$  &  $\tilde{\Omega}_{\lambda} \leq 1$ &  0  &  1  & $\omega = 0$ & $1-\Omega_{\lambda}$  & $\Omega_{\lambda} \leq 1$  &   0 & 0  &  $\frac{1}{2}$ \\
  $P_-$ &  $\tilde{\Omega}_{\lambda} \leq 1$  &  0  &   -1  & $\omega = 0$  &  $1-\Omega_{\lambda}$  & $\Omega_{\lambda} \leq 1$  & 0  & 0  & $\frac{1}{2}$ \\ \hline
\end{tabular}
}
\end{center}
\end{table}
\begin{table}[H]
\begin{center}
\caption{Eigenvalues and stability of fixed points of the 
dynamical system (\ref{Q})-(\ref{z}) for the inverse square 
potential. In the formulas below we have defined $\xi_{\pm}(\zeta) 
= \zeta^2 \pm \zeta \sqrt{\zeta^2+12} +12 \omega$.} \label{tabla4}
\resizebox{\columnwidth}{!}{
\begin{tabular}{|c|c|c|c|c|} 
\hline
Label & $\lambda_1$ & $\lambda_2$ & $\lambda_3$ &  Dynamical Character \\
\hline
$C_{1+}$ & $-\frac{3 \left((1-\omega)\zeta+\sqrt{24(1+\omega)^3+(1+3 \omega)^2 \zeta^2}\right)}{4\zeta}$ &  $-\frac{3 \left((1-\omega)\zeta-\sqrt{24(1+\omega)^3+(1+3 \omega)^2 \zeta^2}\right)}{4\zeta}$   & $1+3 \omega$ &  saddle   \\ 
$C_{1-}$ & $\frac{3 \left((1-\omega)\zeta+\sqrt{24(1+\omega)^3+(1+3 \omega)^2 \zeta^2}\right)}{4\zeta}$ &  $\frac{3 \left((1-\omega)\zeta-\sqrt{24(1+\omega)^3+(1+3 \omega)^2 \zeta^2}\right)}{4\zeta}$   &  $-(1+3 \omega)$ &  saddle   \\
 $C_{2+}$ & -$\frac{3}{8} \left( \xi_-(\zeta) + 8 (1- \omega) +\frac{1}{3} |\xi_-(\zeta)| \right) $ &  -$\frac{3}{8} \left( \xi_-(\zeta) + 8 (1- \omega) -  \frac{1}{3} |\xi_-(\zeta)| \right)$  &   $ -2 + \frac{1}{2} \zeta (-\zeta + \sqrt{\zeta^2 +12}) $ & attractor  ($\zeta<2$), saddle ($\zeta>2$) \\ 
 $C_{2-}$ &  $\frac{3}{8} \left( \xi_+ (\zeta) + 8 (1- \omega) - \frac{1}{3} |\xi_+(\zeta_*)| \right) $  & $\frac{3}{8} \left( \xi_+(\zeta) + 8 (1- \omega) + \frac{1}{3} |\xi_+(\zeta)| \right) $  &  $ 2 + \frac{1}{2} \zeta (\zeta + \sqrt{\zeta^2 +12}) $   & repulsor\\ 
 $D$ & $1-\frac{\zeta}{2}$  & $-1-\frac{\zeta}{2}$ & $-\frac{1}{2} \zeta (1+3\omega)$  &  saddle for $-2 < \zeta < 2$ \\ 
 $L_{1+}$ & $3 \omega$ & $\frac{3 (1+\omega)}{2}$ & $1+3\omega$  & unstable \\ 
 $L_{2+}$ & 1  &  $-3 \omega$ & $\frac{3}{2}$ & unstable for $\omega = 0$, saddle for $\omega \neq 0$ \\ 
 $L_{1-}$ & $-3 \omega$ & $-\frac{3}{2}(1+\omega)$ & $-(1+ 3\omega)$ &  \\ 
 $L_{2-}$ & -1 & $3 \omega$ & $-\frac{3}{2}$ & saddle for $\omega \neq 0$ \\  
  $P_+$ & 0  & $\frac{3}{2}$ &  1  & unstable  \\ 
 $P_-$ & 0  & $-\frac{3}{2}$ &  -1  &  \\ 
 \hline
\end{tabular}
}
\end{center}
\end{table}

To identify an attractor, it should be noted that the points $C_{2+},L_{1-},P_{-}$, and, under certain circumstances, some points at infinity can act as attractors.

To understand the global nature of the model, it is 
necessary essentially to make a conformal compactification 
of the entire phase space of the solutions. The critical 
points at infinity can be identified by projecting onto the 
Poincar\'e sphere, as outlined in appendix \ref{appendixc}. In this case, the given projection is necessary, as it reveals significant contributions to the system's dynamics, including predominantly saddle behavior and the potential presence of attractors.

Now, it is convenient to write the dynamical equations as:
\begin{equation}
    \tilde{\Omega}_{\lambda}' = \hat{P}(\tilde{\Omega}_{\lambda}, \beta, Q) \,, \,\,\,\,\, \beta'
 = \hat{Q}(\tilde{\Omega}_{\lambda}, \beta, Q) \,, \,\,\,\, Q' = \hat{R}(\tilde{\Omega}_{\lambda}, \beta, Q) \,,
\end{equation}
where $\hat{P}$, $\hat{Q}$ and $\hat{R}$ are polynomial 
functions of the dynamical variables. Since $Q$ is compact, 
it is enough to project only the variables 
$\tilde{\Omega}_{\lambda}$ and $\beta$ on the Poincar\'e 
sphere. Therefore, we define the following coordinates on 
the Poincar\'e sphere:
\begin{equation}
    X = \frac{\tilde{\Omega}_{\lambda}}{\sqrt{1+ 
\tilde{\Omega}_{\lambda}^2+\beta^2}}\,, \,\,\,\, Y = 
\frac{\beta}{\sqrt{1+ \tilde{\Omega}_{\lambda}^2+\beta^2}} \,, \,\,\,\,   
Z = \frac{1}{\sqrt{1+ \tilde{\Omega}_{\lambda}^2+\beta^2}} \,.
\end{equation}
So
at infinity ($Z \rightarrow 0$),
 the dynamical system has the form:
 \begin{eqnarray}
 \nonumber X' &=& - Y (X \hat{Q}^* - Y \hat{P}^*) \,,  \\
 \nonumber Y' &=& X (X \hat{Q}^* -Y \hat{P}^*) \,, \\
 Q' &=& \hat{R}^*  \,,
\end{eqnarray}
where the polynomials $\hat{P}^*$, $\hat{Q}^*$ and $\hat{R}^*$ 
are given by:
\begin{eqnarray}
\nonumber \hat{P}^* &=& \sqrt{3} z Y^3 - 3 (1+ \omega) Q X Y^2 \,, \\
\nonumber \hat{Q}^* &=& -\frac{3}{2} (1+ \omega) Q Y^3 \,, \\
 \hat{R}^* &=& \frac{3}{2} (1+ \omega) (1-Q^2) Y^2 \,.
\end{eqnarray}
Therefore, the critical points/lines $(X, Y, Q)$ at infinity are:

\noindent $P_1$: $(\pm 1, 0, Q)$, which corresponds to critical lines
\newline
$P_2$: $(-\cos \theta, \sin \theta, -1)$ and its 
antipodal $(\cos \theta, -\sin \theta, -1)$
\newline
$P_3$: $(\cos \theta, \sin \theta, 1)$ and its antipodal 
$(-\cos \theta, -\sin \theta, 1)$
\newline

\noindent where $\theta = \arctan \left( \frac{\sqrt{3} (1+ \omega)}{ 2 \zeta} \right)$. 
However, it is straightforward to verify that only the critical line 
$P_1$: $(-1, 0, Q)$ is located within the physical 
space $\tilde{\Omega}_{\lambda} \leq 1- \beta^2$.
Nevertheless, 
analysis of the dynamics near this 
line is 
rather
difficult because,
owing to either the compactified variables or the time 
rescaling,
 the compactification 
can end up canceling the dynamics of 
those components 
not at infinity.
Graphically, however, the variables not 
projected on the Poincar\'e sphere show dynamics.
This should be interpreted as indicating that the variables 
with canceled dynamics will undergo 
slower evolution 
than those without. Therefore, the equations are analyzed 
with the aid of graphics 
the
better 
to
understand and visualize 
these dynamics.
\newline

\begin{figure}[ht] 
\begin{center}
\includegraphics[width=0.8\textwidth]{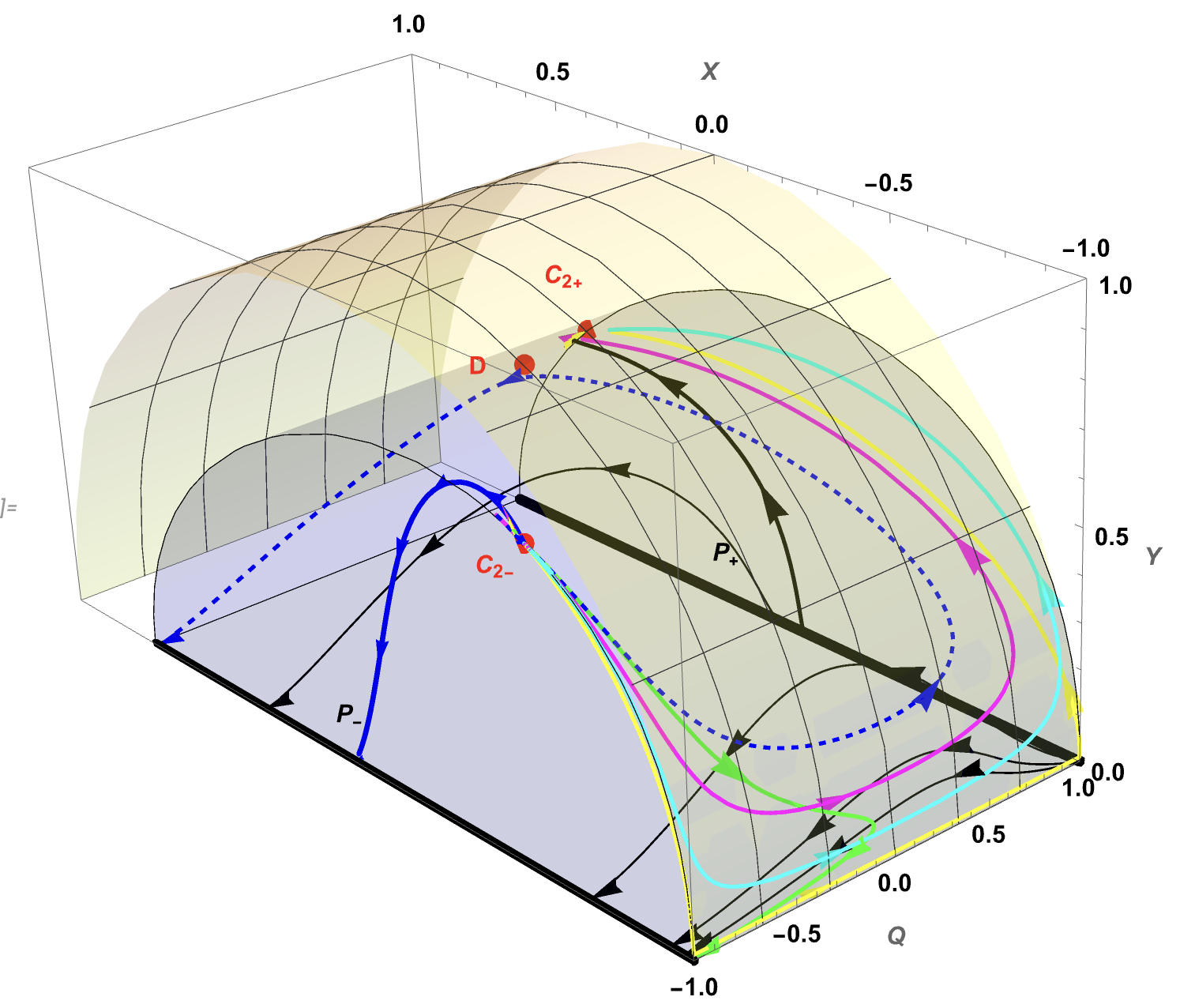}
\caption{Global phase portrait for the inverse square 
potential with $\zeta = 3/2$. The solutions of the system 
are $C_{2+}$, $C_{2-}$, $D$, $P_{+}$, $P_{-}$ and the points 
$X=-1$, $Y=0$, $Q=\pm 1$. Some trajectories starting from the 
matter-dominated critical line $P_+$ converge asymptotically 
to $C_{2+}$,
 which represents a solution dominated by dark
matter and dark energy. We also observe trajectories starting 
from the contracting solution $C_{2-}$ which converge to 
the point $C_{2+}$. In addition, some trajectories after 
some expansion begin contracting and collapse to a big crunch 
at $P_-$.}
\label{Fig4}
\end{center}
\end{figure}

In Fig.~\ref{Fig4} we plot the global phase portrait of 
the dynamical system for $ \kappa \gamma^{1/2} = 4/3$. 
This figure shows that the phase space is divided into 
two halves, one corresponding to a contracting epoch ($Q<0$) 
and the other to an expanding epoch ($Q>0$). The dynamical 
system is bounded by the $Y=0$ ($\beta =0$), $Q=\pm 1$, and 
$\tilde{\Omega}_{\rho}=0$ invariant submanifolds.
Some trajectories which are past asymptotic to the matter 
dominated critical line $P_+$ with a sufficiently small 
value of $\Omega_{\phi}$ after some expansion enter the 
contracting phase space and collapse to a big crunch at 
$P_-$. Conversely, models starting from $P_+$ with a large 
enough value of $\Omega_{\phi}$ evolve to the future 
attractor $C_{2+}$, which represents a solution dominated 
by dark matter and dark energy, exhibiting acceleration 
for $\zeta <2$ (Table \ref{tabla3}).
%
There are trajectories past asymptotic to $C_{2-}$ 
that evolve to $C_{2+}$. Other trajectories past asymptotic 
to $C_{2-}$ enter the contracting phase space and 
recollapse to a big crunch at $P_-$. Additionally, we observe 
trajectories that approximate to the uniformly expanding 
solution, point $D$ with coordinates 
$X=1/2$, $Y=\sqrt{3}/4$, $Q=\zeta / 2$, and then collapse 
to a big crunch.  Note
that points $C_{2+}$, $C_{2-}$ 
and $D$ lie on the $\rho=0$ boundary. On this boundary, 
there exists a trajectory, among others, that is past 
asymptotic to point $C_{2-}$ and evolves to $D$.
\newline

For the case \( \omega = 0 \)  , the analysis presented below, it is concluded that  the system will exhibit saddle behavior with the 
critical point at $(Q,X,Y)=(\pm 1,-1,0)$. However, it is 
important to note that the possibility of a small and constant 
$\beta$ is also contained within this critical point. In particular, 
for $\beta<\frac{1}{\sqrt{3}}$ and under certain circumstances, 
such as $\zeta>0$ and $Q<0$, $Q'<0$ will occur, causing 
$(Q,X,Y)=(-1,-1,0)$ to act as an attractor, while the point 
$(Q,X,Y)=(1,-1,0)$ in this example would require even more 
specific conditions and could act as a repulsor. The necessary 
conditions for the attractor and repulsor behavior will not be 
further explored, but it can be shown from a simple analysis 
of the equations.
It suffices to conclude that the system truly exhibits critical 
behavior at the points $(Q,X,Y)=(\pm 1,-1,0)$ and can exhibit 
saddle behavior or attractor and repulsor behavior, respectively, 
for the points $Q=-1$ and $Q=1$, depending on the initial 
conditions of the solution.
On the other hand, for \( \omega = 1/3 \), the points act as 
saddles, with $Q=\pm 1$ representing the real critical points. 
Based on this observation, it is possible to say that there will 
be two main trajectories, both interacting with the saddle 
points but in different directions. Additionally, there might 
be a third type of trajectory that undergoes a change in direction. 
It is unlikely that trajectories leading to attractor--repulsive 
points will emerge, and the analysis conducted so far lacks the 
capability to demonstrate or visualize such occurrences.
\newline
To analyze the variations, it is convenient to perform the 
projection described in 
appendix  \ref{appendixd}, given 
that $Q$ varies slowly in relation to $\Omega_\lambda$, and 
$\beta$, and $\zeta$ remains constant. The following change 
of variable is considered: \newline
\begin{equation}
    \xi = -\frac{Y}{X} \,, \,\,\,\,\, \chi =-\frac{Z}{X} \,.
\end{equation}
For these variables and $\omega=0$ the system of equations 
is the following:
\begin{equation}
    Q' = \frac{1}{2}(1-Q^2)(3\xi^2-\chi^2) \,, \,\,\,\,\, \xi' 
=\frac{1}{2} \xi\left(3Q(\chi^2+\xi^2)+\sqrt{3} \zeta \xi (2 \xi^2-\chi)\right) \,, \,\,\,\,\, \chi'
=\chi \xi^2 \left(3Q+\sqrt{3}\zeta\xi \right) \,,
\end{equation}
and the variation matrix for the line $(-1,0,Q)$ is null; 
therefore, its eigenvalues are null
and correspond to a 
degenerate line
that
will be analyzed according to the method 
exposed in appendix \ref{appendixe}.
Now
note that the solution is independent of $Q$,
 so it should not show dynamics in the neighborhood (remember that 
in this context it means that it varies more slowly than 
the other variables), so we could consider it as a constant, 
just 
as
$\zeta$ for the potential is constant.
Using the angular equation in the neighborhood of interest, 
it is possible to see that the solutions are $\theta=0,\pi/2$,
 and additionally, for $\frac{\zeta}{Q}<0$, the solution is 
$\theta=\frac{1}{2} \text{arccot} \left(-\frac{\zeta}{2\sqrt{3}Q}\right)$. 
So, reviewing the radial equation in the neighborhood, it is possible
  evaluate the behavior (if the radial component is also 
canceled, the analysis must be discarded as insufficient).
Therefore, for the last solution it 
can be said that it is:
\begin{itemize}
    \item repulsive for $Q>0$ and $\zeta<0$.
    \item attractive for $Q<0$ and $\zeta>0$.
\end{itemize}
the $\theta=0$ solution is:
\begin{itemize}
    \item repulsive for $Q>0$.
    \item attractive for $Q<0$.
\end{itemize}
and the $\theta=\pi/2$ solution does not exhibit radial 
dynamics. At this order the radial behavior is annulled, 
so higher-order
terms must be considered; however, 
graphically it can be seen that the evolution is angular 
and not radial, so it is discarded.

As previously mentioned, it is still necessary to assess 
the dynamics of the variable $Q$ and 
not disregard it. 
It is easy,
then,
 to see that for large $\beta$, $Q' > 0$. 
%
%
\newline

Equivalently, for $\omega=1/3$ , the equations are:
\begin{eqnarray}
\notag Q' &=& \frac{1}{2}(1-Q^2)(4\xi^2-2\chi^2-\chi) \,, \\  
\notag \xi' &=& \frac{1}{2} \xi\left(Q (4\xi^2+2\chi^2-\chi)+\sqrt{3} \zeta \xi (2 \xi^2-\chi)\right) \,, \\
\chi' &=& \sqrt{3}\zeta \xi^3 \chi -Q \chi (\chi+\chi^2-4 \xi^2) \,.
\end{eqnarray}

The variation matrix for the analyzed point have null eigenvalues, 
so its structure is evaluated using polar coordinates (see 
appendix \ref{appendixe}). From this,
 the conclusions are 
slightly more difficult to analyze analytically. Firstly, 
in the plane $(\xi,\chi)$ the angles $\theta=0, \pi/2$ are 
obtained, then its dynamics will be the following:
\newline
1. For $\theta=0$, the solution is:
\begin{itemize}
    \item attractive for $Q>0$.
    \item repulsive for $Q<0$.
    \end{itemize}
2. For $\theta=\pi/2$, the solution is
 \begin{itemize}
     \item repulsive for $Q>0$.
     \item attractive for $Q<0$. 
 \end{itemize}
Secondly,
it is necessary to note that the sign of $Q'$ will be 
described such that $sign(Q')=sign(4\xi^2-2\chi^2-\chi)$. This 
expression also seems to divide the regions of influence of 
the different
behaviors

%
%

\subsubsection{Exponential potential}
\label{potential}

Now, we consider the potential $V(\phi) = V_0 e^{-\alpha \phi}$. 
For this potential we find $\mu(\zeta)-3/2=-1/2$ ; therefore, 
the points $C_{1+}$, $C_{1-}$, $C_{2+}$, $C_{2-}$ and $D$ 
listed in Table \ref{tabla1} do not manifest in the phase 
space for this model. All the remaining points/lines correspond 
to non-accelerating solutions, with the exception of point 
$B_+$,
which represents
a saddle de Sitter point dominated by the 
mimetic field. This point exhibits unstable behavior in the 
$\zeta$ direction.
Point $B_-$ is the contracting analogue of $B_+$. However, 
it is worth 
noting
that the invariant submanifold $\zeta=0$
 divides the phase space into two regions, 
thus
preventing trajectories 
from crossing between them. From Eq.\ (\ref{z}), we deduce that
 if $\mu(\zeta )-3/2 < 0$ as $\zeta \rightarrow 0$, then 
$\zeta' >0$ and point $B_+$ acts as a future attractor for
 trajectories in the region $\zeta <0$.

No attempt will be made to make graphics 
that describe the solutions in 4-dimensional space, since, 
regardless of the method for this, it might not be visual 
enough to be interpreted. However, it is possible to have 
some degree of understanding by analyzing the invariant 
submanifolds
and extrapolating the behavior to their neighborhoods.
That being said, the $\zeta=0$ and $\beta=0$ manifolds have a 
general qualitative behavior independent of the shape of the 
potential, so their behaviors are similar to that described 
in subsection \ref{sub}, added to the analysis of the 
$\tilde{\Omega}_{\rho}=0$ submanifold, which 
it 
is now possible 
to analyze since the potential is known. For this potential, 
the variable $\zeta$ is given by 
$\zeta= \frac{\alpha}{ \kappa \sqrt{V}}$, with $V = V_0 e^{- \alpha t}$.
Therefore,  $\alpha < 0$ implies $\zeta<0$, and 
$\zeta \rightarrow 0$ when $t \rightarrow \infty$. 
However, 
$\alpha > 0$ implies $\zeta>0$, and 
$\zeta \rightarrow \infty$ when $t \rightarrow \infty$, 
it
being necessary to consider the critical points at infinity.
\newline

The
critical points and their respective dynamics in 
the manifold are 
then
detailed in Figs. \ref{Fig7} and \ref{Fig8}:
\begin{itemize}
    \item For both values of $\omega$: $B_+$ exhibits saddle 
behavior for centre manifold, with $\zeta>0$ representing the 
unstable direction, and $B_-$ shows saddle behavior for centre 
manifold, with $\zeta<0$ as the stable direction. However, as
 previously mentioned, the point $B_+$ behaves as an attractor 
for trajectories in the region $\zeta<0$, while manifesting 
saddle behavior for trajectories in the region $\zeta>0$, and 
only the $\zeta$ direction is unstable.
    \item For $\omega=0$: $A_2$ displays saddle behavior, 
    but $A_2$ is a line where one of its ends touches the 
submanifold, since we are interested in the neighborhood it 
must be considered,
    so, the influence on the subsystem is minimal, $P_+$ shows 
repulsive behavior in directions other than the critical axis, 
while $P_-$ is the contracting analogue of $P_+$.
\begin{figure}[H] 
\begin{center}
\includegraphics[width=0.5\textwidth]{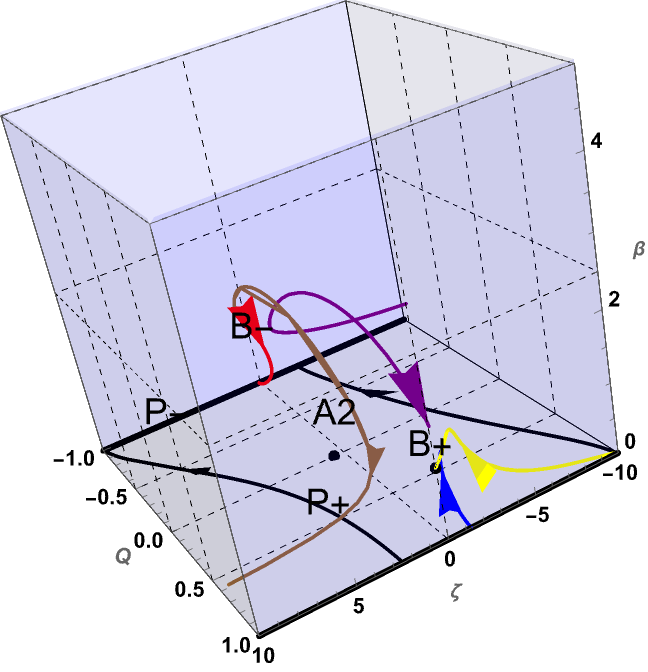}
\caption{Phase portrait of the invariant submanifold 
$\tilde{\Omega}_{\rho}=0$ for $\omega=0$. Some trajectories in 
the region $\zeta <0$ starting from the matter dominated critical 
line $P_+$ as $\tau \rightarrow -\infty$ converge to the dark 
energy dominated point $B_+$ as $\tau \rightarrow \infty$. In the 
region $\zeta >0$ we observe trajectories that evolve to some 
critical point at $\zeta \rightarrow \infty$. Also, there are 
trajectories which are past asymptotic to $P_+$ and collapse to 
a big crunch at $P_-$.}
\label{Fig7}
\end{center}
\end{figure}
    \item For $\omega=1/3$: $L_{2+}$ exhibits repulsive 
behavior in directions other than the critical axis, $L_{2-}$ 
exhibits attractive behavior in directions other than the critical axis,
 and $A_1$ shows saddle behavior. It is important to note 
that $A_1$ is a curve where one of its ends touches the submanifold, 
as we are interested in the neighborhood it must be considered, 
so, the influence on the subsystem is minimal.
\begin{figure}[H] 
\begin{center}
\includegraphics[width=0.5\textwidth]{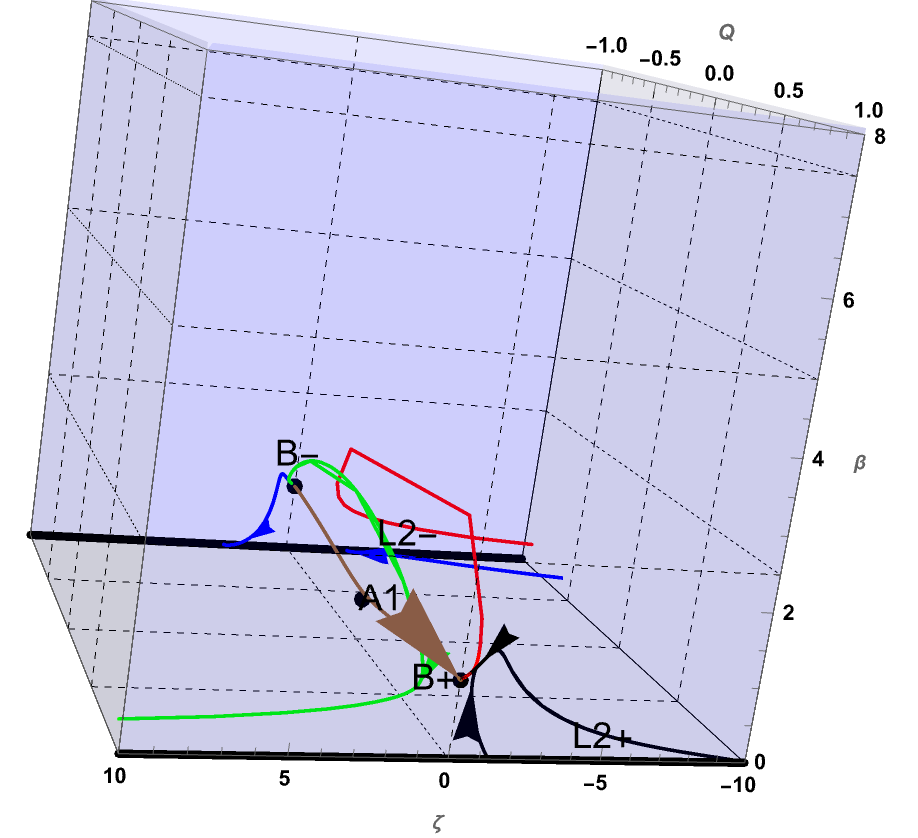}
\caption{Phase portrait of the invariant submanifold 
$\tilde{\Omega}_{\rho}=0$ for $\omega=1/3$. Some trajectories 
in the region $\zeta <0$ starting from the dark matter dominated 
critical line $L_{2+}$ as $\tau \rightarrow -\infty$ converge to 
the dark energy dominated point $B_+$ as $\tau \rightarrow \infty$.  
In the region $\zeta >0$ we observe trajectories that evolve to some 
critical point at $\zeta \rightarrow \infty$.}
\label{Fig8}
\end{center}
\end{figure} 
\end{itemize}

The stability in the $\tilde{\Omega}_{\rho}$ axis does not 
exhibit
a unique behavior throughout the submanifold and, since 
this term depends on $\rho$ and $D$, the analysis is not easy. 
It is evident,
however,
 that the submanifold is 
inaccessible from an external point since $\rho$ must be null 
for this, so it is 
possible 
only
if $\rho$ is null by default. \\

The critical points at infinity can be found by projecting the 
variables $\tilde{\Omega}_{\lambda},\beta, \zeta$ on the 
Poincar\'e sphere, see appendix \ref{appendixc}. In this case 
the procedure applied above is generalizable. Here, the equations 
have the particularity that $\mu(\zeta)=1$, and the coordinates on 
the Poincar\'e sphere are given by:
\begin{eqnarray}
   \notag X &=& \frac{\tilde{\Omega}_{\lambda}}{\sqrt{1+ \tilde{\Omega}_{\lambda}^2+\beta^2+\zeta^2}}\,, \,\,\, Y = \frac{\beta}{\sqrt{1+ \tilde{\Omega}_{\lambda}^2+\beta^2+\zeta^2}} \,, \\
    Z &=& \frac{\zeta}{\sqrt{1+ \tilde{\Omega}_{\lambda}^2+\beta^2+\zeta^2}} \,, \,\,\,  W = \frac{1}{\sqrt{1+ \tilde{\Omega}_{\lambda}^2+\beta^2+\zeta^2}}   \,.
\end{eqnarray}
At infinity ($W \rightarrow 0$), the dynamical system transforms to:
 \begin{eqnarray}
 \nonumber X' &=& \sqrt{3} Y^3 Z (Y^2+Z^2)\,,  \\
 \nonumber Y' &=& -\sqrt{3}X Y^4 Z \,, \\
 \nonumber Z' &=& -\sqrt{3}X Y^3 Z^2 \,, \\
 Q' &=& 0  \,,
\end{eqnarray}
we have omitted showing the form of the functions for each 
derivative after the projection, for simplicity.
In the same way as before, the solution inside of the physical 
space ($X,Y,Z,Q$) can be described as $P_4$ : (X,0,Z,Q), with 
the condition $X^2+Z^2=1$.

Once again, it becomes evident that the variable $Q$ will be 
eliminated in the limit as $W \rightarrow 0$, as it is not 
included in the set of the projected variables. Despite this, 
a thorough analysis of the equations will be conducted and 
supported by graphics for enhanced comprehension.
On the other hand, it is easy to notice that the solution is 
indifferent to the values of $X,Z$,
 which,
 if it is part of the 
coordinates in the Poincar\'e sphere, 
implies a lack of 
dynamics in those directions locally, mainly on $ Z$ since 
$X$ is associated with $\tilde{\Omega}_{\lambda}$ and this 
must still respond to the limitations of the physical phase 
space.

Accordingly, to analyze the variations, it is convenient to 
carry out the projection described in appendix \ref{appendixd}
 since $Q$ varies slowly with respect to $\tilde{\Omega}_{\lambda}$ , 
$\beta$ and $\zeta$. A priori, we considered a new projection tangent to \( (Q,X,Y,Z) = (Q,-1,0,0) \). However, the results were inconclusive, generally indicating some unstable directions and lacking a clear pattern in the relationship between these variables and \( Q \).

However, in this analysis,
taking into consideration
the choice made for the point 
in the 
tangential 
space, the points (Q,X,Y,Z)=(Q,0,0,$\pm$1) have been 
left out, despite the expectation that the behavior described in 
the analysis could be extrapolated to these points. However, it is 
anticipated that this is not the case.


It is possible to make a  projection in the 
tangential space at 
$Z=1$ (it is enough to analyze only one point since $Z=-1$ is 
its antipode). This analysis is important because this point appears to be graphically significant for the system and is also relevant for curves that are  compatible with our universe.
To go directly from the original coordinates to 
the final ones, the following relations are used:
\begin{equation}
    Q = Q\,, \,\,\,\, \Tilde{\Omega}_{\lambda}= 
\chi/\epsilon \,, \,\,\,\,   \beta=\xi/\epsilon \,, \,\,\,\, \zeta=1/\epsilon \,.
\end{equation}

\begin{figure}[H] 
\begin{center}
\includegraphics[width=0.6\textwidth]{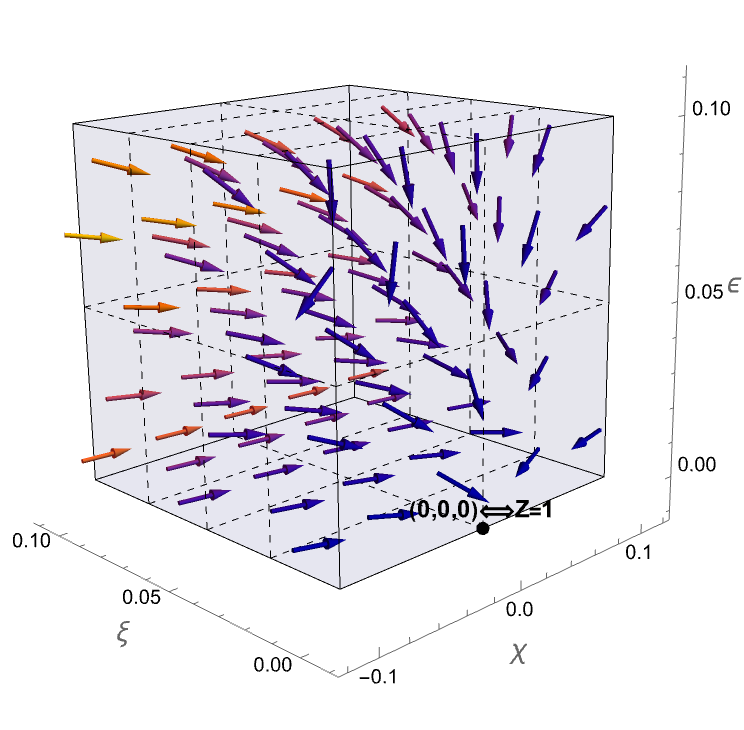}
\caption{Behavior around the point $Z=1$ for $\omega=0,1/3$. 
Note that the graphic is independent of the values of $\omega$ 
and $Q$.}
\label{Z=1 behavior}
\end{center}
\end{figure}

Rewriting the equations in these new variables with the respective 
time rescaling (which is evident when explicitly written), it is 
found that the variation matrix is null, 
which 
indicates
a degenerate point. The procedures detailed in appendix \ref{appendixe} for 
slightly degenerate
points are applied. Unfortunately, these 
procedures do not provide detailed insights into the structures 
around this point; 
we 
therefore 
proceed to show graphics around 
these points, where in the new variables they correspond to the 
points $(Q,\chi,\xi,\epsilon)=(Q,0,0,0)$, see
figure \ref{Z=1 behavior}.

Is possible to see an attractive behavior in all the components, 
except for $Q$, which requires a separate analysis. 
\newline

\begin{figure}[H] 
\begin{center}
\includegraphics[width=0.8\textwidth]{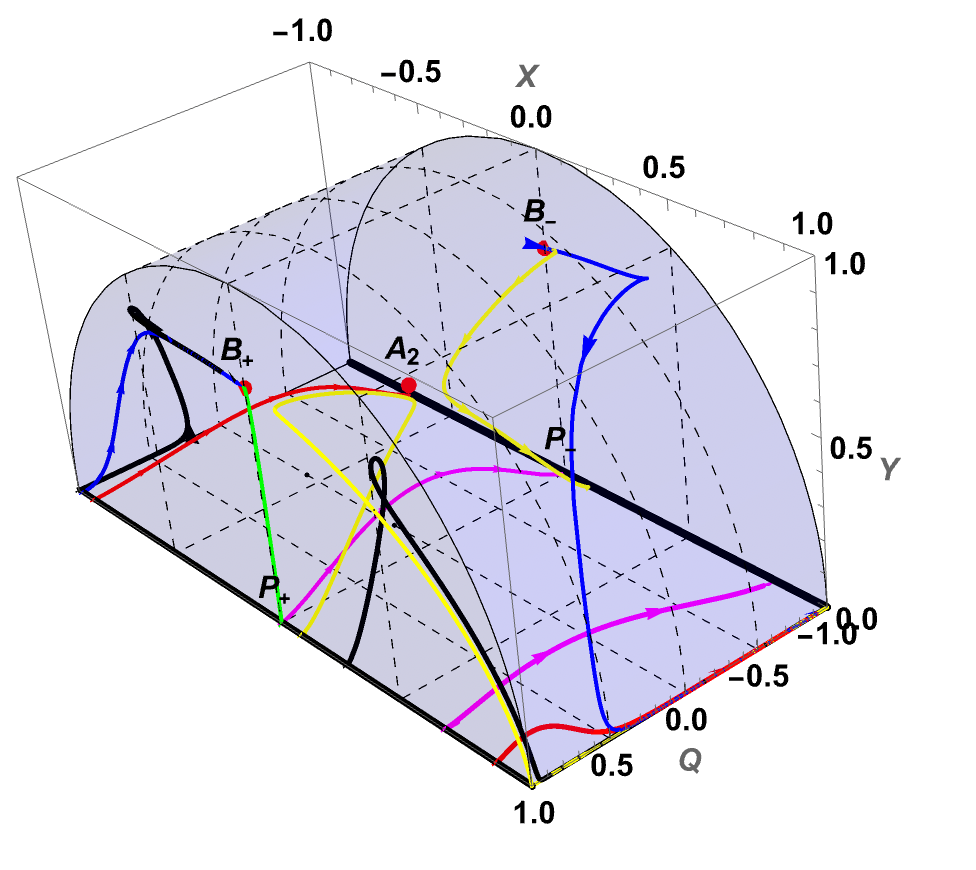}
\caption{Global phase portrait of the reduced dynamical system for 
an exponential potential with $\omega = 0$. The solutions of 
the system are $B_{+}$, $B_{-}$, $A_{2}$, $P_{+}$, $P_{-}$ and 
the points $X=1$, $Y=0$, $Q= \pm 1$. Some trajectories in the 
region $X <0$ ($\zeta <0$) which are past asymptotic to the matter 
dominated critical line $P_+$ evolve to the mimetic field 
dominated point $B_+$. Some trajectories in the region $X>0$ 
($\zeta >0$) which are past asymptotic to the matter-dominated 
critical line $P_{+}$ evolve to the critical point at infinity 
$X=1$, $Y=0$, $Q=1$ and then move to $X=1$, $Y=0$, $Q=-1$. Also, 
there are trajectories that, after some expansion, begin contracting 
and collapse to a big crunch at $P_-$. Moreover, the green curve has been added based on the parameters for the exponential potential obtained from the combined $PaSNe+OHD$ dataset, as listed in Table \ref{tab:results_parameters}. This curve represents accelerated expansion that seems to end in a de Sitter solution. However due to numerical accuracy limitations, it is only possible to represent the segment from $P_+$ to $B_+$. Due to instability in this direction, the curve is expected to move towards $(Q,X,Y) = (1,1,0)$, which represents a similar solution. Finally, the transition to $Q = -1$ is not observed, and based on the evaluation of Figure \ref{NumI}, it does not appear to occur.}
\label{phaseportrait1}
\end{center}
\end{figure}

It is worth noting
from Eqs.\ (\ref{Q})-(\ref{z}) with 
$\omega = 0$ that the $\tilde{\Omega}_{\lambda}'$ equation 
decouples. This leads to a 
three-dimensional 
reduced dynamical system for the variables $Q$, $\zeta$ and $\beta$. So
it is 
possible to graph the global phase space of this reduced dynamical 
system. By projecting the coordinates $\zeta$ and $\beta$ on the 
Poincar\'e sphere we define:
\begin{equation}
    X = \frac{\zeta}{\sqrt{1+\beta^2 + \zeta^2}}\,, \,\,\,\, Y = 
\frac{\beta}{\sqrt{1+\beta^2 +\zeta^2}} \,, \,\,\,\,   Z = \frac{1}{\sqrt{1+\beta^2 +\zeta^2}} \,.
\end{equation}
In Fig. \ref{phaseportrait1} we plot the global phase space 
of the reduced dynamical system characterized by the 
coordinates $Q$, $X$ and $Y$ for the exponential potential. 
The phase space is divided into two halves, one corresponding 
to a contracting epoch ($Q<0$) and the other representing an 
expanding epoch ($Q>0$). We can 
see
that, for different 
values of $X$, some trajectories which are past asymptotic to 
the matter-dominated critical line $P_+$, with a sufficiently 
small value of $\Omega_{\phi}$ after some expansion enter the 
contracting phase space and collapse to a big crunch at $P_-$.
By way of contrast,
in the region $X < 0$ ($\zeta<0$), there 
are models starting from $P_+$ with a large enough value of 
$\Omega_{\phi}$ that evolve to the future attractor $B_+$, 
which represents an expanding de Sitter accelerated solution 
dominated by the mimetic field. In the region $X>0$ ($\zeta>0$), 
there are models starting from $P_+$ with a large enough value 
of $\Omega_{\phi}$ that approximate to the point $B_+$, 
evolve to the point $X=1$, $Y=0$ $Q=1$ (note that $X=1$ coincides 
with $Z=1$ in the previous analysis) at infinity, and then move 
to the point $X=1$, $Y=0$, $Q=-1$.
Also, in $X>0$, we observe trajectories that are past asymptotic 
to $B_-$ and evolve to the point $X=1$, $Y=0$, $Q=-1$ at infinity. 
There are trajectories which are past asymptotic, as well as 
future asymptotic to the Einstein static solution $A_2$, 
characterized by coordinates $X=0$, $Y=1/2$, $Q=0$.

In this context, it can be demonstrated that in the vicinity 
of infinity (where $X=1$ in the last-defined coordinates), 
$Q'<0$ holds for any $Q$, indicating that only $(Q,X,Y)=(-1,1,0)$ 
would act as an attractor, while $(Q,X,Y)=(1,1,0)$ would be a 
saddle point. 
There are special conditions 
in which
$Q'>0$ can be added, 
but they are transient. To understand this, it must be taken 
into account that the form of $Q$ and the temporal rescaling 
suggest that the submanifolds $Q= \pm 1$ are inaccessible, so 
we eventually return to the first case. In summary, some curves 
may experience a transitory asymptotic approach to $(Q,X,Y)=(1,1,0)$ 
of long duration.
However, this may not be numerically appreciable since, as 
previously mentioned, $Q=1$ is an invariant submanifold, and, 
owing 
to  lack of precision, a point in the vicinity might be
approached as one of the submanifolds. The next analysis provides 
a thorough exploration of the behavior of the system:
\begin{align*}
    Q'=(1-Q^2)\frac{3\beta^2-1}{2} \quad &\triangle_Q \quad 0 \\
    3\beta^2-1 \quad &\triangle_Q \quad 0 \\
    \beta^2 \quad &\triangle_Q \quad 1/3
\end{align*}
    Here, $\triangle$ denotes some inequality sign, and 
$\nabla$ its inversion, and
\begin{align*}
    \beta'=-\frac{\sqrt{3}}{2}\beta^2 \zeta - \frac{3}{2}\beta Q (\beta^2-1) \quad &\triangle_\beta \quad 0 \\
    -\beta \zeta -\sqrt{3}Q(\beta^2-1)\quad &\triangle_\beta \quad 0\\
    \beta \zeta +\sqrt{3}Q(\beta^2-1)\quad &\nabla_\beta \quad 0\\
    \zeta \quad &\nabla_\beta \quad -\sqrt{3}Q\frac{\beta^2-1}{\beta} \,.
\end{align*}
We indicate that:
\begin{itemize}
    \item Assuming 
that
$\beta$ is finite (in this context, 
$\beta \ll \zeta$), then $\nabla_\beta \rightarrow >$ implies 
$\triangle_\beta \rightarrow <$, which in turn implies 
$\beta' < 0$. Therefore, with sufficient time at infinity, 
$\beta^2 < 1/3 \Rightarrow \triangle_Q \rightarrow <$, leading 
to $Q' < 0$.
    \item If we consider $\beta$ as a variable at infinity 
(assuming it is in the vicinity of $(X,Y)=(1,0)$, so in the 
best-case scenario $\beta \gg 0$ and $\beta \ll \zeta$), then:
\begin{align*}
    \zeta \quad &\nabla_\beta \quad -\sqrt{3}Q \beta + ...
\end{align*}
Then, under the assumption of being in the mentioned neighborhood, 
it must hold that $\nabla_\beta \rightarrow >$. Following the 
same reasoning as in the previous point, this implies 
$\beta' < 0$. With sufficient time, $\beta^2<1/3$, and 
consequently, $Q'<0$. Note that the second condition may not 
be very evident, but by contradiction, if it is assumed that 
$\zeta \propto \beta$, it is straightforward to conclude that 
$(X,Y) \neq (1,0)$.
\item Still, there appears to be one apparent option, and it 
arises when noticing that the inequality expression to evaluate 
for $\zeta$ diverges at $\beta=0$. However, if enough time has 
elapsed
for $Q<0$, then the expression will transform back to 
$\zeta>-\infty$, thus recovering the previous conclusions.
If the assumption about $Q$ is not true, perhaps some point in 
finite space could be dominating the trajectory; however, the 
curves seem to indicate the opposite.
\end{itemize}

This brief analysis does not constitute a formal proof but 
rather provides arguments to estimate its behavior. Therefore, 
the possibility that some trajectory asymptotically approaches 
$Q=1$ should not be ruled out. Unfortunately, numerical evaluation 
does not yield significantly better results.

\subsubsection{Bounce behavior}
Regarding the possibility of finding bounce-type solutions, we could consider universes that contract and then expand. It is necessary to evaluate the possibility of this behavior. To do so, one can pay special attention to the transitions from contraction to expansion solutions; several examples of this can be found in Article \cite{Leon_2013}. Specifically, these universes would have \( (a, \dot{a}, \ddot{a}) = ( \text{positive}, \text{negative or } -\infty, \text{preferably positive}) \) initially and end with \(  (a, \dot{a}, \ddot{a}) = (\text{positive} ,\text{positive or } \infty, \text{preferably positive}) \). In some specific cases, we observe curves satisfying \( Q=-1 \rightarrow Q=1 \), which are good candidates for the aforementioned conditions and will be analyzed in conjunction with the other variables they might imply.

It is important to note that the time transformation \( d\tau = D dt \) allows the parameter \( \tau \) to potentially  diverge when \( a=0 \) or \( \dot{a}=\pm \infty \). Consequently, a bouncing solution might not be fully represented if any of the above conditions are met. Moreover, it is known that bounce-type solutions are not possible in certain scenarios, such as those discussed in the article \cite{Fadragas_2014}. However, in this specific case, the presence of positive curvature and the possibility of the scalar field (acting as dark matter) to violate the weak energy condition are considered. These conditions make the manifestation of bounce behaviour possible.

In addition to considering the variable \( Q(\tau) \), we should also examine the variable \( \beta(\tau) \). Assuming that we know both \( Q(\tau) \) and \( \beta(\tau) \) for a given curve for all \( \tau \) --at least numerically--  it is possible to determine the behaviour of the scale factor \( a(\tau) \) for a general potential as follows:

\begin{equation}
    a(\tau)=\frac{\sqrt{3k}}{\kappa}\frac{1}{\sqrt{V(t(\tau))}}\frac{\beta}{\sqrt{1-Q^2}}
\end{equation}
where \( t = t(\tau) \) can be obtained by isolating \( D \) from the definition of \( \beta \) and then using it to solve \( d\tau = D dt \). This allows \( t = t(\tau) \) to be determined. Consequently, the next step involves solving:
\begin{equation}
    \int{\frac{\sqrt{3}\beta}{\kappa} d\tau}=\int{\sqrt{V(t)} dt}
\end{equation}
In these expressions, the assumption that \( a > 0 \) has been made, allowing \( \dot{a} = \dot{a}(Q) \) to be determined from the definition of \( Q \). Although the specific form of \( a(\tau) \) will depend significantly on the potential, it is natural to seek a bouncing solution that does not start at \( a=0 \) (i.e., does not begin in contraction), and therefore, preferably does not start at \( \beta=0 \).

Therefore, for specific cases:
\begin{itemize}
    \item For the inverse square potential, the curves satisfying the conditions are the pink, cyan, and yellow curves. All of these curves transition from the points \( C_{2-} \) to the point \( C_{2+} \).
    \item For the exponential potential, although it is more general, these curves are not directly visible. However, in the submanifold \( \tilde{\Omega_{\lambda}} = 0 \) (see Figs. 7 and 8), both curves going from point \( B_- \) to point \( B_+ \) and curves that approach or depart from infinity while passing near these points can be observed. In this case, it is necessary to evaluate \( a(\tau) \) to confirm the behavior.
\end{itemize}
\section{Mimetic field potential analysis}
\label{sec:potential}

In this section we obtain numerical solutions for the differential 
equations, and explore the dynamics of the Hubble parameter and 
the scale factor for different mimetic field potentials. Moreover, 
we compare these solutions with those of the non-flat 
$\Lambda$CDM model.

It is useful to write the cosmological equations as function of 
redshift $z$ to facilitate the comparison of the models with 
the observational data. The relationship between redshift and 
scale factor is given by:
\begin{equation}
1+ z = \frac{a(t_0)}{a(t)} \,,
\end{equation}
where $t_0$ refers to the present time which will be set to zero.
The derivative with respect to cosmic time of this expression results in:
\begin{equation} \label{zt}
\frac{dz}{dt} = -(1+z)H \,.
\end{equation}

Expressed in terms of the parameter $z$, the second Friedmann 
equation can be written as:
\begin{equation}
\label{eq:Hz}
\frac{dH}{dz} = \frac{3H}{2(1+z)} - \frac{\kappa^2}{2(1+z)H} V(t) - 
\frac{\Omega_{k} H_0^2 (1+z)}{2 H} \,,
\end{equation}
where $t = t(z)$ and the potential $V(t)$ is parametrically 
considered as $V(z)$. The function $t(z)$ is determined from 
the equation:
\begin{equation} 
\label{eq:tz}
\frac{dt}{dz}= -\frac{1}{(1+z)H}
\end{equation}
with the initial condition $t(0)=0$. Additionally, the first 
Friedman equation (\ref{density}) evaluated at the present time 
yields the following constraint among the density parameters:
\begin{equation}
1 = \Omega_{\rho,0}+ \Omega_{\lambda,0}+ \Omega_{\phi,0}+ \Omega_{k,0} \,.
\end{equation}
Here, the subscript $0$ indicates that the respective density
 parameter is evaluated at the present time.
\newline

\begin{figure}[H] 
\begin{center}
\includegraphics[width=0.45\textwidth]{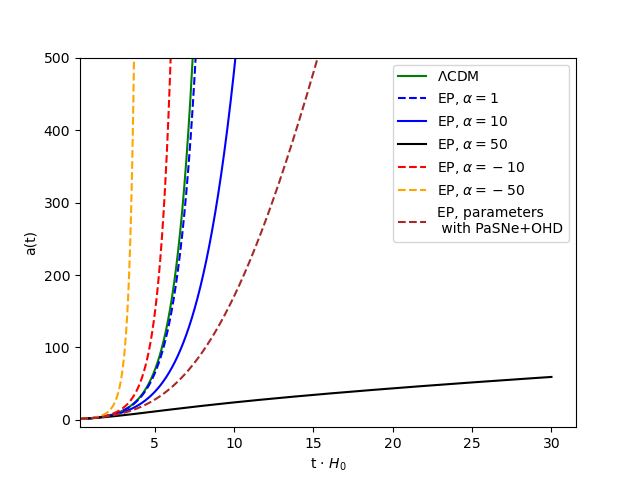}
\includegraphics[width=0.45\textwidth]{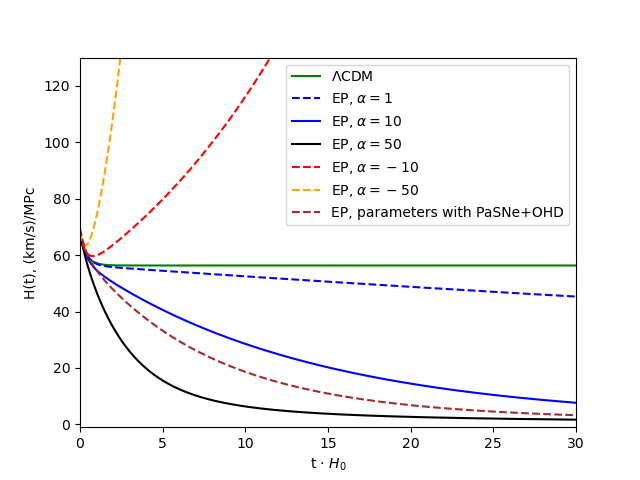}
\caption{$H(t)$ vs $t \cdot H_0$ (left panel) and $a(t)$ vs 
$t \cdot H_0$ (right panel) for the exponential potential 
with $\alpha= 1, 10, 50, -10, -50$. The dashed brown curves have been fully evaluated using the parameters from Table \ref{tab:results_parameters}, i.e., \( H_0 = 69.49 \), \( \alpha = 18.48 \), \( \Omega_{\phi,0} = 0.699 \), \( \Omega_{k,0} = -0.02 \).}
\label{NumI}
\end{center}
\end{figure}

\begin{figure}[H] 
\begin{center}
\includegraphics[width=0.45\textwidth]{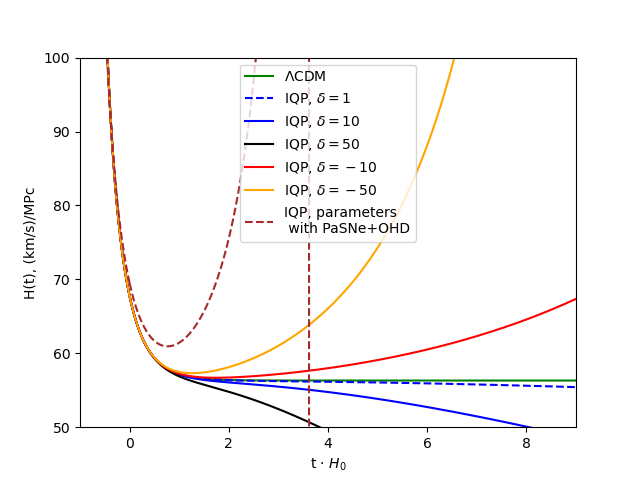}
\includegraphics[width=0.45\textwidth]{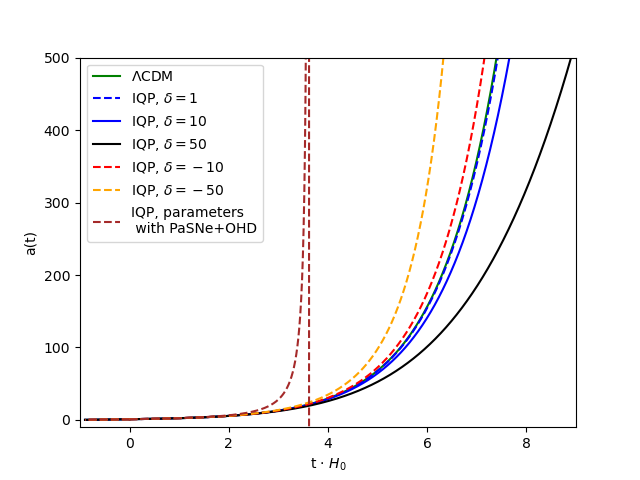}
\caption{$H(t)$ vs $t \cdot H_0$ (left panel) and $a(t)$ 
vs $t \cdot H_0$ (right panel) for the inverse quartic 
potential with $\delta = 1, 10, 50, -10, -50$. The dashed brown curves have been fully evaluated using the parameters from Table \ref{tab:results_parameters}, i.e., \( H_0 = 69.41 \), \( \delta = -368,8 \), \( \Omega_{\phi,0} = 0.696 \), \( \Omega_{k,0} = 0.035 \).}
\label{NumII}
\end{center}
\end{figure}

We 
analyze two family of models for the mimetic field 
potential. The first family comprises exponential potentials 
(EP) defined as $V(\phi) = V_0 e^{-\alpha \phi}$. The second 
involves inverse quartic potentials (IQP) expressed as 
$V(\phi) = V_0 (1+ \delta \phi^2)^{-2}$. 
This potential with a fixed $\delta=1$ was previously considered 
in Ref.\ \cite{Chamseddine:2014vna} to construct solutions 
featuring a non-singular bounce in a contracting flat universe. 
By utilizing the definition of the density parameter 
$\Omega_{\phi}$, we establish the relationship between $V_0$ and 
the parameters $H_0$ and $\Omega_{\phi,0}$ as 
$V_0= 3 H_0^2 \Omega_{\phi,0}$. In Figs.~\ref{NumI}~and~\ref{NumII} 
we respectively plot the behavior of the Hubble parameter and 
the scale factor as a function of the dimensionless cosmic 
time $t \cdot H_0$.
As initial conditions, we used $a_0=1$ and 
$H_0 = 67.74\, \rm km \, s^{-1} \, Mpc^{-1}$, and
$\Omega_{\phi,0} = 0.6911$ and $\Omega_{k,0} = -0.001$ as the 
values of the density parameters at the present time. For 
illustration, we consider different values of $\alpha$ and $\delta$ 
for each family. 
We have also included the non-flat $\Lambda$CDM 
solution for comparison, which is analogous to setting $\alpha=0$ 
of $\delta=0$ in the potentials under consideration. The density 
parameter associated 
with 
radiation was not considered in the
 analysis. We observe that the solutions are consistent with 
late-time acceleration, and the deviation from the $\Lambda$CDM 
model increases when $\alpha$ or $\delta$ increases. 
The 
deviation of the models from $\Lambda$CDM is more sensitive to 
changes of the parameter $\alpha$ than $\delta$. For the exponential 
potential, we see that the dynamical evolution of the system 
depends on the sign of the parameter $\alpha$ (Fig.~\ref{NumI}), 
which agrees with the dynamical system analysis performed in 
Sec.\ (\ref{potential}). The numerical solutions of Fig.\ \ref{NumI} 
suggests that $H \rightarrow \infty$ for 
$\alpha < 0$, $H \rightarrow$ constant for $\alpha = 0$ and 
$H \rightarrow 0$ for $\alpha > 0$. In fact, for $\alpha < 0$, 
we observe that several trajectories in the region $X<0$ evolve 
to the point $B_+$ (see Fig.\ \ref{phaseportrait1}), which corresponds 
to an accelerating solution dominated by the mimetic field 
$\Omega_{\phi} = 1$ and $H \rightarrow \infty$ when the trajectories 
approximate to $B_+$ in this region. For $\alpha > 0$ there are 
also trajectories in the region $X>0$ that approximate to the 
point $B_+$, but in this case $H \rightarrow 0$; however, 
the trajectories 
then
approximate to the point $X=1$, $Y=0$, $Q=1$ at 
infinity, and then go to the point $X=1$, $Y=0$, $Q=-1$.
Analogously, for the inverse quartic potential, the evolution of 
the system depends on the sign of the parameter $\delta$
(see Fig. \ref{NumII}). The numerical solutions of Fig.\ \ref{NumII} 
suggest
that $H \rightarrow \infty$ for $\delta < 0$, $H \rightarrow$ 
constant for $\delta = 0$ and $H \rightarrow 0$ for $\delta > 0$. 
For this potential we obtain 
$\zeta = \frac{4 \delta \phi}{\kappa V_0^{1/2}}$ with 
$1+ \delta \phi^2 >0$, which shows that $\delta > 0$ implies
 $\zeta>0$ and the potential $V(\phi) = V_0 (1+ \delta \phi^2)^{-2}$ 
tends to zero and $\zeta \rightarrow \infty$ when $t \rightarrow \infty$. 
For $\delta < 0$ we note that the variable $\zeta < 0$ and the potential 
diverges at the times $t = \pm\sqrt{-1/\delta}$.  We 
may 
 infer from the general analysis of the dynamical system of 
section \ref{dyn} for arbitrary potential that the trajectories 
evolve to the point $C_{2+}$, where $\Omega_{\phi}=1$ and 
$H \rightarrow \infty$, and the solution represents a big rip.
\newline



\section{Observational analysis}
\label{sec:description_model_Hz}
This section is devoted to exploring
the parameter space of 
the cosmological models proposed in section~\ref{sec:potential} 
using public available observational data.
To this end, the Observational Hubble Database from the 
cosmic chronometers \citep[OHD,][]{Magana:2017nfs} and the 
Pantheon database 
for Type Ia supernovae
\cite[PaSN,][]{Pan-STARRS1:2017jku} are considered.
The fit of the data is performed with a Bayesian statistical 
analysis using the algorithm called \textsc{emcee} 
\cite{Foreman-Mackey:2012any}, which implements an 
affine-invariant Markov Chain Monte Carlo (MCMC) sampler 
of the posterior distribution function (PDF) 
to minimize
the following expression:
\begin{equation}
\ln\left( \mathcal{L}(\theta) \ \prod_j \mathcal{P}(\theta_j) \right)  \propto -\dfrac{\chi^2(\theta)}{2}
+ \sum_{j} \ln\left(\mathcal{P}(\theta_j)\right)
\label{eq:likelihood_sed}
\end{equation}
where $\mathcal{L}(\theta)$ is the likelihood and 
$\mathcal{P}(\theta_j)$ identifies the prior imposed 
on the $j$th parameter.
$\theta$ identifies the set of free parameters considered in 
the MCMC approach.
The merit function, $\chi^2(\theta)$, depends on the differences 
between observed data ($\textbf{D}$) and model predictions 
($\textbf{m}(\theta)$) over the redshift:
\begin{equation}
\chi^2(\theta) = \big(\textbf{D} - \textbf{m}(\theta) \big)^T \textbf{C}^{-1} \big(\textbf{D} 
- \textbf{m}(\theta) \big)
\label{eq:chi2_sed}
\end{equation}
$\textbf{C}$ being the noise covariance matrix of the observed data.
In this notation, $\textbf{D}$ and $\textbf{m}(\theta)$ 
are vectors while $\textbf{C}^{-1}$ is the inverse matrix 
of $\textbf{C}$.
The value of each parameter and its uncertainty are obtained 
thought marginalization of the posterior distribution function 
of each parameter, as the median (the 50th percentile) and 
recovering the 16th and 84th percentiles to provide respectively 
the upper and lower uncertainties.

Each cosmological model is tested with three merit functions that
depends on the data set considered, which are based on the 
observational Hubble data only (OHD, $\chi^2_{\rm OHD}$), on the 
Pantheon data set data only (PaSNe, $\chi^2_{\rm SNIa}$) and on a 
combination of the two (OHD+PaSNe, $\chi^2_{\rm OHD+PaSNe}$).
Concerning the simultaneous fit, the merit function is written as:
\begin{equation}
\label{eq:xi_joint}
\chi^2_{\rm OHD+PaSNe}(\theta) = \chi^2_{\rm OHD}(\theta) + 
\chi^2_{\rm PaSNe}(\theta)  
\end{equation}
for which we assume that measurements from different databases 
are uncorrelated.
Details about $\chi^2_{\rm OHD}$ and $\chi^2_{\rm PaSNe}$ are 
presented in the next sections.

\subsection{Data and modelling}

\subsubsection{The Observational Hubble Database}

The Observational Hubble Database, OHD, consists of 51 measurements 
of the Hubble parameter that covers the redshift range 
0.07--2.36 \citep[][]{Magana:2017nfs,Moresco:2022phi},
which are obtained from the differential age method 
\citep[DA,][]{Jimenez:2001gg,Moresco:2022phi} and from the 
baryon acoustic oscillation measurements.
These OHD values can be compared directly with our proposed 
cosmological models that yield $H(z;\Theta)$ solving equations 
(\ref{eq:Hz}) and (\ref{eq:tz}) (section~\ref{sec:potential}), 
with $\Theta$ being the set of cosmological parameters that define 
our proposed models.
Although some correlation 
is
expected for the BAO measurements
owing to 
overlapping among the galaxy samples or the 
cosmological model dependence in the clustering estimation 
\citep[see discussion in][]{Magana:2017nfs}, we follow 
previous work that assume no correlation between these 
measurements \citep[e.g.][]{Corral:2020lxt}. 
Therefore, the covariance noise matrix is diagonal 
(i.e.\ uncorrelated noise, $({\bf C}^{-1})_{i,j} = \delta_{i,j}/\sigma_i^2$), 
and consequently the maximization of the posterior distribution 
is driven by the merit function:
\begin{equation}
    \chi_{\rm OHD}^2(\theta) = \sum_{i = 1}^{51} \left (\frac{ H_{\textrm{obs},i} - 
H(z_i;\theta)}{\sigma_{H_{\rm obs,i}}} \right)^2 \,,
\end{equation}
where $H_{\textrm{obs},i}$ and $\sigma_{H_i}$ are respectively 
the observed Hubble parameter and its uncertainty at redshift 
$z_i$ (OHD values).
$H(z_i;\theta)$ is the Hubble parameter provided by the model 
at the redshift $z_i$, 
and
$\theta$ 
is
the set of free 
parameters. In this case, $\theta$ corresponds to the set of 
cosmological parameters, i.e.\ $\theta = \Theta$.
In our implementation, we applied the explicit Runge-Kutta method 
of order 8 to obtain the solution of $H(z_i;\theta)$ for each point 
explored in the parameter space.

\subsubsection{The Pantheon Supernovae Database}

The Pantheon data set, PaSNe, contains the apparent magnitude of 
1048 
SNe\,Ia
in
the redshift range 
0.01--2.3
\cite{Pan-STARRS1:2017jku}.
These observational measurements are fitted to the apparent magnitude 
model, $m_b(z_i;\theta)$, which is written as 
\citep{SNLS:2011lii, Deng:2018jrp, Asvesta:2022fts, Cao:2021ldv}:
\begin{equation}
\label{eq:mb_aparente}
 m_b(z_i;\theta) = 5 \log_{10} \left( D_{\rm L}(z_i;\Theta) \right) + \mathcal{M} \,,
\end{equation}
where $\mathcal{M}$ is considered a nuisance parameter that 
encompasses elements such as the absolute magnitude $\rm M_{\rm abs}$, 
which acts as a calibration term of the apparent magnitude of the 
SNIa sample 
\citep[see details in][]{SNLS:2011lii, Deng:2018jrp, Asvesta:2022fts, Cao:2021ldv}.
The (dimensionless) luminosity distance, $D_{\rm L} (z_i;\Theta)$, 
depends on the cosmological model studied:
\begin{equation}
D_{\rm L}(z_i;\Theta)= 
(1+z_i) \ S_k\left( H_0 \int_0^{z_i} \frac{dz'}{H(z';\Theta)} \right) \,,
\label{eq:luminosityDistance}
\end{equation}
where
\begin{equation} \label{spatial}
S_k (x)  = \left\{ \begin{array} {rcl} 
\frac{1}{\sqrt{-\Omega_{k,0}}} \sin ( \sqrt{-\Omega_{k,0}} x) \quad \text{for} \quad k  &=& +1 \\
x \quad \text{for} \quad k &=& 0 \\
\frac{1}{\sqrt{\Omega_{k,0}}} \sinh (\sqrt{\Omega_{k,0}} x) \quad \text{for} \quad k &=& -1  \end{array}  \right.
\end{equation}
The variable $\Theta$ identifies the set of cosmological 
parameters of the proposed model. 
%
As in
the previous section, a numerical approach is 
taken into consideration to generate the model. We compute 
the numerical solution of $D_{\rm L}(z_i;\Theta)$ for each 
$\Theta$ and $z_i$, for which the explicit Runge-Kutta method 
of order 8 is applied 
to
Eqs.\,(\ref{eq:Hz}), (\ref{eq:tz}) 
and 
to
the derivative of the luminosity distance:
\begin{equation} 
\frac{d}{dz} D_{\rm L}(z;\Theta)= \frac{D_{\rm L}(z;\Theta)}{1+z} + 
\frac{H_0 \ (1+z)}{H(z;\Theta)} \sqrt{1+ \Omega_{k,0} \left( \frac{ \ D_{\rm L}(z;\Theta)}{ \ (1+z)} \right)^2} \,,
\label{eq:derivate_luminosityDistance}
\end{equation}
which
is valid for every value of $k$ ($k= -1,0,1$).
Once 
the solution of $D_{\rm L}(z_i;\Theta)$ 
for each $z_i$
is achieved,
the apparent magnitude model is established 
with the set of free parameters $\theta = \{\Theta,\mathcal{M}\}$.
The
merit function is 
then 
given by:
\begin{equation}
\label{eq:chi2_pasne_matrix}
\chi^2_{\rm PaSNe}(\theta) = \big( \mathbf{m}_{b,\rm obs} - 
\mathbf{m}_{b}(\theta) \big)^T \mathbf{C}^{-1} \big( \mathbf{m}_{b,\rm obs} - {\bf m}_{b}(\theta) \big) ,
\end{equation}
where ${\bf m}_{b,\rm obs}$ and ${\bf m}_{b}(\theta)$ are 
vectors containing respectively the observed and theoretical 
apparent magnitudes of the whole SNe\,Ia sample. 
For the Pantheon measurements, the covariance matrix 
$\mathbf{C}$ depends on 
both
a statistical and 
a systematic 
component describing the uncertainties; i.e.\ $\mathbf{C} 
= \mathbf{C}_{\rm stat}+\mathbf{C}_{\rm sys}$.
The systematic uncertainties arise from the BEAMS method, 
which is applied to correct the bias 
generated by 
fits 
of both
nuisance parameters from the light curve of SNe\,Ia 
and the cosmological parameters \cite{Kessler:2016uwi}. 
Therefore, $\mathbf{C}_{\rm sys}$ also accounts for the 
correlation among the apparent magnitudes of the Pantheon data set;
i.e., the $\mathbf{C}_{\rm sys}$ is not a diagonal matrix: 
$({\bf C}_{\rm sys})_{i,j} \neq 0 \  \forall \ i,j$. 
For an ideal case, if we assume that uncertainties are uncorrelated 
the noise covariance matrix becomes diagonal, so the merit 
function comes to take a simple shape:
\begin{equation}
    \chi_{\rm PaSNe}^2(\theta) = 
\sum_{i = 1}^{1048} \left (\frac{ m_{\textrm{obs},i} - 
m_{b}(z_i;\theta)}{\sigma_{m_{\textrm{obs},i}}} \right)^2 \,,
\end{equation}
where the uncertainty of each measurement is the contribution 
of the statistical and systematic uncertainties;
i.e.,  
$\sigma_{m_{\textrm{obs},i}}^2 =  \sigma_{m_{\textrm{stat},i}}^2 
+ \sigma_{m_{\textrm{sys},i}}^2$.

\subsubsection{Remarks on the fitting and analysis strategies}
\label{sec:model_results}

We investigate the cosmological models associated with the 
mimetic field potentials, as outlined in section~\ref{sec:potential}.
The first family of models is characterized by an exponential 
potential $V(\phi) = V_0 \, \rm{e}^{- \alpha \cdot \phi}$ (EP). 
We also consider a specific case  with a fixed value of 
$\alpha=1$, $V(\phi) = V_0 \, \rm{e}^{-\phi}$, which is
we label
EPf.
The second family considers an inverse quartic potential 
$V(\phi) = V_0 \, \left( 1+ \delta \cdot \phi^2 \right)^{-2}$ 
(IQP). In particular, the case $\delta = 1$ is identified as IQPf 
($V(\phi) = V_0 \, \left( 1+\phi^2 \right)^{-2}$).
Table~\ref{tab:models_priors} provides a summary of the models,
 parameter space,
 and data sets employed for each fit. 
The numerical solutions in section \ref{sec:potential} indicate 
that for small values of $\alpha$ and $\delta$, the curves 
closely resemble the $\Lambda$CDM model for early times, as 
expected. As a preliminary approximation, we consider $\alpha=1$ 
and $\delta=1$ to assess how well these models align with the 
data and analyze the differences compared to $\Lambda$CDM. In 
this case, the reduced parameter space also leads to a shorter 
computational time. However, it is acknowledged that for extended 
times periods, all curves will deviate from $\Lambda$CDM. 
Subsequently, we 
shall
allow these parameters to be free,
thus
 enabling a data-driven fitting process.

Since the MCMC performance requires of several iterations, 
the estimation of the all $H(z_i;\theta)$ and $m_{b}(z_i;\theta)$ 
models is computationally expensive.
We therefore consider priors to expedite the characterization 
of the posterior distribution with the MCMC sampling approach.
We consider two type of priors: $i)$ uniform priors, 
$\mathcal{U}(a,b)$, that follow a top-hat function in the 
interval $[a,b]$, and $ii)$ Gaussian priors,  
$\mathcal{N}(\overline{p},\sigma_p)$, that follow a normal 
Gaussian distribution
described by the mean and standard 
deviation of the $p$-parameter.
The priors imposed on the free parameters are shown in 
Table~\ref{tab:models_priors}.
For instance, the uniform priors over $\Omega_{\phi,0}$ and 
$\Omega_{k,0}$ guarantee the physical meaning of these densities.
In other cases, the Gaussian and uniform priors are useful 
for
alleviating
the exploration of the parameter space. For example, 
we use the same constraint imposed in Ref.\ \cite{Asvesta:2022fts} 
for $\mathcal{M}$;
i.e., 
$\mathcal{P}(\mathcal{M}) = \mathcal{U}(23,24)$.

Once the priors are established, we run the {\sc emcee} with 
50 chains and 14000 iteration steps to characterize the posterior 
distributions 
for
each case. The first 3500 steps are excluded 
from
each chain (the \emph{burn-in} sample). 
Overall, the parameters from the cosmological modeling are 
computed from the marginalized the posterior distribution function, 
PDF($\theta_{i}$), over each parameter $\theta_i$.
Table~\ref{tab:results_parameters} presents the parameters and their 
uncertainties, 
i.e., respectively the median and uncertainty (the 16th 
and 84th percentiles) of its PDF($\theta_{i}$), for all cases studied.

To assess the goodness of fits we provide the 
chi-squared 
($\chi^2$) 
values
in the parameter space recovered and 
the reduced chi-squared $\chi_{\rm red}^2$ 
values 
(accounting for the degrees 
of freedom).
In addition, we can use 
criteria based on 
statistical information 
to determine 
whether the data favors one model over another. We selected two 
standard statistical indicators: 1) the Akaike Information Criterion 
\cite[AIC,][]{akaike1974new}
and 2) the Bayesian Information 
Criterion \cite[BIC,][]{Schwarz:1978tpv}. They are computed from 
the following expressions:
\begin{equation}
\label{eq:aic_bic}
 {\rm AIC} = 2 \, N_\theta - 2 \ln\,(\mathcal{L}_{\rm max}) \ \ {\rm and} \ \ {\rm BIC} 
= N_\theta \ln\,(n) - 2 \ln\,(\mathcal{L}_{\rm max}),
\end{equation}
where $N_\theta$ corresponds to the number of free parameters of 
the model, $\mathcal{L}_{\rm max}$ is the value of the likelihood 
function evaluated at the best fit, and $n$ corresponds to the sample size.
In the case of asymmetric distributions, where the median departs 
from the peak of the posterior, we use the latter to maintain the 
meaning of the definition of AIC and BIC.
In our case, we have respectively $n_{\rm OHD}=51$ and
 $n_{\rm PaSNe}=1048$ for the OHD and Pantheon cases, which corresponds 
to $n_{\rm OHD+PaSNe} = 1099$ for the joint analysis.
These statistical quantities are confronted with those values 
obtained for the $\Lambda$CDM fits with curvature assumption cases, 
which are computed as:
\begin{equation}
\label{eq:delta_aic_bic}
  \rm \Delta\,AIC_{model} = AIC_{model} - AIC_{\Lambda CDM} \ \ 
and \ \ \Delta\,BIC_{model} = BIC_{model} - BIC_{\Lambda CDM},
\end{equation}
with ``model'' being the identification name of the model considered. 
$\rm \Delta\,AIC_{model}$ and $\rm \Delta\,BIC_{model}$ are 
computed with respect to fits that use the same database.
The AIC, BIC, $\rm \Delta\,AIC_{model}$ and $\rm \Delta\,BIC_{model}$ 
are also listed in Table~\ref{tab:results_parameters}.

As a consistent test, the $\Lambda$CDM model for a flat universe is 
studied in order to compare the performance of our methodology with 
respect to 
analysis
in the literature.
We follow \citep{Corral:2020lxt} and fit the three data sets to the 
$\Lambda$CDM model (assuming a flat universe) with the prior 
$\mathcal{P}(H_0) = \mathcal{N}(73.24,1.74)$, where the parameter 
space consists of $\theta = \{\Omega_{\phi,0},H_0\}$ for the OHD case, 
and $\theta = \{\Omega_{\phi,0},H_0,\mathcal{M} \}$ for the PaSNe and 
OHD+PaSNe cases. For all analyses, the $H_0$ and $\Omega_{\phi,0}$ 
recovered are in full agreement with the results obtained by 
\citep{Corral:2020lxt}.
As an example, we obtained 
$H_0 = 70.51^{+0.85}_{-0.86}$\,$\rm km\,s^{-1}\,Mpc^{-1}$ and 
$\Omega_{\phi,0} = 0.265\pm0.013$ for the OHD+PaSNe case, while 
they reported $H_0 = 70.5\pm0.9$\,$\rm km\,s^{-1}\,Mpc^{-1}$ and 
$\Omega_{\phi,0} = 0.265^{+0.013}_{-0.012}$ (see Table\,I in 

\citep{Corral:2020lxt}).
This confirms that our methodology is a reliable and robust approach.

\begin{table}[t]
    \centering
    \begin{tabular}{c c l}
    \hline
    Model ID & Potential & Parameter space for non-flat models\\
    \hline
     & & $\theta_{\rm OHD} = \{H_0, \Omega_{\phi,0}, \Omega_{k,0}, \alpha \}$\\ [1.0 ex]
    EP & $V(\phi) = V_0 \, \rm{e}^{- \alpha \cdot \phi}$ & $\theta_{\rm PaSNe} = \{H_0, \Omega_{\phi,0}, \Omega_{k,0}, \alpha, \mathcal{M} \}$ \\ [1.0 ex]
    &  & $\theta_{\rm OHD+PaSNe} = \{H_0, \Omega_{\phi,0}, \Omega_{k,0}, \alpha, \mathcal{M} \}$ \\ [1.0 ex]    
    \hline
     & & $\theta_{\rm OHD} = \{H_0, \Omega_{\phi,0}, \Omega_{k,0} \}$ \\ [1.0 ex]
    EPf & $V(\phi) = V_0 \, \rm{e}^{-\phi}$ & $\theta_{\rm PaSNe} = \{H_0, \Omega_{\phi,0}, \Omega_{k,0}, \mathcal{M} \}$ \\ [1.0 ex]
    &  & $\theta_{\rm OHD+PaSNe} = \{H_0, \Omega_{\phi,0}, \Omega_{k,0}, \mathcal{M} \}$ \\ [1.0 ex]
    \hline
     & & $\theta_{\rm OHD} = \{H_0, \Omega_{\phi,0}, \Omega_{k,0}, \delta \}$\\ [1.0 ex]
    IQP & $V(\phi) = V_0 \, \left( 1+ \delta \cdot \phi ^2 \right)^{-2}$ & $\theta_{\rm PaSNe} = \{H_0, \Omega_{\phi,0}, \Omega_{k,0}, \delta, \mathcal{M} \}$ \\ [1.0 ex]
    &  & $\theta_{\rm OHD+PaSNe} = \{H_0, \Omega_{\phi,0}, \Omega_{k,0}, \delta, \mathcal{M} \}$\\ [1.0 ex]    
    \hline
     & & $\theta_{\rm OHD} = \{H_0, \Omega_{\phi,0}, \Omega_{k,0} \}$ \\ [1.0 ex]
    IQPf & $V(\phi) = V_0 \, \left( 1+ \phi^2 \right)^{-2}$ & $\theta_{\rm PaSNe} = \{H_0, \Omega_{\phi,0}, \Omega_{k,0}, \mathcal{M} \}$ \\ [1.0 ex]
    &  & $\theta_{\rm OHD+PaSNe} = \{H_0, \Omega_{\phi,0}, \Omega_{k,0}, \mathcal{M} \}$ \\ [1.0 ex]    
    \hline
    {\bf Priors} & \multicolumn{2}{l}{$\mathcal{P}(H_0): \ \mathcal{U}(0,100)$ and $\mathcal{N}(73.24,5)$ $\rm (km\,s^{-1} Mpc^{-1})$} \\ [1.0 ex]
    & \multicolumn{2}{l}{$\mathcal{P}(\Omega_{\phi,0})= \mathcal{U}(0,1)$ and $\mathcal{N}(0.73,0.05)$} \\[1.0 ex]
    & \multicolumn{2}{l}{$\mathcal{P}(\Omega_{k,0}): \ \mathcal{U}(-0.2,0.2)$, $\mathcal{P}(\alpha)= \mathcal{U}(-50,50)$} \\ [1.0 ex]
    & \multicolumn{2}{l}{
    $\mathcal{P}(\delta)= \mathcal{U}(-1000,\infty)$, and $\mathcal{P}(\mathcal{M})= \mathcal{U}(23,24)$} \\
    \hline
    \end{tabular}
    \caption{Models, parameter spaces and priors considered in this 
study.}
    \label{tab:models_priors}
\end{table}

\begin{table}[h!!!!!]
\begin{center}
\resizebox{\columnwidth}{!}{
\begin{tabular}{| c | c  c  c   c  c  c |  c   c c  c c  c |}
\hline
Dataset & $\Omega_{\phi,0}$ & $\alpha$ & $\delta$ & $\Omega_{k,0}$ &  $H_0$ & $\mathcal{M}$ & $\chi_{\rm min}^2$ & $\chi_{\rm red}^2$ & AIC & BIC & $\Delta\,{\rm AIC}$ & $\Delta\,{\rm BIC}$\\ 
\hline
\multicolumn{13}{ |c| }{$\Lambda$CDM} \\
\hline
OHD & $0.707\pm 0.045$ & $-$ & $-$ & $0.047\pm 0.058$ & $70.23^{+1.26}_{-1.21}$ & $-$ & $26.5$ & $0.55$ & $32.5$ & $38.3$ & $-$ & $-$  \\ 
PaSNe & $0.730^{+0.043}_{-0.045} $ & $-$ & $-$ & $-0.047^{+0.077}_{-0.075}$ & $73.41^{+4.89}_{-5.04}$ & $23.81\pm 0.01$ & $1027.2$ & $0.98$& $1035.2$ & $1055.0$ & $-$ & $-$ \\
OHD+PaSNe & $0.691\pm 0.039$ & $-$ & $-$ & $0.051^{+0.055}_{-0.054}$ & $69.39^{+1.00}_{-0.98} $ & $23.80\pm 0.01$ & $1055.0$ & $0.96$ & $1063.0$ & $1083.0$ & $-$ & $-$ \\
\hline
\multicolumn{13}{|c|}{EP}  \\
\hline
OHD & $0.715\pm 0.042$ & $-11.94^{+34.53}_{-25.31}$ & $-$ & $0.082^{+0.078}_{-0.129}$ & $70.34^{+1.35}_{-1.29}$ & $-$ & $26.4$ &  $0.56$ & $34.4$ & $42.1$ & $1.9$ & $3.8$\\
PaSNe & $0.727^{+0.041}_{-0.042}$ & 
$-2.54^{+32.09}_{-31.21}$ & $-$ & $-0.041^{+0.099}_{-0.093}$ & $73.35^{+4.88}_{-5.06}$ & $23.81\pm 0.01$ & $1026.9$ & $0.98$ &$1036.9$ & $1061.7$ & $1.7$ & $6.7$\\
OHD+PaSNe & $0.699^{+0.037}_{-0.038}$ & $18.08^{+19.00}_{-27.50}$ & $-$ & $-0.020^{+0.117}_{-0.104}$ & $69.49^{+1.01}_{-0.99}$ & $23.81\pm 0.01$ & $1055.0$ & $0.96$ &$1065.0$ & $1090.0$ & $2.0$ & $7.0$ \\\hline
\multicolumn{13}{|c|}{EPf} \\
\hline
    OHD & $ 0.706^{+0.045}_{-0.044}$ & $-$ & $-$ & $0.045\pm 0.058$ & $70.20^{+1.24}_{-1.21}$ & $-$ & $26.5$ & $0.55$ & $32.5$ & $38.3$ & $0$ & $0$ \\
    PaSNe  & $0.730^{+0.043}_{-0.044}$ & $-$ & $-$ & $-0.051^{+0.077}_{-0.072}$ & $73.29^{+4.88}_{-5.01}$ & $23.81 \pm 0.01$ & $1027.0$ & $0.98$ & $1035.0$ & $1054.8$ & $-0.2$ & $-0.2$\\
OHD+PaSNe & $0.692\pm 0.039$ & $-$ & $-$ & $0.047^{+0.054}_{-0.055}$ & $69.41^{+1.01}_{-0.98}$ & $23.80\pm 0.01$ & $1054.9$ & $0.96$ & $1062.9$ & $1082.9$ & $-0.1$ & $-0.1$\\\hline
\multicolumn{13}{| c| }{IQP} \\
\hline
    OHD & $0.717^{+0.042}_{-0.043}$ & $-$ & $1568.9^{+2644.8}_{-1859.9}$ & $0.116^{+0.061}_{-0.095}$ & $71.03^{+1.65}_{-1.53}$ & $-$ & $29.1$ & $0.62$  &$37.1$ & $44.8$ & $4.6$ & $6.5$ \\
PaSNe & $0.725^{+0.042}_{-0.041}$ & $-$ & $7012.1^{+10114.6}_{-5753.0}$ & $0.063^{+0.092}_{-0.110}$ & $73.88^{+4.89}_{-4.95}$ & $23.80\pm0.01$ & $1027.8$ &  $0.99$  &$1037.8$ & $1062.6$ & $2.6$ & $7.6$ \\
OHD+PaSNe & $0.696\pm 0.041$ & $-$ & $-368.8^{+970.0}_{-469.5}$ & $0.035^{+0.074}_{-0.069}$ & $69.41^{+1.00}_{-0.97}$ & $23.80\pm 0.01$ & $1054.5$ & $0.96$ &$1064.5$ & $1089.5$ & $1.5$ & $6.5$\\
\hline 
\multicolumn{13}{|c|}{IQPf } \\ 
\hline
OHD & $0.707\pm 0.045$ & $-$ & $-$ & $0.048^{+0.058}_{-0.057}$ & $70.23^{+1.27}_{-1.22}$ & $-$ & $26.5$ & $0.55$ &$32.5$ & $38.3$ & $0$ & $0$\\
PaSNe & $0.730^{+0.043}_{-0.044}$ & $-$ & $-$ & $-0.048^{+0.077}_{-0.074}$ & $73.27^{+5.01}_{-4.98}$ & $23.81\pm 0.01$ & $1027.1$ & $0.98$ & $1035.1$  &$ 1054.9$ & $-0.1$ & $-0.1$ \\
OHD+PaSNe & $0.692\pm 0.039$ & $-$ & $-$ & $0.050\pm 0.054$ & $69.41^{+1.00}_{-0.99}$ & $23.80\pm 0.01$ & $1054.9$ & $0.96$  &$1062.9$ & $1082.9$ & $-0.1$ & $-0.1$ \\
\hline
\end{tabular}
}
\caption{Parameters and uncertainties obtained for the 
non-flat cosmological models investigated. The identification 
names and field potentials for the $\Lambda$CDM model and mimetic 
gravity models (EP, EPf, IQP and IQPf) are listed in 
Table~\ref{tab:models_priors}. The properties of these 
non-flat cosmological models are obtained using three data 
set: the observational Hubble database from the cosmic 
chronometers (OHD), the Pantheon database 
for Type Ia supernovae (PaSNe),
 and the simultaneous analysis of the OHD 
and PaSNe data sets (OHD+PaSNe). The statistical parameters 
$\Delta \rm AIC$ and $\Delta \rm BIC$ are computed with 
respect to the non-flat $\Lambda$CDM results for the same database.}
\label{tab:results_parameters}
\end{center}
\end{table}


\subsection{Cosmological constraints}

\begin{figure}[t]
    \begin{center} 
    \includegraphics[width=0.8\textwidth]{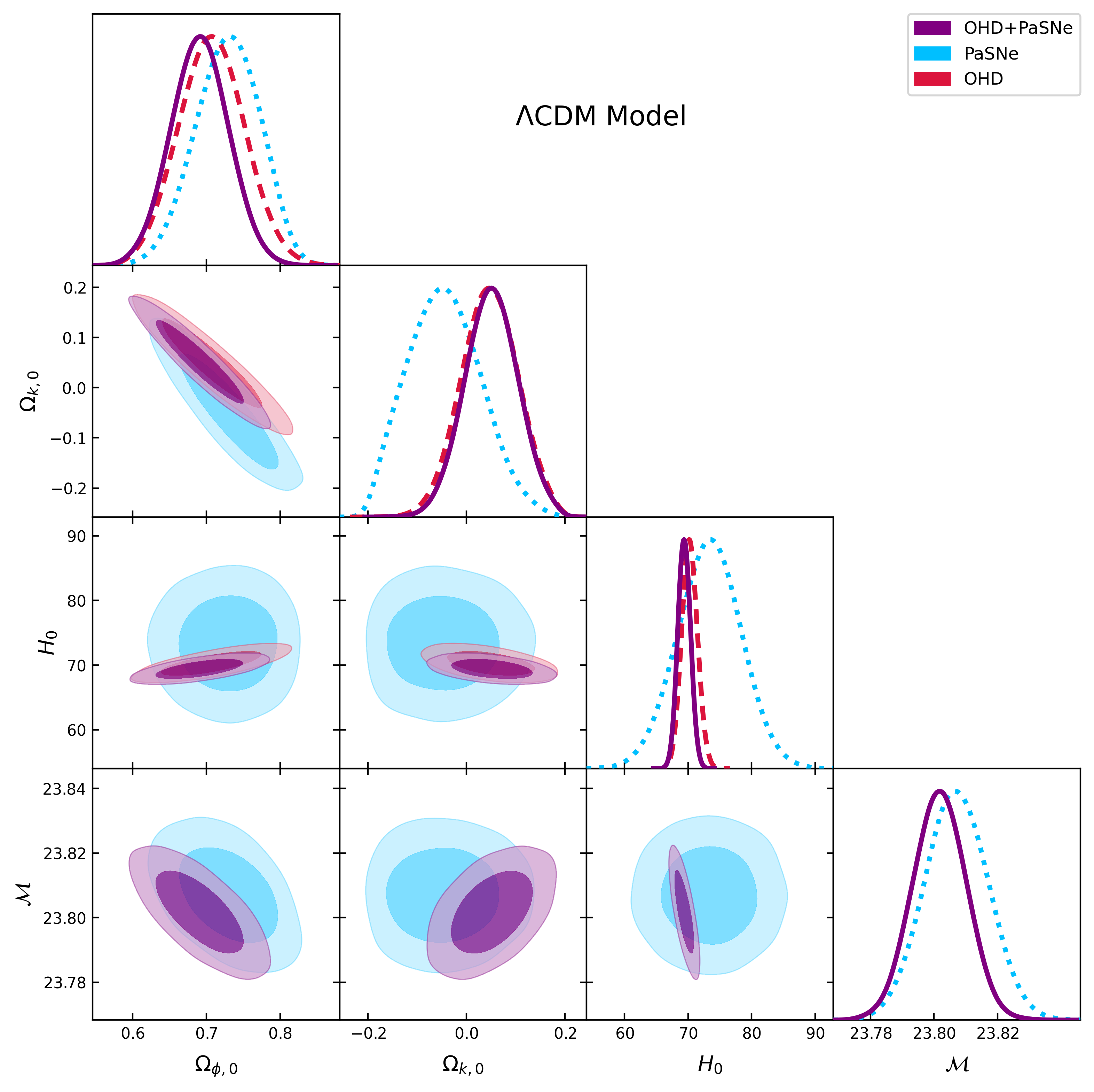}
    \caption{
    Marginalized posteriors and correlation plots for the parameters 
$\Omega_{\phi,0}$, $\Omega_{k,0}$, $H_0$, $\mathcal{M}$ that 
describe the non-flat $\Lambda$CDM model. The contours represent 
the 68.3\% and 95.4\% confidence levels. The results obtained from 
the OHD+PaSNe, PsSNe, and OHD analysis
(Table~\ref{tab:results_parameters})
are identified respectively by the purple, blue, and red colors. 
$H_0$ is in units of $\rm km\,s^{-1}\,Mpc^{-1}$, while $\Omega_{\phi,0}$, 
$\Omega_{k,0}$, and $\mathcal{M}$ are dimensionless units.
    }
    \label{fig:results_pdfs_lambdaCDM}
     \end{center}
\end{figure}

\begin{figure}[ht!!!]
    \begin{center}
    \includegraphics[width=1.0\textwidth]{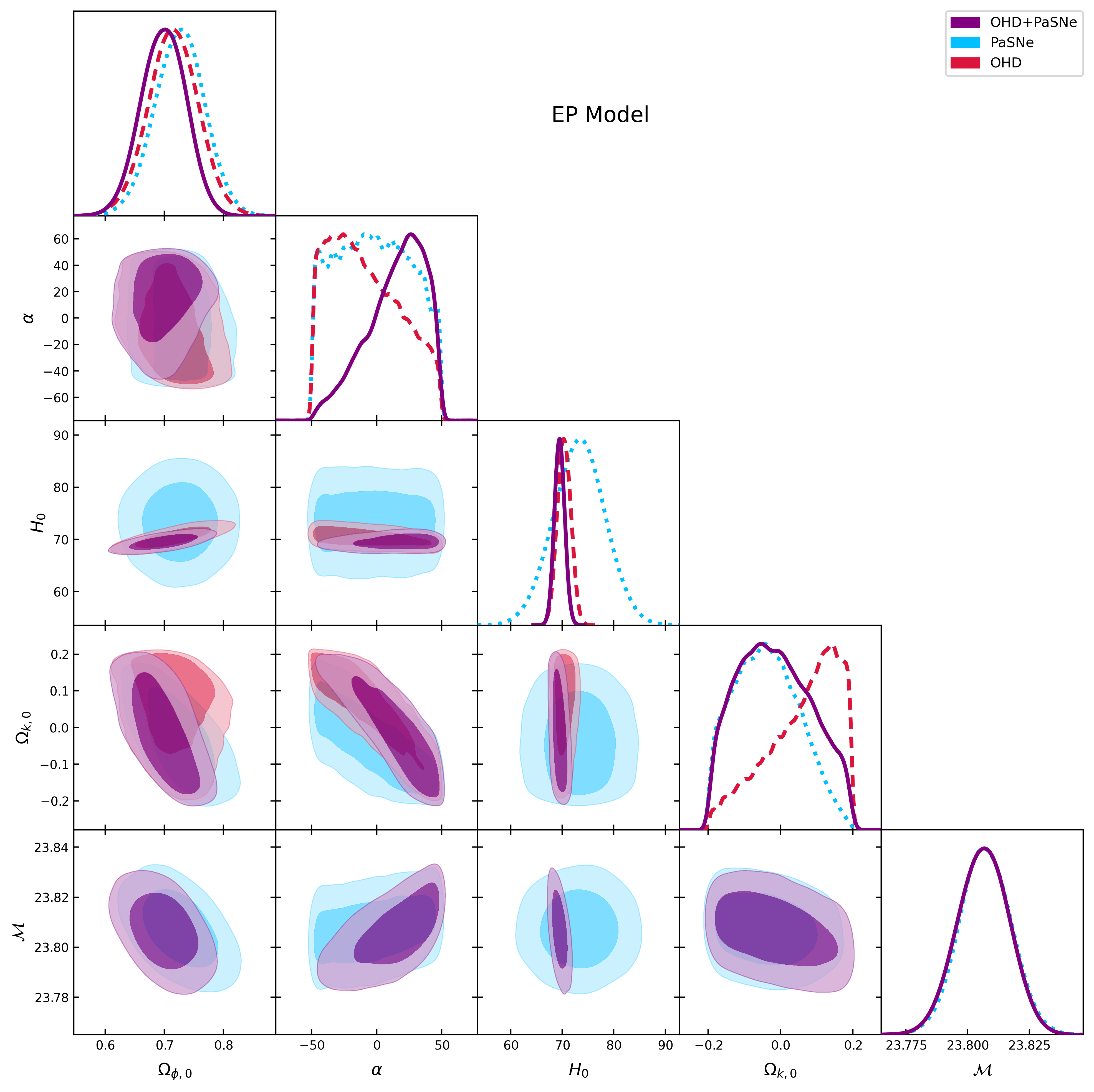}
    \caption{
    Marginalized posteriors and correlation plots for the parameters 
$\Omega_{\phi,0}$, $\alpha$, $\Omega_{k,0}$, $H_0$, $\mathcal{M}$ 
that describe the EP model (the exponential potential) with spatial 
curvature. The contours represent the 68.3\% and 95.4\% confidence 
levels. The results obtained from the OHD+PaSNe, PsSNe and OHD analysis, 
Table~\ref{tab:results_parameters}, are 
identified 
respectively
by the 
purple, blue and red colors. $H_0$ is in units of 
$\rm km\,s^{-1}\,Mpc^{-1}$, while $\Omega_{\phi,0}$, $\alpha$, $\Omega_{k,0}$, 
and $\mathcal{M}$ are dimensionless units.}
    \label{fig:results_pdfs_EP}
     \end{center}
\end{figure}

\begin{figure}[t]
    \begin{center}
    \includegraphics[width=0.8\textwidth]{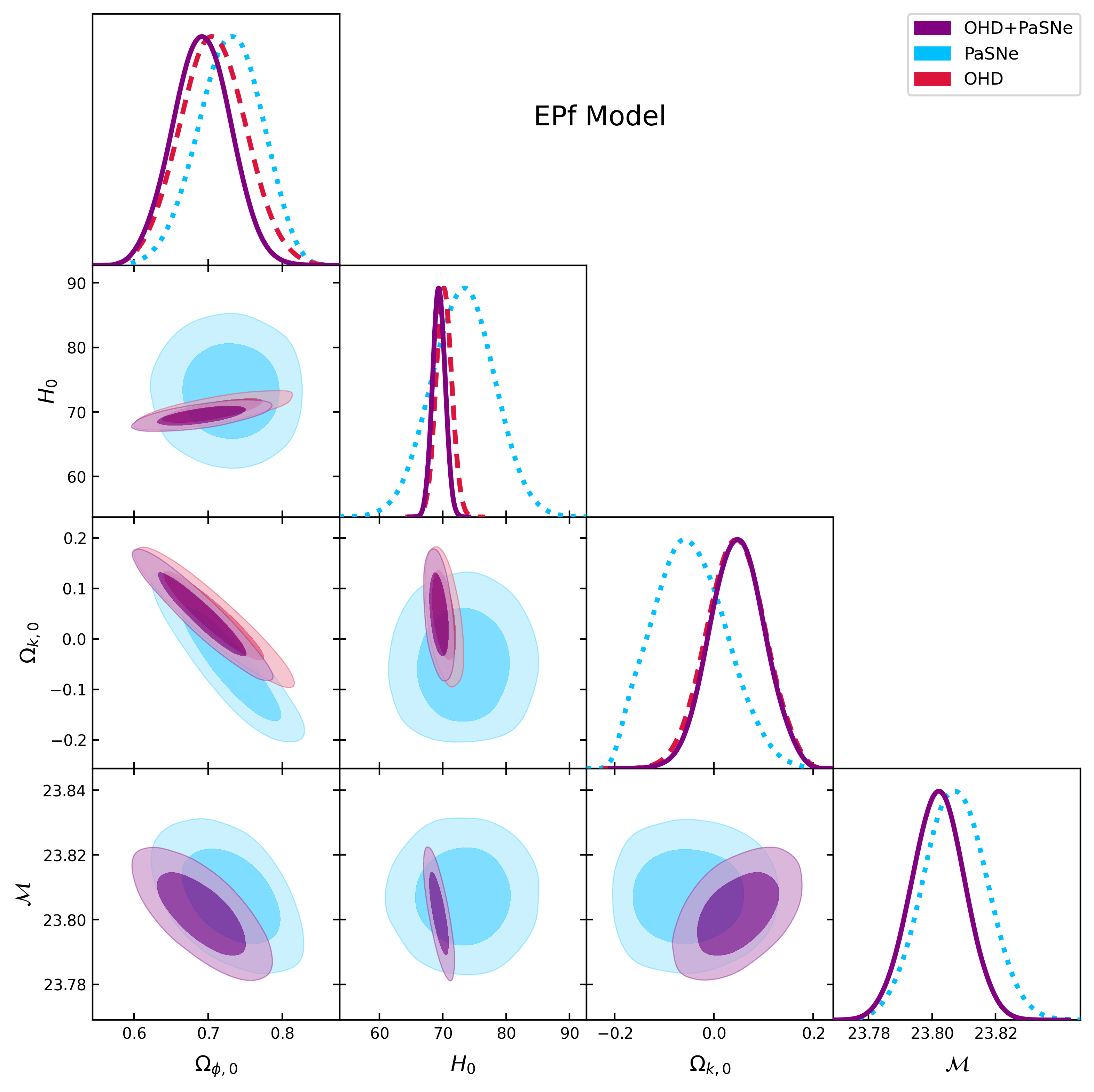}
    \caption{
    Marginalized posteriors and correlation plots for the parameters 
$\Omega_{\phi,0}$, $\alpha$, $\Omega_{k,0}$, $H_0$, $\mathcal{M}$ that 
describe the EPf model (the exponential potential with $\alpha=1$) with 
spatial curvature. The contours represent the 68.3\% and 95.4\% confidence 
levels. The results obtained from the OHD+PaSNe, PsSNe and OHD analysis, 
Table~\ref{tab:results_parameters}, are identified respectively by the 
purple, blue and red colors. $H_0$ is in units of 
$\rm km\,s^{-1}\,Mpc^{-1}$, while $\Omega_{\phi,0}$, $\alpha$, $\Omega_{k,0}$ 
and $\mathcal{M}$ are dimensionless units.}
    \label{fig:results_pdfs_EPf}
     \end{center}
\end{figure}

\begin{figure}[t!!!!] 
\begin{center}
\includegraphics[width=1.0\textwidth]{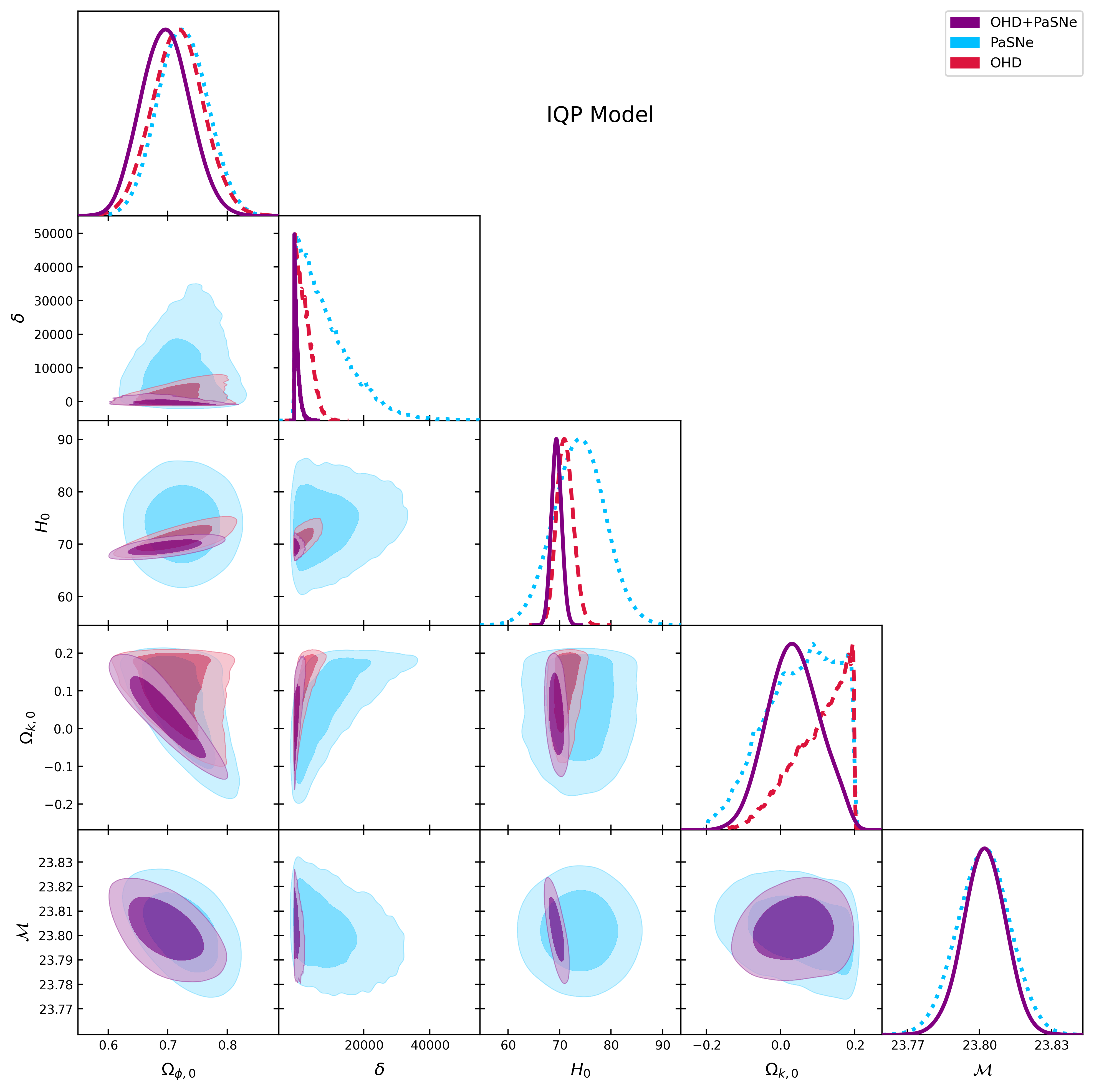}
\caption{
    Marginalized posteriors and correlation plots for the
 parameters $\Omega_{\phi,0}$, $\delta$, $\Omega_{k,0}$, $H_0$, 
$\mathcal{M}$ that describe the IQP model (the inverse quadratic
 potential) with spatial curvature. The contours represent the 
68.3\% and 95.4\% confidence levels. The results obtained from 
the OHD+PaSNe, PsSNe and OHD analysis, 
Table~\ref{tab:results_parameters}, are 
respectively
identified 
 by the purple, blue and red colors. $H_0$ is in units of 
$\rm km\,s^{-1}\,Mpc^{-1}$, while $\Omega_{\phi,0}$, $\delta$,
 $\Omega_{k,0}$ and $\mathcal{M}$ are dimensionless units.}
\label{fig:results_pdfs_QRP}
\end{center}
\end{figure}

\begin{figure}[h] 
\begin{center}
\includegraphics[width=0.8\textwidth]{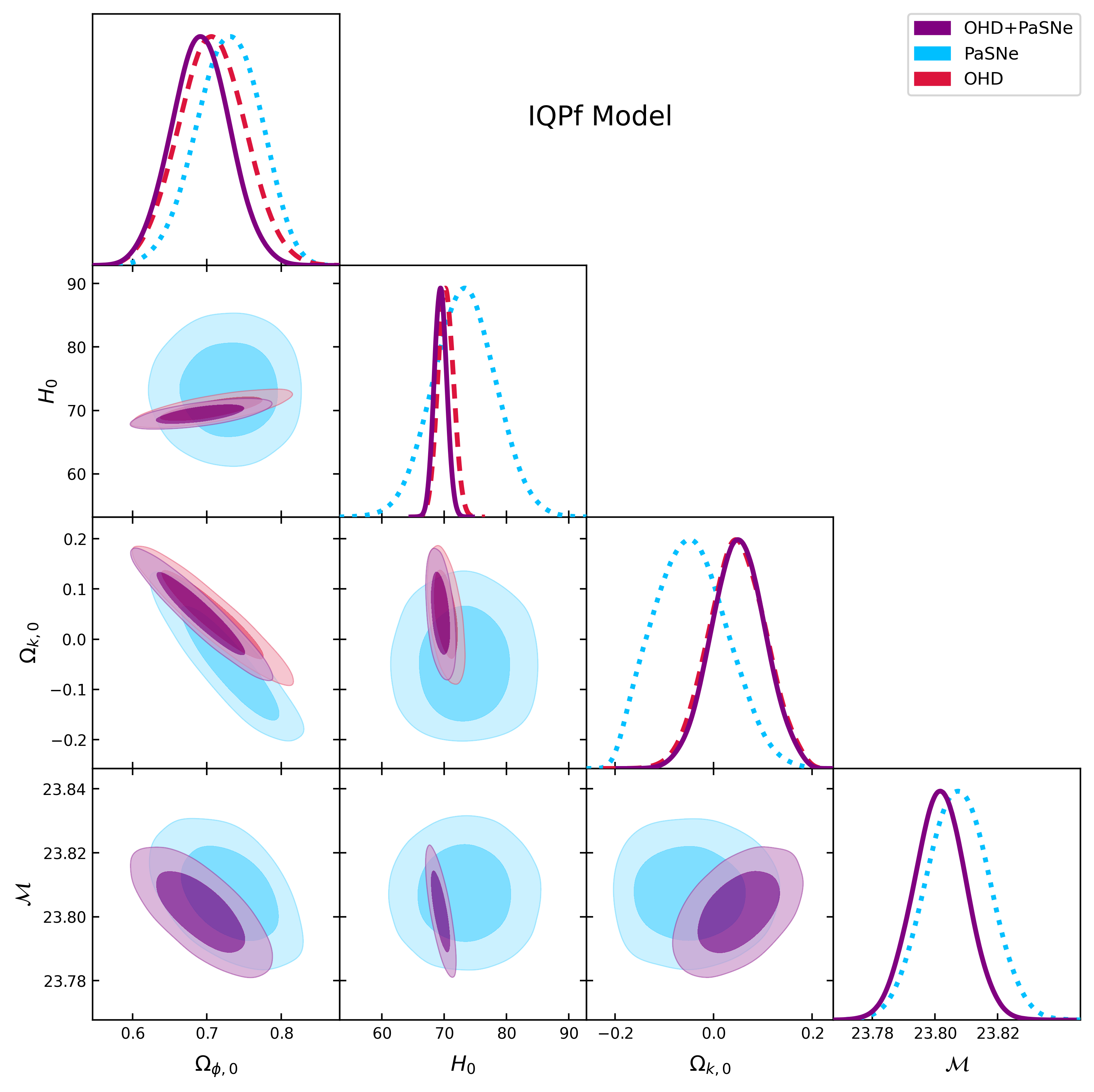}
\caption{
    Marginalized posteriors and correlation plots for the 
parameters $\Omega_{\phi,0}$, $\Omega_{k,0}$, $H_0$, $\mathcal{M}$ 
that describe the IQPf model (the inverse quadratic potential with 
$\delta=1$) with spatial curvature. The contours represent the 68.3\% 
and 95.4\% confidence levels. The results obtained from the OHD+PaSNe, 
PsSNe and OHD analysis, Table~\ref{tab:results_parameters}, are 
respectively
identified 
by the purple, blue and red colors. $H_0$ 
is in units of $\rm km\,s^{-1}\,Mpc^{-1}$, while $\Omega_{\phi,0}$,
 $\Omega_{k,0}$ and $\mathcal{M}$ are dimensionless units.}
\label{fig:results_pdfs_QRPf}
\end{center}
\end{figure}

We provide the properties of non-flat cosmological models 
obtained with the three data sets (Table~\ref{tab:results_parameters}). 
We investigate the marginalized posteriors and corner plots 
(the 2D contours of the PDFs) for the $\Lambda$CDM 
(Fig.~\ref{fig:results_pdfs_lambdaCDM}), EP 
(Figs.~\ref{fig:results_pdfs_EP}~and~\ref{fig:results_pdfs_EPf}), 
and IQP (Figs.~\ref{fig:results_pdfs_QRP}~and~\ref{fig:results_pdfs_QRPf}) 
cosmological models with spatial curvature.
Overall, the non-flat cosmological models do not show substantial 
statistical preference over the base $\Lambda$CDM, EP, IQP models 
with the assumption of a flat universe,  $\Omega_{k,0}=0$. For 
example, the OHD+PaSNe analysis 
provides 
$\Omega_{k,0}$ values of $0.051^{+0.055}_{-0.054}$ ($\Lambda$CDM), $-0.020^{+0.117}_{-0.104}$ 
(EP), and $0.035^{+0.074}_{-0.069}$ (IQP) which are consistent with 
a flat universe. 
For
reference, the results for the flat universe 
scenario are presented in appendix~\ref{appendix:flat_universe} (see Table \ref{tab:results_LCDM}). 

We consider the OHD+PaSNe analysis as our reference model
because it considers both data sets at the same time, which allow 
us to access more cosmological information.
In spite of this, the 
goodness of fit 
appears to be better for 
the OHD cases, with $\chi^2_{\rm red}$ around 0.55 for all models, 
tested (Table~\ref{tab:results_parameters}),
although
this low 
value also could reflect 
an underestimation of the
uncertainties of the 
observational measurements are underestimated.
In contrast, the OHD+PaSNe and PaSNe analysis respectively provide 
$\chi^2_{\rm red}$ of 0.96 and $\sim 0.98$. 
Close to unity.
We therefore confirm 
that 
the performance of the fits is robust and 
similar when the same data set is considered. 

%

Regarding the curved universe scenarios, the $\Omega_{k,0}$ values 
are compatible with zero for all
the
 cases studied, with the significance 
of detection being around one (see Table~\ref{tab:results_parameters}).
Overall, the OHD and OHD+PaSNe analysis prefer positive curvature in 
contrast to the PaSNe fits. 
Only the IQP modeling shows positive curvature values for the three data sets.
The non-flat $\Lambda$CDM scenario yields $\Omega_{k,0}$ 
values
of $0.047\pm0.058$, $-0.047^{+0.077}_{-0.075}$, and $0.051^{+0.055}_{-0.054}$ 
for the OHD, PaSNe and OHD+PaSNe fits, respectively.
Despite our higher uncertainties, these values are compatible with 
the
literature 
\citep[for example, see][]{Yang:2022kho,DiValentino:2019qzk,Handley:2019tkm,Vagnozzi:2020dfn,Vagnozzi:2020rcz,Dhawan:2021mel,Abdalla:2022yfr,Planck:2018vyg,Rosenberg:2022sdy,Tristram:2023haj}.  
For instance, \cite{Yang:2022kho} obtained $\Omega_{k,0}$ between 
$-0.043$ up to 0.0008 using the CMB
temperature and polarization power spectra (TT, TE and EE), BAO, and 
the Pantheon sample for the base $\Lambda$CDM with spatial curvature. 
The CMB data alone have provided significance between 2 up to $3 \sigma$ 
that depart from the $\Omega_{k,0}=0$ scenario 
\cite[][and references therein]{DiValentino:2019qzk,Handley:2019tkm,Abdalla:2022yfr}; 
for example, $\Omega_{k,0} = -0.044^{+0.018}_{-0.019}$ was obtained 
from the {\it Planck} data \cite{Planck:2018vyg}. Although the 
new analysis of the {\it Planck} data have allowed 
stronger constraints such as $-0.025^{+0.013}_{-0.010}$ 
\citep{Rosenberg:2022sdy} and $-0.012\pm0.010$ \cite{Tristram:2023haj}
to be achieved. 
Including the CMB data is beyond the scope of this 
paper, 
although this is one of the goals of a future 
paper.

For the PaSNe fits, we also observe that negative values of 
$\Omega_{k,0}$ are associated 
with 
high values of the 
$H_0$ ($\sim 73.4\,\rm km\,s^{-1}\,Mpc^{-1}$) and $\Omega_{\phi,0}$ 
($\sim 0.73$) parameters, except 
for 
the EPf model where $\Omega_{k,0}>0$.
The correlation between them is not clear in the corner plots (see 
the blue 2D-PDFs of $\Omega_{k,0}$ versus $H_0$ and $\Omega_{\phi,0}$).
However, the OHD+PaSNe analysis show 
an
 anti-correlation between 
$\Omega_{k,0}$ and $H_0$. This is in 
the 
opposite direction when the 
CMB data is considered, which provides 
higher 
values of $\Omega_{k,0}$ 
as $H_0$ increases; 
see, for example, 
\citep{Planck:2013pxb,Rosenberg:2022sdy,Tristram:2023haj} for only 
the 
CMB data.
Previous 
studies have 
also reported higher values of $H_0$ when only the 
Type Ia supernovae data 
are
considered \cite[e.g.,][]{Corral:2020lxt,Brout:2022vxf,Riess:2021fzl,Riess:2022jrx}. 
For instance, \cite{Corral:2020lxt} obtained 
values of $H_0$ of $73.2\pm1.7$ and $70.5\pm0.9\,\rm km\,s^{-1}\,Mpc^{-1}$ 
respectively 
for their SNe\,Ia and Joint (similar to our PSNe and OHD+PSNe) 
fits for the $\Lambda$CDM model. They also reported this trend 
for their four models 
\citep[see Table~I in][]{Corral:2020lxt}.
For non-flat Universes, for example, \cite{Yang:2022kho},
considering the CMB+BAO and CMB+PaSNe+BAO for all model investigated, 
observed lower values of $H_0$ for the CMB+PaSNe data set with respect to 
fits  
where the CMB+PaSNe cases  always provided 
$\Omega_{k} < 0$. This change in trend could be connected with the 
inclusion of the CMB data (temperature and polarization power spectra). \\

Regarding the possibility of alleviating the \( H_0 \) tension, this work has not compared the model to \( \Lambda CDM \) in much detail for intermediate and early times, and this comparison will likely depend on the specific potential in most cases.

On the other hand, there does not appear to be a reduction in \( H_0 \) that can be attributed to the model itself rather than to the data. In all cases, the values obtained are lower in the combined data, followed by the \( OHD \) values, which are lower than those reported in \cite{Riess_2019}. In any case, to gain certainty, the model should be compared with CMB measurements, such as those reported in \cite{Planck:2018vyg}.

That said, it is not possible to make a comparison, since no comparison was made at intermediate redshifts, as done in \cite{Petronikolou:2021shp}, and it was not evaluated for CMB measurements either. All of this could constitute future research.


\subsubsection{Non-flat Exponential Potential models}

Regarding the EP model, the $\Omega_{\phi,0}$, $H_0$,
 and $\mathcal{M}$ 
values
recovered are in agreement with 
those
obtained for the non-flat $\Lambda$CDM analyses (for all cases). These 
are performed with very similar 
goodness of fit 
(see Table~\ref{tab:results_parameters}), 
producing
$\chi^2$ 
 =
0.96 for both 
the
non-flat $\Lambda$CDM and 
the
EP models (OHD+PaSNe case),
although 
we obtained a relatively high $\Delta \rm BIC=7$ and 
moderate $\Delta \rm AIC = 2$, which 
favor
 the non-flat $\Lambda$CDM over the EP models.
The OHD+PaSNe analysis 
produces 
$\Omega_{k,0}=-0.02^{+0.12}_{-0.104}$,  
which
contrasts with the positive curvature from the 
non-flat $\Lambda$CDM model ($\Omega^{\rm \Lambda CDM}_{k,0}=
0.051^{+0.055}_{-0.054}$), although 
still consistent 
and compatible with zero at one-$\sigma$.
In addition, we observed 
that 
asymmetric PDFs for $\Omega_{k,0}$ and 
their behaviors depend on the data used. The marginalized PDFs 
reach their maximum at around $-0.01$ (OHD+PaSNe and PaSNe) and 
0.013 (OHD), respectively.
We also 
found 
that $\alpha$ is described by asymmetric 
marginalized PDFs, and 
that 
the high dispersion does not allow 
good constraints on $\alpha$. For the OHD+PaSNe and OHD cases, their 
PDFs are peaking at around 25 and $-30$, respectively. In contrast, 
the PDF of $\alpha$ is poorly characterized for the PaSNe data, 
yielding $\alpha = 3 \pm  32$. 
We therefore obtained a wide range of values of $\alpha$ that 
follow
the imposed uniform prior ranging from $-50$ to 50, with 
$\alpha = 18^{+19}_{-28}$ (OHD+PaSNe) being the most restrictive 
value. This is a poor constraint on $\alpha$, which can increase 
the degeneracy between other parameters of the EP model.

The contour plot between $\alpha$ and $\Omega_{k,0}$ shows an 
anti-correlation (see Fig.~\ref{fig:results_pdfs_EP}), where the 
extended tails are slightly shifted to negative and positive values 
of $\Omega_{k,0}$ and $\alpha$, respectively. 
For other 2D contours, we 
observed no 
significant change 
with respect to the non-flat $\Lambda$CDM. 
For instance, the $\Omega_{\phi,0}-\Omega_{k,0}$ PDFs appear to be 
broader than the the non-flat $\Lambda$CDM cases. This suggests 
that the $\alpha-\Omega_{k,0}$ anti-correlation could be independent 
of other parameters, although we must keep in mind that $\alpha$ 
is poorly characterized. 

To avoid the effects of 
considering 
$\alpha$ as 
a
free parameter, we explored the EP model with $\alpha=1$ (called EPf, 
Fig.~\ref{fig:results_pdfs_EPf}). This case provides results more 
consistent with the $\Lambda$CDM than 
with
those from the EP model. 
For instance, the OHD+PaSNe case is performed with 
$\chi^2_{\rm red}=0.96$ for both 
the
non-flat $\Lambda$CDM and 
the
EPf models, 
with $\Delta \rm AIC=-0.1$ and $\Delta \rm BIC =-0.1$,
which, on a
negligible level,
 favor the EPf results.
We recover a positive-curvature $\Omega_{k,0} =0.047^{+0.054}_{-0.055}$, 
$H_0 = 69.4\pm 1.0\,\rm km\,s^{-1}\,Mpc^{-1}$, and 
$\Omega_{\phi,0}=0.692\pm0.039$.
In addition, the EPf model prefers positive $\Omega_{k,0}$ 
at 0.8$\sigma$, in contrast to the negative curvature at $1.7\sigma$ 
from the EP model which is slightly more shifted towards $\Omega_{k,0}=0$.
Moreover, the marginalized PDFs of $\Omega_{k,0}$ are symmetrical
 and become more sharpened than in the cases of the EP analysis. 
These two features are also extended to the behavior of the 2D 
contours for $\Omega_{\phi,0}-\Omega_{k,0}$, which now are more 
compatible with the corner plots from the non-flat $\Lambda$CDM 
analysis.

%
%

\subsubsection{Non-flat Inverse Quartic Potential models}

As for 
the EP model, the parameters ($\Omega_{\phi,0}$, 
$H_0$ and $\mathcal{M}$) and 
the goodness
of
the 
fits obtained from 
the IQP models agree with the results from the non-flat 
$\Lambda$CDM analyses (see Fig.~\ref{fig:results_pdfs_QRP}). 
For the OHD+PaSNe, the $\chi^2=0.96$ is the same for both non-flat 
$\Lambda$CDM and IQP models, yielding 
$\Omega_{k,0}=0.035^{+0.074}_{-0.069}$, 
$H_0 = 69.4\pm 1.0\,\rm km\,s^{-1}\,Mpc^{-1}$,
 and $\Omega_{\phi,0}=0.696\pm0.041$. The positive curvature moves 
slightly (with respect to the non-flat $\Lambda$CDM model) towards 
$\Omega_{k,0}=0$ at a level of $0.5\sigma$, although the uncertainty 
is broader in the IQP model. In addition, the $\Delta \rm BIC=6.6$ 
and moderate $\Delta \rm AIC = 1.5$ favor
the non-flat $\Lambda$CDM over the IQP models with freely sampled of $\delta$.

The marginalized PDFs of $\delta$ are asymmetrical, particularly 
for the OHD and PaSNe cases, with tails extending up to 
unreasonable values of 40000 in the worst case. We 
can
therefore 
provide only a weak description of the $\delta$. The sharper PDF 
results from the OHD+PaSNe case, even though its peak is located 
near 
the lower bound of the uniform prior imposed on $\delta$. 
%
The OHD and PaSNe analysis also show a significant 
impact on the shape of the $\Omega_{k,0}$ PDFs, mainly owing to 
the irregular description of the $\delta$ parameter. For the OHD 
analysis, the peak of the PDF reaches the upper bound of the uniform 
prior imposed on $\Omega_{k,0}$, while the curve is wider for the 
PaSNe case, 
peaking around 0.1.
We need a more restrictive prior to evaluate possible correlation 
between
$\delta$ and other parameters such 
as
$\Omega_{k,0}$ (as for
the EP analysis).

For the case with $\delta=1$ (IQPf), we obtained a parameter space 
in full agreement with the non-flat $\Lambda$CDM results, where the 
goodness
of
fit and comparison parameters ($\Delta \rm BIC$ and 
$\Delta \rm AIC$) are practically equal
for each database that 
explores
both models (Table~\ref{tab:results_parameters}). For example, 
the OHD+PaSNe case is performed with $\chi^2_{\rm red}=0.96$ for 
both non-flat $\Lambda$CDM and IQPf models, with $\Delta \rm AIC=-0.1$ 
and $\Delta \rm BIC =-0.1$ favoring (to
a negligible 
degree) the IQPf results.
The IQPf analysis provide sharper and more symmetrical PDF
behavior 
than the IQP model analysis (see Fig.~\ref{fig:results_pdfs_QRPf}). 
The correlation plots are 
also
now well delimited (with respect to the IQP models). 
The PDFs of $\Omega_{k,0}$ are described 
within
the range of the 
uniform prior considered for $\Omega_{k,0}$. In particular, the OHD 
and PaSNe curves 
do
not 
reach
the prior upper limits as their maximum. 
For example, in the OHD case, the peak is now at 
$\Omega_{k,0} \approx 0.48$ in contrasted to the value of $-$0.2 from 
the IQP model.
For the OHD+PaSNe analysis, we obtained 
$\Omega_{k,0}=0.050\pm0.054$, $H_0 = 69.4\pm 1.0\,\rm km\,s^{-1}\,Mpc^{-1}$ 
and $\Omega_{\phi,0}=0.692\pm0.039$, which is compatible with 
the non-flat $\Lambda$CDM results and with a flat universe.
%

\section{Conclusions}
\label{sec:conclusion} 

We presented a dynamic description of 
isotropic and homogeneous Friedmann-Lematre-Robertson-Walker cosmological 
models, with positive spatial curvature within the framework 
of mimetic gravity theory. We consider families of specific 
mimetic field potentials: the exponential potential
(EP), the inverse square potential (ISP),
 and the inverse quartic 
potential (IQP). In addition, the parameters of
the EP and IQP models were fitted with observational data from the 
Pantheon 
Type Ia Supernovae
database 
and the Observational Hubble 
database from cosmic chronometers.

First, we considered a generic potential for the mimetic field 
and determined the critical points of the dynamical system. We 
analyzed the stability of each of them using linear stability and 
the centre manifold method for cases with null eigenvalues of the 
variation matrix. To understand the dynamics of the system at 
infinity, we studied the global phase space of the dynamical 
system by projecting on the Poincar\'e sphere. The dynamical system 
is four-dimensional, with the exception of the ISP 
($V(\phi) \propto 1/\phi^2$) for which the dimension reduces to 
three. We explored the dynamics on invariant submanifolds
and revealed
trajectories that can evolve to a de Sitter universe and a division 
of the phase space in contracting and expanding epochs.

For the ISP ($V= \gamma / \phi^2$), the dynamical system reduced 
to a three-dimensional autonomous system. Trajectories starting 
from the matter-dominated critical line could evolve towards a 
future attractor representing a solution dominated by dark matter 
and dark energy, exhibiting acceleration under certain conditions. 
Additionally, we identified trajectories that can enter the 
contracting phase space and recollapse to a big crunch. The phase 
space of this potential 
does
 not include the solution representing 
Einstein static universe. 
For the EP ($V=V_0 e^{-\alpha \phi}$),
however,
 we found that one equation decouples 
from the others, resulting in a three dimensional reduced dynamical 
system. The dynamical evolution of the system depends on the sign of 
$\alpha$. The invariant submanifold $\zeta=0$ divides the phase 
space 
into
 two regions, such that trajectories with different signs 
of $\alpha$ only have dynamics in their respective regions, 
$\alpha<0$ ($\alpha >0 $) associated with the region 
$\zeta <0$ ($\zeta >0$). When $\alpha =0$, the trajectories are 
confined to the invariant submanifold $\zeta =0$. We mainly found 
trajectories with $\alpha <0$ that are past asympotic to the 
matter-dominated 
critical line, evolving towards a point corresponding to 
an expanding de Sitter accelerated solution dominated by the mimetic 
field $\Omega_{\phi} = 1$. 
We also
observed trajectories with 
$\alpha>0$ that are past asymptotic to the matter-dominated 
critical line, approximating the point dominated by the mimetic 
field, but then evolving to a point at infinity, and then going 
to another point, also at infinity. 
We also
found trajectories 
that
are 
both
past asymptotic 
and
future asymptotic to the 
Einstein static solution, 
as well as 
trajectories starting from the 
matter-dominated critical line that, after some expansion, begin 
to
contract
 and collapse to a big crunch. Therefore, by applying 
dynamical system techniques we found that cosmological models in 
mimetic gravity can support physically relevant solutions representing 
different stages of cosmic evolution.

Utilizing data from the Observational Hubble Dataset (OHD) 
and the Pantheon database of 
Type Ia Supernovae
(PaSNe),
observational constraints on the parameters of non-flat cosmological 
models (EP and IQP models) were obtained through Bayesian statistical 
analysis. In all cases, 
the $\Omega_{k,0}$ values are consistent with a flat Universe. In 
fact, the parameter space of non-flat models are in agreement with 
the parameter space obtained for models assuming flat curvature. For 
non-flat curvature cases, both the EP and IQP analyses produce results 
in agreement with the non-flat $\Lambda$CDM model. The current data 
provide weak constraints on the main parameters describing the EP 
and IQP models. For the EP model and the OHD+PaSNe case, we find 
$\alpha=18^{+19}_{–28}$,
 which shows an anticorrelation with $\Omega_{k,0}$. 
For the IQP model, $ \beta = –369^{+970}_{–470}$ is a weak constraint 
(the OHD+PaSNe case). Despite the provided observational constraints, 
the viability of mimetic gravity models is supported by the data sets, 
and these models 
can
effectively 
describe the late-time accelerated
 expansion of the universe. Comparison with the $\Lambda$CDM model 
using the AIC and BIC statistical indicators provided insights into 
the observational support of the models. In this context, the CMB 
data could improve our current 
constraints 
on the EP and IQP models.

\acknowledgments

We thank Terry Mahoney (Scientific Editorial Service of the IAC) 
for proofreading this manuscript.
This work is partially supported by ANID Chile through FONDECYT 
Grants Nº 1201673 (A.F. and J.C.H.) and Nº 1220871  (A. F. and Y. V.). 
Y.V. acknowledges the financial support of DIDULS/ULS, through the
 project PTE2121313. This work is also funded by ANID$-$Millen\-nium Program$-$ICN2019\_044 (Chile).
The authors wish to thank the FIULS 2030 project 18ENI2-104235 -- CORFO 
for providing computing resources. 
CHLC acknowledges financial support by the Spanish Ministry of Science 
and Innovation under the projects AYA2017-84185-P and PID2020-120514GB-I00, 
and the ACIISI, Consejeria de Economia, Conocimiento y Empleo del 
Gobierno de Canarias and the European Regional Development Fund (ERDF) 
under grant with reference ProID2020010108.

\appendix
\section{Dynamical character of points $C_{2+}$ and $C_{2-}$}
\label{appendixa}

In this appendix we show the dynamical character of the critical 
points $C_{2+}$ and $C_{2-}$ for the different values of $\zeta$ and $\mu'(\zeta)$.

\begin{figure}[H] 
\centering
\resizebox{1\columnwidth}{!}{
\tikzstyle{level 1}=[level distance=15mm, sibling distance=18mm]
\tikzstyle{level 2}=[level distance=20mm, sibling distance=5mm]
\tikzstyle{level 3}=[level distance=30mm]
\begin{tikzpicture}[grow=right,->,>=angle 60]
  \node {$C_{2+}$}
    child {node {$\zeta>2$}
        child[-] {node{saddle}}
        }
    child {node {$\zeta=2$}
      child {node {$\mu'(\zeta)>0$}
        child[-] {node{non-hyperbolic}}  
      }
            child {node {$\mu'(\zeta) =0$}
        child[-] {node{non-hyperbolic}}  
      }
      child {node {$\mu'(\zeta)<0$}
          child[-] {node{non-hyperbolic, saddle}}  
      }
      }
    child {node {$\zeta<2$}
        child[-] {node{$\mu'(\zeta)>0$}
        child[-] {node{hyperbolic, attractor}}}  
     child {node {$\mu'(\zeta)=0$}
        child[-] {node{non-hyperbolic}}  
      }
      child {node{$\mu'(\zeta)<0$}
        child[-] {node{hyperbolic, saddle}}  
      }
    }
    ;
\end{tikzpicture}
\tikzstyle{level 1}=[level distance=10mm, sibling distance=35mm]
\tikzstyle{level 2}=[level distance=20mm, sibling distance=15mm]
\tikzstyle{level 3}=[level distance=30mm, sibling distance= 5mm]
\begin{tikzpicture}[grow=right,->,>=angle 60]
  \node {$C_{2-}$}
    child {node {$\omega = 1/3$}{
    child {node {$\zeta<-2$}
      child {node {$\mu'(\zeta)>0$}
        child[-] {node{hyperbolic, atractor}}  
      }
            child {node {$\mu'(\zeta) =0$}
        child[-] {node{non-hyperbolic}}  
      }
      child {node {$\mu'(\zeta)<0$}
          child[-] {node{hyperbolic, saddle}}  
      }
      }
    child {node {$\zeta=-2$}
      child {node {$\mu'(\zeta)>0$}
        child[-] {node{non-hyperbolic}}  
      }
            child {node {$\mu'(\zeta) =0$}
        child[-] {node{non-hyperbolic}}  
      }
      child {node {$\mu'(\zeta)<0$}
          child[-] {node{non-hyperbolic, saddle}}  
      }
      }
    child {node {$\zeta>-2$}
        child[-] {node{saddle}  
      }}
    }}
    child {node {$\omega = 0$}{
    child {node {$\zeta<-2$}
        child[-] {node{saddle}}
        }
    child {node {$\zeta=-2$}
      child {node {$\mu'(\zeta)>0$}
        child[-] {node{non-hyperbolic, saddle}}  
      }
            child {node {$\mu'(\zeta) =0$}
        child[-] {node{unstable}}  
      }
      child {node {$\mu'(\zeta)<0$}
          child[-] {node{unstable}}  
      }
      }
    child {node {$\zeta>-2$}
        child[-] {node{$\mu'(\zeta)>0$}
        child[-] {node{hyperbolic, saddle}}}  
     child {node {$\mu'(\zeta)=0$}
        child[-] {node{non-hyperbolic, unstable}}  
      }
      child {node{$\mu'(\zeta)<0$}
        child[-] {node{hyperbolic, repulsor}}  
      }}
    }
    }
    ;
\end{tikzpicture}
}
\label{points}
\end{figure}

\section{Centre manifold method}
\label{appendixb}

In this appendix we describe the centre manifold method for 
analyzing the stability of critical points that cannot be 
categorized as hyperbolic stable or unstable 
owing 
to the presence 
of null eigenvalues in the variation matrix. We illustrate the method 
by 
specifically focusing on the critical point $B_+$ for 
$\omega = 0$, $(\zeta^2 \mu(\zeta))|_{\zeta=0}$ and 
$\frac{d}{d\zeta}(\zeta^2 \mu(\zeta))|_{\zeta=0}$.  To enhance the 
clarity and conciseness in our expressions, we define 
$M(\zeta) = -\sqrt{3}\zeta^2 \mu(\zeta)$.
\newline

The critical point $B_+$ with coordinates $(1, 0, 1, 0)$ can be 
shifted to the origin by applying the transformation $Q \rightarrow Q+1$ 
and $\beta \rightarrow \beta+1$. After this transformation, the 
critical point is located at the origin and the Jacobian matrix reads
\begin{equation}
J_{B_+} = \begin{pmatrix}
  -2  & 0 & 0 & 0  \\
   0 & -3   &  0  & \sqrt{3} \\
   0  &  0 & -3 & -\frac{\sqrt{3}}{2}  \\
   0  &  0 & 0 &  0
\end{pmatrix} \,.
\end{equation}

Now, 
diagonalizing 
the Jacobian matrix with the matrix of 
eigenvectors, given by
\begin{equation}
 S = \begin{pmatrix}
  1  & 0 & 0 & 0  \\
   0 & 1   &  0  & 1 \\
   0  &  0 & 1 & -\frac{1}{2}  \\
   0  &  0 & 0 &  \sqrt{3}
\end{pmatrix} \,,
\end{equation}
and introducing the new set of variables
\begin{equation}
\begin{pmatrix}
y_1 \\
y_2 \\
y_3 \\
x
\end{pmatrix} = S^{-1} \begin{pmatrix}
Q \\ \tilde{\Omega}_{\lambda}  \\  \beta  \\ \zeta
\end{pmatrix} \,,
\end{equation}
the system of equations transforms to
\begin{eqnarray} \label{center}
 x' &=& A x + f(x, \mathbf{y})   \\
 \mathbf{y}' &=& \mathbf{B} \mathbf{y} + \mathbf{g}(x, \mathbf{y})
\end{eqnarray}
where
\begin{equation}
 \mathbf{y} = \begin{pmatrix}
y_1 \\
y_2 \\
y_3
\end{pmatrix} \,, \,\,\,  A = 0\,, \,\,\,   \mathbf{B} = \begin{pmatrix}
  -2  & 0 & 0  \\
   0 & -3   &  0 \\
   0  &  0 & -3
\end{pmatrix} \,, \,\,\,  \mathbf{g} = \begin{pmatrix}
g_1 \\
g_2 \\
g_3 \
\end{pmatrix} \,,
\end{equation}
and $f$, $\mathbf{g}$ are nonlinear functions of the coordinates $x$ 
and $\mathbf{y}$. For brevity, we omit explicit expressions for $f$ 
and $\mathbf{g}$. 
\newline

The center manifold for (\ref{center}) can be locally represented as
\begin{equation}
\{(x, \mathbf{y}) | \mathbf{y} = \mathbf{h}(x), \mathbf{h}(0) = 0, 
\mathbf{\nabla} \mathbf{h}(0) = 0  \} \,,
\end{equation}

\noindent where $\mathbf{h}(x) = \begin{pmatrix}  h_1(x)   \\ h_2(x)  \\ h_3(x) \end{pmatrix}$ 
satisfies the following quasilinear differential equation
\begin{equation} \label{h}
\mathbf{\nabla} \mathbf{h}(x) (A x + f(x, \mathbf{h}(x))) - \mathbf{B} \mathbf{h}(x) - \mathbf{g}(x, \mathbf{h}(x)) 
= 0 \,.
\end{equation}
\newline

For the purpose of stability, a series expansion of $\mathbf{h}(x)$ 
is considered, where
\begin{equation}
h_i(x) = a_i x^2 + b_i x^3 +\mathcal{O}(x^4)  \,\, \text{for} \,\, i = 1,2,3 \,.
\end{equation}
Additionally, the function $M(x)$ is assumed to have a Taylor series 
expansion around $x=0$, with $M(0)=0$ and $M'(0)=0$.
By inserting these expressions into Eq.\ (\ref{h}), the coefficients 
$a_i$ and $b_i$ can be obtained. Consequently, the dynamics of the 
system reduced to the centre manifold is given by
\begin{equation}
u' = A u + f(u, \mathbf{h}(u)) \,.
\end{equation}
Upon inserting the functions $h_i$, the expression becomes
\begin{equation}
u' = \frac{M''(0)}{2 \sqrt{3}} x^2 + \left( -\frac{M''(0)}{4 \sqrt{3}} 
+ \frac{M^{(3)}(0)}{6 \sqrt{3}}  \right) x^3 + \mathcal{O}(x^4) \,.
\end{equation}

\section{Poincar\'e projection method}
\label{appendixc}

In this appendix we review the Poincar\'e projection method that we used 
to analyze the critical points at infinity. Here,
we apply the method for a general three-dimensional dynamical system; 
however, it can be straightforwardly generalized to higher-dimensional 
system.
\newline

Consider the following autonomous dynamical system described by the set of equations
\begin{equation}
    \dot{x} = P(x, y, w) \,, \,\,\,\, \dot{y} = Q(x, y, w) \,,  \,\,\,\,   \dot{w} = R(x, y, w)\,.
\end{equation}

We 
 assume that the variable $w$ is compact, and
that 
 $P$, $Q$ and $R$ 
are polynomials functions of $x$, $y$ and $w$. Our interest is to 
describe the behavior of the system at infinity, $x \rightarrow \infty$ 
and/or $y \rightarrow \infty$. Also, we consider $P$ and $Q$ polynomials 
of maximum degree $m$ of the non-compact variables $x$ and $y$, whereas $R$ 
is 
a polynomial of maximun degree $m-1$ of the non-compact variables, this is 
the case of section \ref{section} with $m = 3$. Then, compactifying the 
phase space by means of the following change of variable
\begin{equation}
    X = \frac{x}{\sqrt{1+x^2+y^2}} \,, \,\,\,\, Y = 
\frac{y}{\sqrt{1+x^2+y^2}} \,, \,\,\,\, Z = \frac{1}{\sqrt{1+x^2+y^2}} \,,
\end{equation}
we obtain
\begin{eqnarray}
 \nonumber \dot{X} &=& Z \left( (1-X^2)P -X YQ \right) \,, \\
 \nonumber \dot{Y} &=& Z \left( -X Y P +(1-Y^2) Q \right) \,, \\
  \dot{Z} &=& - Z^2 (XP+YQ) \,.
\end{eqnarray}
Now, the behavior near $Z = 0$ is given by:
\begin{eqnarray}
\nonumber P(x ,y ,w) &=& P\left( \frac{X}{Z}, \frac{Y}{Z}, w \right) \sim Z^{-m} \,, \\
\nonumber Q(x ,y ,w) &=& Q\left( \frac{X}{Z}, \frac{Y}{Z}, w \right) \sim Z^{-m} \,, \\
 R(x, y, w) &=& R\left( \frac{X}{Z}, \frac{Y}{Z}, w \right) \sim Z^{-m+1} \,.
\end{eqnarray}
Defining $P^* = Z^m P$, $Q^* = Z^m Q$ and $R^* = Z^{m-1}R$, 
performing the change of variable $d \tau = Z^{1-m} dt$, and setting $Z=0$, we 
then 
arrive at
\begin{eqnarray}
 \nonumber X' &=& - Y (X Q^* - Y P^*) \,,  \\
 \nonumber Y' &=& X (X Q^* -Y P^*) \,, \\
 w' &=& R^*  \,.
\end{eqnarray}

\section{Stability at infinity}
\label{appendixd}
In this appendix we show a way of analyzing the stability at 
infinity using a second projection of any point 
in
the space 
described by the Poincaré projection to a plane 
tangential 
to it, 
in this way avoiding the divergences of the eigenvalues.
First of all, it should be said that it is necessary for some 
free axes to coincide locally in direction and sign with the 
projection, so the signs are adjusted manually. Below, without 
loss of generality, we consider a 2-dimensional example to 
understand this concept. This is easily extensible to more dimensions.
Consider the following autonomous dynamical system described by 
the set of equations:
\begin{equation}
    \dot{x} = P(x, y) \,, \,\,\,\, \dot{y} = Q(x, y) \,. 
\end{equation}
%
We can 
then 
compactify the phase space, as explained in the 
previous appendix, 
by
using the following change of variables:
\begin{equation}
    X = \frac{x}{\sqrt{1+x^2+y^2}} \,, \,\,\,\, Y = 
\frac{y}{\sqrt{1+x^2+y^2}} \,, \,\,\,\, Z = \frac{1}{\sqrt{1+x^2+y^2}} \,.
\end{equation}
Now
suppose we want to analyze critical points for $X<1$.
We
choose $X<1$ to note the choice of sign, therefore:
\begin{equation}
    \xi = -\frac{Y}{X}=-\frac{y}{x} \,, \,\,\,\, \zeta 
= -\frac{Z}{X}=-\frac{1}{x} \,. 
\end{equation}
Suppose that the order of the largest homogeneous polynomial 
for $P(x,y)$ and $Q(x,y)$ in $x,y$ is $m$. 
By performing 
a change in the time coordinate such that
$dt=\zeta^{m-1} d\tau$, 
the resulting system will
then 
 be:
\begin{equation}
    \xi' = \zeta^m \left(Q(\frac{-1}{\zeta},\frac{\xi}{\zeta})
+\xi P(\frac{-1}{\zeta},\frac{\xi}{\zeta}) \right)\,, \,\,\,\, \zeta' 
= \zeta^{m+1} P(\frac{-1}{\zeta},\frac{\xi}{\zeta}) \,. 
\end{equation}

\section{Degenerate points}
\label{appendixe}
In this appendix we present an analysis for degenerate points. 
A fixed point is called degenerate if all the eigenvalues of the 
variation matrix at the 
point
analyzed 
are zero, 
thus
rendering the 
centre manifold analysis impractical.
The chosen method for addressing this is a change of coordinates 
to 
a
polar
system.
 Tipically it is referred to as a little degenerate point 
if the behavior of the point becomes evident through this 
(or another)
simple
transformation.
It is worth noting
that this approach can be readily generalized to higher dimensions.
A
two-dimensional 
dynamical system is 
then 
considered:
\begin{equation}
    \dot{x} = P(x, y) \,, \,\,\,\, \dot{y} = Q(x, y) \,. 
\end{equation}
In addition, the origin is considered as a critical point to 
be analyzed. 
The
change from coordinates to polar is then made:
\begin{equation}
    r = \sqrt{x^2+y^2} \,, \,\,\,\, \theta = \arctan(y/x) \,. 
\end{equation}
Therefore, the dynamical system can be expressed as follows:
\begin{eqnarray}
    r' &=& c P(r c,r s)+s Q(r c,r s) \,, \\
        \theta'&=& c^2 \left( \frac{Q(r c,r s)}{r c}-
    \frac{s}{r c^2} P(r c, r s) \right) \,,
\end{eqnarray}
where, we have defined $c=\cos(\theta)$ and $ s=\sin(\theta)$ 
for convenience.
\newline

For purposes of analyzing in the neighborhood of the origin, it 
is convenient to separate $(x,y)$ into homogeneous polynomials, 
which in turn can be expressed as polynomials in $r$ in polar terms, 
to finally stay with those of lower order:
%
\begin{eqnarray}
        P(x, y) && \thickapprox P_n (x,y)=r^n P_n(c,s) \,,  \\
        Q(x, y) && \thickapprox Q_n (x,y)=r^n Q_n(c,s) \,,
\end{eqnarray}
where $n$ is the smallest order between $P$ and $Q$, and at least 
one of them not null. The
objective is 
then 
to search for directions
 that locally behave as an invariant local manifold or,
 equivalently, 
to find a curve that reaches or departs from $0$ with a slope of 
$\tan \theta_0$, where $\theta_0$ is the value that cancels $\theta '$.
This idea is encapsulated in the following theorem: 
\cite{aranda_iriarte_1998}\newline \newline
``Suppose $\theta '=0$ have $q \neq 0$ real zeros $\theta_1,...,\theta_q$ close to the origin in $]–\pi /2,\pi /2]$. If all of them 
are simple or if, for
those $\theta_i$
 that are multiple, the term that accompanies the smallest order 
of $r'$ is non-zero, there exists for each 
 $i=1,...,q$ at least
one manifold that goes to 0 with a slope of $\tan \theta_i$. The local 
flux over each of them is determined by the sign of the term that 
accompanies the lowest order of $r'$  if 
it is 
non-zero or
can be specified by calculating some term of the development of the 
manifold if it is zero. On the contrary, these flows and the sign of $\theta'$ in
 $[0,2 \pi]$ determine the local structure of the lowest-order term 
of $\theta'$'' .\newline

Having said this, to determine the 
above-mentioned 
manifold, we use 
approximations for small $r$. Consequently,
\begin{equation}
    \theta' \thickapprox r^{n-1} \left(c Q_n(c,s)-s P_n(c,s) \right) 
\end{equation}
for some $\theta_ 0$. 
It then
becomes necessary to evaluate the 
sign of the following expression
(note that $\theta_0 + \pi$ is also 
a solution and should be evaluated if necessary):
\begin{equation}
    r' \thickapprox r^n \left( c P_n(c,s)+s Q_n(c,s)\right) \,.
\end{equation}
The sign of this expression, whether positive or negative, indicates 
whether the flow is repulsive or attractive in this direction. If the 
analysis proves to be inconclusive using this method, a graphical 
analysis will be carried out, as other types of analysis 
are beyond 
the intended scope of this 
paper.

\section{Flat cosmological models}
\label{appendix:flat_universe}
In this appendix we present the results obtained 
by
considering the 
flat case for the 
various
models presented in 
section \ref{sec:description_model_Hz}.

\begin{table}[h!!!!!]
\begin{center}
\resizebox{\columnwidth}{!}{
\begin{tabular}{| c | c  c  c   c  c | c  c   c c  c c | c |}
\hline
Dataset & $\Omega_{\phi,0}$ & $\alpha$ & $\delta$  &  $H_0$ & $\mathcal{M}$ & $\chi_{\rm min}^2$ & $\chi_{\rm red}^2$ & AIC & BIC & $\Delta\,{\rm AIC}$ & $\Delta\,{\rm BIC}$\\ 
\hline
\multicolumn{12}{ |c| }{$\Lambda$CDM} \\
\hline
OHD & $0.741^{+0.016}_{-0.017}$ & $-$ & $-$ &  $70.69\pm 1.13$ & $-$ & $27.5$ & $0.56$ & $31.5$ & $35.4$ & $-1.0$ & $-2.9$  \\ 
PaSNe & $0.706^{+0.019}_{-0.020} $ & $-$ & $-$ &$73.20^{+4.99}_{-5.00}$ & $23.81\pm 0.01$ & $1027.4$ & $0.98$& $1033.4$ & $1048.3$ & $-1.8$ & $-6.7$ \\
PaSNe+OHD & $0.726^{+0.013}_{-0.014}$ & $-$ & $-$ & $69.78^{+0.95}_{-0.94} $ & $23.80\pm 0.01$ & $1056.6$ & $0.96$ & $1064.6$ & $1084.6$ & $1.6$ & $1.6$ \\
\hline
\multicolumn{12}{ |c| }{EP}  \\
\hline
OHD & $0.717\pm 0.046$ & $8.64^{+15.23}_{-15.30}$ & $-$ & $70.24^{+1.38}_{-1.34}$ & $-$ & $26.9$ &  $0.56$ & $32.9$ & $38.7$ & $0.4$ & $0.4$\\
PaSNe & $0.720^{+0.033}_{-0.037}$ & 
$-13.48^{+26.18}_{-22.49}$ & $-$ & $73.39^{+4.97}_{-4.88}$ & $23.81\pm 0.01$ & $1027.4$ & $0.98$ &$1035.4$ & $1055.2$ & $0.2$ & $0.2$\\
PaSNe+OHD & $0.692^{+0.031}_{-0.030}$ & $14.82^{+11.95}_{-12.57}$ & $-$ &  $69.44\pm 0.96$ & $23.81\pm 0.01$ & $1054.8$ & $0.96$ &$1062.8$ & $1082.8$ & $-0.2$ & $-0.2$ \\\hline
\multicolumn{12}{ |c| }{EPf} \\
\hline
    OHD & $ 0.739^{+0.016}_{-0.017}$ & $-$ & $-$ & $70.68^{+1.11}_{-1.15}$ & $-$ & $27.4$ & $0.56$ & $31.4$ & $35.3$ & $-1.1$ & $-3.0$ \\
    PaSNe  & $0.705^{+0.019}_{-0.020}$ & $-$ & $-$ &  $73.23^{+4.96}_{-4.98}$ & $23.81 \pm 0.01$ & $1027.4$ & $0.98$ & $1033.4$ & $1048.3$ & $-1.8$ & $-6.7$\\
PaSNe+OHD & $0.724^{+0.013}_{-0.014}$ & $-$ & $-$ & $69.79^{+0.91}_{-0.95}$ & $23.80\pm 0.01$ & $1056.3$ & $0.96$ & $1062.3$ & $1077.3$ & $-0.7$ & $-5.7$\\\hline
\multicolumn{12}{ |c| }{IQP} \\
\hline
    OHD & $0.734^{+0.031}_{-0.025}$ & $-$ & $-267.6^{+816.0}_{-514.1}$ &  $70.52^{+1.35}_{-1.28}$ & $-$ & $27.0$ & $0.56$  &$33.0$ & $38.8$ & $0.5$ & $0.5$ \\
PaSNe & $0.732^{+0.032}_{-0.029}$ & $-$ & $3765.2^{+4633.7}_{-3047.7}$ & $73.85^{+4.98}_{-4.91}$ & $23.80\pm0.01$ & $1027.8$ &  $0.98$  &$1035.8$ & $1055.6$ & $0.6$ & $0.6$ \\
PaSNe+OHD & $0.713^{+0.017}_{-0.016}$ & $-$ & $-575.3^{+517.0}_{-301.6}$ & $69.56^{+0.97}_{-0.95}$ & $23.80\pm 0.01$ & $1054.5$ & $0.96$ &$1062.5$ & $1082.5$ & $-0.5$ & $-0.5$\\
\hline
\multicolumn{12}{ |c| }{IQPf} \\
\hline
OHD & $0.742^{+0.016}_{-0.017}$ & $-$ & $-$ & $70.72^{+1.12}_{-1.13}$ & $-$ & $27.5$ & $0.56$ &$31.5$ & $35.4$ & $-1.0$ & $-2.9$\\
PaSNe & $0.706^{+0.019}_{-0.020}$ & $-$ & $-$ &  $73.24^{+4.99}_{-4.96}$ & $23.81\pm 0.01$ & $1027.4$ & $0.98$ & $1033.4$  &$ 1048.3$ & $-1.8$ & $-6.7$ \\
PaSNe+OHD & $0.726^{+0.013}_{-0.014}$ & $-$ & $-$ &  $69.78\pm0.93$ & $23.80\pm 0.01$ & $1056.6$ & $0.96$  &$1062.6$ & $1077.6$ & $-0.4$ & $-5.4$ \\
\hline
\end{tabular}
}
\caption{Parameters and uncertainties obtained for the cosmological 
models investigated assuming a spatial flat universe. The 
identification names and field potentials for the $\Lambda$CDM 
model and mimetic gravity models (EP, EPf, IQP and IQPf) are 
listed in Table~\ref{tab:models_priors}. The properties of these
cosmological models (with $\Omega_{k,0}=0$) are also obtained 
using three data sets: the Observational Hubble database from 
the cosmic chronometers (OHD), the Pantheon database from 
Type Ia 
supernovae 
(PaSNe) and the simultaneous analysis of the OHD and 
PaSNe data sets (OHD+PaSNe). The statistical parameters $\Delta \rm AIC$ 
and $\Delta \rm BIC$ are computed with respect to the non-flat 
cosmological models (presented in Table~\ref{tab:results_parameters}).}
\label{tab:results_LCDM}
\end{center}
\end{table}

\bibliography{Cosmology_biblio}

\begin{thebibliography}{73}
\expandafter\ifx\csname natexlab\endcsname\relax\def\natexlab#1{#1}\fi
\expandafter\ifx\csname bibnamefont\endcsname\relax
  \def\bibnamefont#1{#1}\fi
\expandafter\ifx\csname bibfnamefont\endcsname\relax
  \def\bibfnamefont#1{#1}\fi
\expandafter\ifx\csname citenamefont\endcsname\relax
  \def\citenamefont#1{#1}\fi
\expandafter\ifx\csname url\endcsname\relax
  \def\url#1{\texttt{#1}}\fi
\expandafter\ifx\csname urlprefix\endcsname\relax\def\urlprefix{URL }\fi
\providecommand{\bibinfo}[2]{#2}
\providecommand{\eprint}[2][]{\url{#2}}

\bibitem[{\citenamefont{Riess et~al.}(1998)}]{SupernovaSearchTeam:1998fmf}
\bibinfo{author}{\bibfnamefont{A.~G.} \bibnamefont{Riess}} \bibnamefont{et~al.}
  (\bibinfo{collaboration}{Supernova Search Team}), \bibinfo{journal}{Astron.
  J.} \textbf{\bibinfo{volume}{116}}, \bibinfo{pages}{1009}
  (\bibinfo{year}{1998}), \eprint{astro-ph/9805201}.

\bibitem[{\citenamefont{Perlmutter
  et~al.}(1999)}]{SupernovaCosmologyProject:1998vns}
\bibinfo{author}{\bibfnamefont{S.}~\bibnamefont{Perlmutter}}
  \bibnamefont{et~al.} (\bibinfo{collaboration}{Supernova Cosmology Project}),
  \bibinfo{journal}{Astrophys. J.} \textbf{\bibinfo{volume}{517}},
  \bibinfo{pages}{565} (\bibinfo{year}{1999}), \eprint{astro-ph/9812133}.

\bibitem[{\citenamefont{Spergel et~al.}(2003)}]{WMAP:2003elm}
\bibinfo{author}{\bibfnamefont{D.~N.} \bibnamefont{Spergel}}
  \bibnamefont{et~al.} (\bibinfo{collaboration}{WMAP}),
  \bibinfo{journal}{Astrophys. J. Suppl.} \textbf{\bibinfo{volume}{148}},
  \bibinfo{pages}{175} (\bibinfo{year}{2003}), \eprint{astro-ph/0302209}.

\bibitem[{\citenamefont{Ade et~al.}(2014)}]{Planck:2013pxb}
\bibinfo{author}{\bibfnamefont{P.~A.~R.} \bibnamefont{Ade}}
  \bibnamefont{et~al.} (\bibinfo{collaboration}{Planck}),
  \bibinfo{journal}{Astron. Astrophys.} \textbf{\bibinfo{volume}{571}},
  \bibinfo{pages}{A16} (\bibinfo{year}{2014}), \eprint{1303.5076}.

\bibitem[{\citenamefont{Eisenstein et~al.}(2005)}]{SDSS:2005xqv}
\bibinfo{author}{\bibfnamefont{D.~J.} \bibnamefont{Eisenstein}}
  \bibnamefont{et~al.} (\bibinfo{collaboration}{SDSS}),
  \bibinfo{journal}{Astrophys. J.} \textbf{\bibinfo{volume}{633}},
  \bibinfo{pages}{560} (\bibinfo{year}{2005}), \eprint{astro-ph/0501171}.

\bibitem[{\citenamefont{Sotiriou and Faraoni}(2010)}]{Sotiriou:2008rp}
\bibinfo{author}{\bibfnamefont{T.~P.} \bibnamefont{Sotiriou}} \bibnamefont{and}
  \bibinfo{author}{\bibfnamefont{V.}~\bibnamefont{Faraoni}},
  \bibinfo{journal}{Rev. Mod. Phys.} \textbf{\bibinfo{volume}{82}},
  \bibinfo{pages}{451} (\bibinfo{year}{2010}), \eprint{0805.1726}.

\bibitem[{\citenamefont{Starobinsky}(1980)}]{Starobinsky:1980te}
\bibinfo{author}{\bibfnamefont{A.~A.} \bibnamefont{Starobinsky}},
  \bibinfo{journal}{Phys. Lett. B} \textbf{\bibinfo{volume}{91}},
  \bibinfo{pages}{99} (\bibinfo{year}{1980}).

\bibitem[{\citenamefont{Nojiri et~al.}(2017{\natexlab{a}})\citenamefont{Nojiri,
  Odintsov, and Oikonomou}}]{Nojiri:2017ncd}
\bibinfo{author}{\bibfnamefont{S.}~\bibnamefont{Nojiri}},
  \bibinfo{author}{\bibfnamefont{S.~D.} \bibnamefont{Odintsov}},
  \bibnamefont{and} \bibinfo{author}{\bibfnamefont{V.~K.}
  \bibnamefont{Oikonomou}}, \bibinfo{journal}{Phys. Rept.}
  \textbf{\bibinfo{volume}{692}}, \bibinfo{pages}{1}
  (\bibinfo{year}{2017}{\natexlab{a}}), \eprint{1705.11098}.

\bibitem[{\citenamefont{Nojiri and Odintsov}(2007)}]{Nojiri:2006be}
\bibinfo{author}{\bibfnamefont{S.}~\bibnamefont{Nojiri}} \bibnamefont{and}
  \bibinfo{author}{\bibfnamefont{S.~D.} \bibnamefont{Odintsov}},
  \bibinfo{journal}{J. Phys. Conf. Ser.} \textbf{\bibinfo{volume}{66}},
  \bibinfo{pages}{012005} (\bibinfo{year}{2007}), \eprint{hep-th/0611071}.

\bibitem[{\citenamefont{Chamseddine and Mukhanov}(2013)}]{Chamseddine:2013kea}
\bibinfo{author}{\bibfnamefont{A.~H.} \bibnamefont{Chamseddine}}
  \bibnamefont{and} \bibinfo{author}{\bibfnamefont{V.}~\bibnamefont{Mukhanov}},
  \bibinfo{journal}{JHEP} \textbf{\bibinfo{volume}{11}}, \bibinfo{pages}{135}
  (\bibinfo{year}{2013}), \eprint{1308.5410}.

\bibitem[{\citenamefont{Lim et~al.}(2010)\citenamefont{Lim, Sawicki, and
  Vikman}}]{Lim:2010yk}
\bibinfo{author}{\bibfnamefont{E.~A.} \bibnamefont{Lim}},
  \bibinfo{author}{\bibfnamefont{I.}~\bibnamefont{Sawicki}}, \bibnamefont{and}
  \bibinfo{author}{\bibfnamefont{A.}~\bibnamefont{Vikman}},
  \bibinfo{journal}{JCAP} \textbf{\bibinfo{volume}{05}}, \bibinfo{pages}{012}
  (\bibinfo{year}{2010}), \eprint{1003.5751}.

\bibitem[{\citenamefont{Jirou\v{s}ek et~al.}(2022)\citenamefont{Jirou\v{s}ek,
  Shimada, Vikman, and Yamaguchi}}]{Jirousek:2022kli}
\bibinfo{author}{\bibfnamefont{P.}~\bibnamefont{Jirou\v{s}ek}},
  \bibinfo{author}{\bibfnamefont{K.}~\bibnamefont{Shimada}},
  \bibinfo{author}{\bibfnamefont{A.}~\bibnamefont{Vikman}}, \bibnamefont{and}
  \bibinfo{author}{\bibfnamefont{M.}~\bibnamefont{Yamaguchi}}
  (\bibinfo{year}{2022}), \eprint{2212.14867}.

\bibitem[{\citenamefont{Sebastiani et~al.}(2017)\citenamefont{Sebastiani,
  Vagnozzi, and Myrzakulov}}]{Sebastiani:2016ras}
\bibinfo{author}{\bibfnamefont{L.}~\bibnamefont{Sebastiani}},
  \bibinfo{author}{\bibfnamefont{S.}~\bibnamefont{Vagnozzi}}, \bibnamefont{and}
  \bibinfo{author}{\bibfnamefont{R.}~\bibnamefont{Myrzakulov}},
  \bibinfo{journal}{Adv. High Energy Phys.} \textbf{\bibinfo{volume}{2017}},
  \bibinfo{pages}{3156915} (\bibinfo{year}{2017}), \eprint{1612.08661}.

\bibitem[{\citenamefont{Chamseddine et~al.}(2014)\citenamefont{Chamseddine,
  Mukhanov, and Vikman}}]{Chamseddine:2014vna}
\bibinfo{author}{\bibfnamefont{A.~H.} \bibnamefont{Chamseddine}},
  \bibinfo{author}{\bibfnamefont{V.}~\bibnamefont{Mukhanov}}, \bibnamefont{and}
  \bibinfo{author}{\bibfnamefont{A.}~\bibnamefont{Vikman}},
  \bibinfo{journal}{JCAP} \textbf{\bibinfo{volume}{06}}, \bibinfo{pages}{017}
  (\bibinfo{year}{2014}), \eprint{1403.3961}.

\bibitem[{\citenamefont{Nojiri and Odintsov}(2014)}]{Nojiri:2014zqa}
\bibinfo{author}{\bibfnamefont{S.}~\bibnamefont{Nojiri}} \bibnamefont{and}
  \bibinfo{author}{\bibfnamefont{S.~D.} \bibnamefont{Odintsov}}
  (\bibinfo{year}{2014}), \bibinfo{note}{[Erratum: Mod.Phys.Lett.A 29, 1450211
  (2014)]}, \eprint{1408.3561}.

\bibitem[{\citenamefont{Myrzakulov et~al.}(2015)\citenamefont{Myrzakulov,
  Sebastiani, and Vagnozzi}}]{Myrzakulov:2015qaa}
\bibinfo{author}{\bibfnamefont{R.}~\bibnamefont{Myrzakulov}},
  \bibinfo{author}{\bibfnamefont{L.}~\bibnamefont{Sebastiani}},
  \bibnamefont{and} \bibinfo{author}{\bibfnamefont{S.}~\bibnamefont{Vagnozzi}},
  \bibinfo{journal}{Eur. Phys. J. C} \textbf{\bibinfo{volume}{75}},
  \bibinfo{pages}{444} (\bibinfo{year}{2015}), \eprint{1504.07984}.

\bibitem[{\citenamefont{Astashenok et~al.}(2015)\citenamefont{Astashenok,
  Odintsov, and Oikonomou}}]{Astashenok:2015haa}
\bibinfo{author}{\bibfnamefont{A.~V.} \bibnamefont{Astashenok}},
  \bibinfo{author}{\bibfnamefont{S.~D.} \bibnamefont{Odintsov}},
  \bibnamefont{and} \bibinfo{author}{\bibfnamefont{V.~K.}
  \bibnamefont{Oikonomou}}, \bibinfo{journal}{Class. Quant. Grav.}
  \textbf{\bibinfo{volume}{32}}, \bibinfo{pages}{185007}
  (\bibinfo{year}{2015}), \eprint{1504.04861}.

\bibitem[{\citenamefont{Nojiri et~al.}(2016)\citenamefont{Nojiri, Odintsov, and
  Oikonomou}}]{Nojiri:2016vhu}
\bibinfo{author}{\bibfnamefont{S.}~\bibnamefont{Nojiri}},
  \bibinfo{author}{\bibfnamefont{S.~D.} \bibnamefont{Odintsov}},
  \bibnamefont{and} \bibinfo{author}{\bibfnamefont{V.~K.}
  \bibnamefont{Oikonomou}}, \bibinfo{journal}{Phys. Rev. D}
  \textbf{\bibinfo{volume}{94}}, \bibinfo{pages}{104050}
  (\bibinfo{year}{2016}), \eprint{1608.07806}.

\bibitem[{\citenamefont{Nojiri et~al.}(2017{\natexlab{b}})\citenamefont{Nojiri,
  Odintsov, and Oikonomou}}]{Nojiri:2017ygt}
\bibinfo{author}{\bibfnamefont{S.}~\bibnamefont{Nojiri}},
  \bibinfo{author}{\bibfnamefont{S.~D.} \bibnamefont{Odintsov}},
  \bibnamefont{and} \bibinfo{author}{\bibfnamefont{V.~K.}
  \bibnamefont{Oikonomou}}, \bibinfo{journal}{Phys. Lett. B}
  \textbf{\bibinfo{volume}{775}}, \bibinfo{pages}{44}
  (\bibinfo{year}{2017}{\natexlab{b}}), \eprint{1710.07838}.

\bibitem[{\citenamefont{Kaczmarek and Szczke\'sniak}(2021)}]{Kaczmarek:2021psy}
\bibinfo{author}{\bibfnamefont{A.~Z.} \bibnamefont{Kaczmarek}}
  \bibnamefont{and}
  \bibinfo{author}{\bibfnamefont{D.}~\bibnamefont{Szczke\'sniak}},
  \bibinfo{journal}{Sci. Rep.} \textbf{\bibinfo{volume}{11}},
  \bibinfo{pages}{18363} (\bibinfo{year}{2021}), \eprint{2105.05050}.

\bibitem[{\citenamefont{Cognola et~al.}(2016)\citenamefont{Cognola, Myrzakulov,
  Sebastiani, Vagnozzi, and Zerbini}}]{Cognola:2016gjy}
\bibinfo{author}{\bibfnamefont{G.}~\bibnamefont{Cognola}},
  \bibinfo{author}{\bibfnamefont{R.}~\bibnamefont{Myrzakulov}},
  \bibinfo{author}{\bibfnamefont{L.}~\bibnamefont{Sebastiani}},
  \bibinfo{author}{\bibfnamefont{S.}~\bibnamefont{Vagnozzi}}, \bibnamefont{and}
  \bibinfo{author}{\bibfnamefont{S.}~\bibnamefont{Zerbini}},
  \bibinfo{journal}{Class. Quant. Grav.} \textbf{\bibinfo{volume}{33}},
  \bibinfo{pages}{225014} (\bibinfo{year}{2016}), \eprint{1601.00102}.

\bibitem[{\citenamefont{Casalino et~al.}(2019)\citenamefont{Casalino, Rinaldi,
  Sebastiani, and Vagnozzi}}]{Casalino:2018wnc}
\bibinfo{author}{\bibfnamefont{A.}~\bibnamefont{Casalino}},
  \bibinfo{author}{\bibfnamefont{M.}~\bibnamefont{Rinaldi}},
  \bibinfo{author}{\bibfnamefont{L.}~\bibnamefont{Sebastiani}},
  \bibnamefont{and} \bibinfo{author}{\bibfnamefont{S.}~\bibnamefont{Vagnozzi}},
  \bibinfo{journal}{Class. Quant. Grav.} \textbf{\bibinfo{volume}{36}},
  \bibinfo{pages}{017001} (\bibinfo{year}{2019}), \eprint{1811.06830}.

\bibitem[{\citenamefont{Casalino et~al.}(2018)\citenamefont{Casalino, Rinaldi,
  Sebastiani, and Vagnozzi}}]{Casalino:2018tcd}
\bibinfo{author}{\bibfnamefont{A.}~\bibnamefont{Casalino}},
  \bibinfo{author}{\bibfnamefont{M.}~\bibnamefont{Rinaldi}},
  \bibinfo{author}{\bibfnamefont{L.}~\bibnamefont{Sebastiani}},
  \bibnamefont{and} \bibinfo{author}{\bibfnamefont{S.}~\bibnamefont{Vagnozzi}},
  \bibinfo{journal}{Phys. Dark Univ.} \textbf{\bibinfo{volume}{22}},
  \bibinfo{pages}{108} (\bibinfo{year}{2018}), \eprint{1803.02620}.

\bibitem[{\citenamefont{Tsujikawa}(2013)}]{Tsujikawa:2013fta}
\bibinfo{author}{\bibfnamefont{S.}~\bibnamefont{Tsujikawa}},
  \bibinfo{journal}{Class. Quant. Grav.} \textbf{\bibinfo{volume}{30}},
  \bibinfo{pages}{214003} (\bibinfo{year}{2013}), \eprint{1304.1961}.

\bibitem[{\citenamefont{Faraoni}(2004)}]{Faraoni:2004pi}
\bibinfo{author}{\bibfnamefont{V.}~\bibnamefont{Faraoni}},
  \emph{\bibinfo{title}{{Cosmology in scalar tensor gravity}}}
  (\bibinfo{year}{2004}), ISBN \bibinfo{isbn}{978-1-4020-1988-3}.

\bibitem[{\citenamefont{Kobayashi}(2019)}]{Kobayashi:2019hrl}
\bibinfo{author}{\bibfnamefont{T.}~\bibnamefont{Kobayashi}},
  \bibinfo{journal}{Rept. Prog. Phys.} \textbf{\bibinfo{volume}{82}},
  \bibinfo{pages}{086901} (\bibinfo{year}{2019}), \eprint{1901.07183}.

\bibitem[{\citenamefont{Capozziello et~al.}(1993)\citenamefont{Capozziello,
  Occhionero, and Amendola}}]{Capozziello:1993xn}
\bibinfo{author}{\bibfnamefont{S.}~\bibnamefont{Capozziello}},
  \bibinfo{author}{\bibfnamefont{F.}~\bibnamefont{Occhionero}},
  \bibnamefont{and} \bibinfo{author}{\bibfnamefont{L.}~\bibnamefont{Amendola}},
  \bibinfo{journal}{Int. J. Mod. Phys. D} \textbf{\bibinfo{volume}{1}},
  \bibinfo{pages}{615} (\bibinfo{year}{1993}).

\bibitem[{\citenamefont{Odintsov and Oikonomou}(2017)}]{Odintsov:2017tbc}
\bibinfo{author}{\bibfnamefont{S.~D.} \bibnamefont{Odintsov}} \bibnamefont{and}
  \bibinfo{author}{\bibfnamefont{V.~K.} \bibnamefont{Oikonomou}},
  \bibinfo{journal}{Phys. Rev. D} \textbf{\bibinfo{volume}{96}},
  \bibinfo{pages}{104049} (\bibinfo{year}{2017}), \eprint{1711.02230}.

\bibitem[{\citenamefont{Copeland et~al.}(1998)\citenamefont{Copeland, Liddle,
  and Wands}}]{Copeland:1997et}
\bibinfo{author}{\bibfnamefont{E.~J.} \bibnamefont{Copeland}},
  \bibinfo{author}{\bibfnamefont{A.~R.} \bibnamefont{Liddle}},
  \bibnamefont{and} \bibinfo{author}{\bibfnamefont{D.}~\bibnamefont{Wands}},
  \bibinfo{journal}{Phys. Rev. D} \textbf{\bibinfo{volume}{57}},
  \bibinfo{pages}{4686} (\bibinfo{year}{1998}), \eprint{gr-qc/9711068}.

\bibitem[{\citenamefont{Roy and Banerjee}(2014)}]{Roy:2014yta}
\bibinfo{author}{\bibfnamefont{N.}~\bibnamefont{Roy}} \bibnamefont{and}
  \bibinfo{author}{\bibfnamefont{N.}~\bibnamefont{Banerjee}},
  \bibinfo{journal}{Eur. Phys. J. Plus} \textbf{\bibinfo{volume}{129}},
  \bibinfo{pages}{162} (\bibinfo{year}{2014}), \eprint{1402.6821}.

\bibitem[{\citenamefont{Hao and Li}(2003)}]{Hao:2003ww}
\bibinfo{author}{\bibfnamefont{J.-g.} \bibnamefont{Hao}} \bibnamefont{and}
  \bibinfo{author}{\bibfnamefont{X.-z.} \bibnamefont{Li}},
  \bibinfo{journal}{Phys. Rev. D} \textbf{\bibinfo{volume}{67}},
  \bibinfo{pages}{107303} (\bibinfo{year}{2003}), \eprint{gr-qc/0302100}.

\bibitem[{\citenamefont{Odintsov and Oikonomou}(2016)}]{Odintsov:2015wwp}
\bibinfo{author}{\bibfnamefont{S.~D.} \bibnamefont{Odintsov}} \bibnamefont{and}
  \bibinfo{author}{\bibfnamefont{V.~K.} \bibnamefont{Oikonomou}},
  \bibinfo{journal}{Phys. Rev. D} \textbf{\bibinfo{volume}{93}},
  \bibinfo{pages}{023517} (\bibinfo{year}{2016}), \eprint{1511.04559}.

\bibitem[{\citenamefont{Dutta et~al.}(2018)\citenamefont{Dutta, Khyllep,
  Saridakis, Tamanini, and Vagnozzi}}]{Dutta:2017fjw}
\bibinfo{author}{\bibfnamefont{J.}~\bibnamefont{Dutta}},
  \bibinfo{author}{\bibfnamefont{W.}~\bibnamefont{Khyllep}},
  \bibinfo{author}{\bibfnamefont{E.~N.} \bibnamefont{Saridakis}},
  \bibinfo{author}{\bibfnamefont{N.}~\bibnamefont{Tamanini}}, \bibnamefont{and}
  \bibinfo{author}{\bibfnamefont{S.}~\bibnamefont{Vagnozzi}},
  \bibinfo{journal}{JCAP} \textbf{\bibinfo{volume}{02}}, \bibinfo{pages}{041}
  (\bibinfo{year}{2018}), \eprint{1711.07290}.

\bibitem[{\citenamefont{Leon and Saridakis}(2015)}]{Leon:2014yua}
\bibinfo{author}{\bibfnamefont{G.}~\bibnamefont{Leon}} \bibnamefont{and}
  \bibinfo{author}{\bibfnamefont{E.~N.} \bibnamefont{Saridakis}},
  \bibinfo{journal}{JCAP} \textbf{\bibinfo{volume}{04}}, \bibinfo{pages}{031}
  (\bibinfo{year}{2015}), \eprint{1501.00488}.

\bibitem[{\citenamefont{Hrycyna and Szydlowski}(2013)}]{Hrycyna:2013hla}
\bibinfo{author}{\bibfnamefont{O.}~\bibnamefont{Hrycyna}} \bibnamefont{and}
  \bibinfo{author}{\bibfnamefont{M.}~\bibnamefont{Szydlowski}},
  \bibinfo{journal}{Phys. Rev. D} \textbf{\bibinfo{volume}{88}},
  \bibinfo{pages}{064018} (\bibinfo{year}{2013}), \eprint{1304.3300}.

\bibitem[{\citenamefont{Boehmer et~al.}(2022)\citenamefont{Boehmer, Jensko, and
  Lazkoz}}]{Boehmer:2022wln}
\bibinfo{author}{\bibfnamefont{C.~G.} \bibnamefont{Boehmer}},
  \bibinfo{author}{\bibfnamefont{E.}~\bibnamefont{Jensko}}, \bibnamefont{and}
  \bibinfo{author}{\bibfnamefont{R.}~\bibnamefont{Lazkoz}},
  \bibinfo{journal}{Eur. Phys. J. C} \textbf{\bibinfo{volume}{82}},
  \bibinfo{pages}{500} (\bibinfo{year}{2022}), \eprint{2201.09588}.

\bibitem[{\citenamefont{Carloni and Mimoso}(2017)}]{Carloni:2017ucm}
\bibinfo{author}{\bibfnamefont{S.}~\bibnamefont{Carloni}} \bibnamefont{and}
  \bibinfo{author}{\bibfnamefont{J.~P.} \bibnamefont{Mimoso}},
  \bibinfo{journal}{Eur. Phys. J. C} \textbf{\bibinfo{volume}{77}},
  \bibinfo{pages}{547} (\bibinfo{year}{2017}), \eprint{1701.00231}.

\bibitem[{\citenamefont{Goliath and Ellis}(1999)}]{Goliath:1998na}
\bibinfo{author}{\bibfnamefont{M.}~\bibnamefont{Goliath}} \bibnamefont{and}
  \bibinfo{author}{\bibfnamefont{G.~F.~R.} \bibnamefont{Ellis}},
  \bibinfo{journal}{Phys. Rev. D} \textbf{\bibinfo{volume}{60}},
  \bibinfo{pages}{023502} (\bibinfo{year}{1999}), \eprint{gr-qc/9811068}.

\bibitem[{\citenamefont{Y\i{}lmaz and G\"udekli}(2021)}]{Yilmaz:2021rsc}
\bibinfo{author}{\bibfnamefont{A.~O.} \bibnamefont{Y\i{}lmaz}}
  \bibnamefont{and}
  \bibinfo{author}{\bibfnamefont{E.}~\bibnamefont{G\"udekli}},
  \bibinfo{journal}{Sci. Rep.} \textbf{\bibinfo{volume}{11}},
  \bibinfo{pages}{2750} (\bibinfo{year}{2021}).

\bibitem[{\citenamefont{Boehmer and Chan}(2014)}]{Boehmer:2014vea}
\bibinfo{author}{\bibfnamefont{C.~G.} \bibnamefont{Boehmer}} \bibnamefont{and}
  \bibinfo{author}{\bibfnamefont{N.}~\bibnamefont{Chan}}
  (\bibinfo{year}{2014}), \eprint{1409.5585}.

\bibitem[{\citenamefont{Bahamonde et~al.}(2018)\citenamefont{Bahamonde,
  B\"ohmer, Carloni, Copeland, Fang, and Tamanini}}]{Bahamonde:2017ize}
\bibinfo{author}{\bibfnamefont{S.}~\bibnamefont{Bahamonde}},
  \bibinfo{author}{\bibfnamefont{C.~G.} \bibnamefont{B\"ohmer}},
  \bibinfo{author}{\bibfnamefont{S.}~\bibnamefont{Carloni}},
  \bibinfo{author}{\bibfnamefont{E.~J.} \bibnamefont{Copeland}},
  \bibinfo{author}{\bibfnamefont{W.}~\bibnamefont{Fang}}, \bibnamefont{and}
  \bibinfo{author}{\bibfnamefont{N.}~\bibnamefont{Tamanini}},
  \bibinfo{journal}{Phys. Rept.} \textbf{\bibinfo{volume}{775-777}},
  \bibinfo{pages}{1} (\bibinfo{year}{2018}), \eprint{1712.03107}.

\bibitem[{\citenamefont{Di~Valentino et~al.}(2019)\citenamefont{Di~Valentino,
  Melchiorri, and Silk}}]{DiValentino:2019qzk}
\bibinfo{author}{\bibfnamefont{E.}~\bibnamefont{Di~Valentino}},
  \bibinfo{author}{\bibfnamefont{A.}~\bibnamefont{Melchiorri}},
  \bibnamefont{and} \bibinfo{author}{\bibfnamefont{J.}~\bibnamefont{Silk}},
  \bibinfo{journal}{Nature Astron.} \textbf{\bibinfo{volume}{4}},
  \bibinfo{pages}{196} (\bibinfo{year}{2019}), \eprint{1911.02087}.

\bibitem[{\citenamefont{Handley}(2021)}]{Handley:2019tkm}
\bibinfo{author}{\bibfnamefont{W.}~\bibnamefont{Handley}},
  \bibinfo{journal}{Phys. Rev. D} \textbf{\bibinfo{volume}{103}},
  \bibinfo{pages}{L041301} (\bibinfo{year}{2021}), \eprint{1908.09139}.

\bibitem[{\citenamefont{Di~Valentino et~al.}(2021)\citenamefont{Di~Valentino,
  Melchiorri, and Silk}}]{DiValentino:2020hov}
\bibinfo{author}{\bibfnamefont{E.}~\bibnamefont{Di~Valentino}},
  \bibinfo{author}{\bibfnamefont{A.}~\bibnamefont{Melchiorri}},
  \bibnamefont{and} \bibinfo{author}{\bibfnamefont{J.}~\bibnamefont{Silk}},
  \bibinfo{journal}{Astrophys. J. Lett.} \textbf{\bibinfo{volume}{908}},
  \bibinfo{pages}{L9} (\bibinfo{year}{2021}), \eprint{2003.04935}.

\bibitem[{\citenamefont{Leon et~al.}(2013)\citenamefont{Leon, Saavedra, and
  Saridakis}}]{Leon_2013}
\bibinfo{author}{\bibfnamefont{G.}~\bibnamefont{Leon}},
  \bibinfo{author}{\bibfnamefont{J.}~\bibnamefont{Saavedra}}, \bibnamefont{and}
  \bibinfo{author}{\bibfnamefont{E.~N.} \bibnamefont{Saridakis}},
  \bibinfo{journal}{Classical and Quantum Gravity}
  \textbf{\bibinfo{volume}{30}}, \bibinfo{pages}{135001}
  (\bibinfo{year}{2013}).

\bibitem[{\citenamefont{Fadragas et~al.}(2014)\citenamefont{Fadragas, Leon, and
  Saridakis}}]{Fadragas_2014}
\bibinfo{author}{\bibfnamefont{C.~R.} \bibnamefont{Fadragas}},
  \bibinfo{author}{\bibfnamefont{G.}~\bibnamefont{Leon}}, \bibnamefont{and}
  \bibinfo{author}{\bibfnamefont{E.~N.} \bibnamefont{Saridakis}},
  \bibinfo{journal}{Classical and Quantum Gravity}
  \textbf{\bibinfo{volume}{31}}, \bibinfo{pages}{075018}
  (\bibinfo{year}{2014}).

\bibitem[{\citenamefont{Magana et~al.}(2018)\citenamefont{Magana, Amante,
  Garcia-Aspeitia, and Motta}}]{Magana:2017nfs}
\bibinfo{author}{\bibfnamefont{J.}~\bibnamefont{Magana}},
  \bibinfo{author}{\bibfnamefont{M.~H.} \bibnamefont{Amante}},
  \bibinfo{author}{\bibfnamefont{M.~A.} \bibnamefont{Garcia-Aspeitia}},
  \bibnamefont{and} \bibinfo{author}{\bibfnamefont{V.}~\bibnamefont{Motta}},
  \bibinfo{journal}{Mon. Not. Roy. Astron. Soc.}
  \textbf{\bibinfo{volume}{476}}, \bibinfo{pages}{1036} (\bibinfo{year}{2018}),
  \eprint{1706.09848}.

\bibitem[{\citenamefont{Scolnic et~al.}(2018)}]{Pan-STARRS1:2017jku}
\bibinfo{author}{\bibfnamefont{D.~M.} \bibnamefont{Scolnic}}
  \bibnamefont{et~al.} (\bibinfo{collaboration}{Pan-STARRS1}),
  \bibinfo{journal}{Astrophys. J.} \textbf{\bibinfo{volume}{859}},
  \bibinfo{pages}{101} (\bibinfo{year}{2018}), \eprint{1710.00845}.

\bibitem[{\citenamefont{Foreman-Mackey
  et~al.}(2013)\citenamefont{Foreman-Mackey, Hogg, Lang, and
  Goodman}}]{Foreman-Mackey:2012any}
\bibinfo{author}{\bibfnamefont{D.}~\bibnamefont{Foreman-Mackey}},
  \bibinfo{author}{\bibfnamefont{D.~W.} \bibnamefont{Hogg}},
  \bibinfo{author}{\bibfnamefont{D.}~\bibnamefont{Lang}}, \bibnamefont{and}
  \bibinfo{author}{\bibfnamefont{J.}~\bibnamefont{Goodman}},
  \bibinfo{journal}{Publ. Astron. Soc. Pac.} \textbf{\bibinfo{volume}{125}},
  \bibinfo{pages}{306} (\bibinfo{year}{2013}), \eprint{1202.3665}.

\bibitem[{\citenamefont{Moresco et~al.}(2022)}]{Moresco:2022phi}
\bibinfo{author}{\bibfnamefont{M.}~\bibnamefont{Moresco}} \bibnamefont{et~al.},
  \bibinfo{journal}{Living Rev. Rel.} \textbf{\bibinfo{volume}{25}},
  \bibinfo{pages}{6} (\bibinfo{year}{2022}), \eprint{2201.07241}.

\bibitem[{\citenamefont{Jimenez and Loeb}(2002)}]{Jimenez:2001gg}
\bibinfo{author}{\bibfnamefont{R.}~\bibnamefont{Jimenez}} \bibnamefont{and}
  \bibinfo{author}{\bibfnamefont{A.}~\bibnamefont{Loeb}},
  \bibinfo{journal}{Astrophys. J.} \textbf{\bibinfo{volume}{573}},
  \bibinfo{pages}{37} (\bibinfo{year}{2002}), \eprint{astro-ph/0106145}.

\bibitem[{\citenamefont{Corral et~al.}(2020)\citenamefont{Corral, Cruz, and
  Gonz\'alez}}]{Corral:2020lxt}
\bibinfo{author}{\bibfnamefont{C.}~\bibnamefont{Corral}},
  \bibinfo{author}{\bibfnamefont{N.}~\bibnamefont{Cruz}}, \bibnamefont{and}
  \bibinfo{author}{\bibfnamefont{E.}~\bibnamefont{Gonz\'alez}},
  \bibinfo{journal}{Phys. Rev. D} \textbf{\bibinfo{volume}{102}},
  \bibinfo{pages}{023508} (\bibinfo{year}{2020}), \eprint{2005.06052}.

\bibitem[{\citenamefont{Conley et~al.}(2011)}]{SNLS:2011lii}
\bibinfo{author}{\bibfnamefont{A.}~\bibnamefont{Conley}} \bibnamefont{et~al.}
  (\bibinfo{collaboration}{SNLS}), \bibinfo{journal}{Astrophys. J. Suppl.}
  \textbf{\bibinfo{volume}{192}}, \bibinfo{pages}{1} (\bibinfo{year}{2011}),
  \eprint{1104.1443}.

\bibitem[{\citenamefont{Deng and Wei}(2018)}]{Deng:2018jrp}
\bibinfo{author}{\bibfnamefont{H.-K.} \bibnamefont{Deng}} \bibnamefont{and}
  \bibinfo{author}{\bibfnamefont{H.}~\bibnamefont{Wei}}, \bibinfo{journal}{Eur.
  Phys. J. C} \textbf{\bibinfo{volume}{78}}, \bibinfo{pages}{755}
  (\bibinfo{year}{2018}), \eprint{1806.02773}.

\bibitem[{\citenamefont{Asvesta et~al.}(2022)\citenamefont{Asvesta,
  Kazantzidis, Perivolaropoulos, and Tsagas}}]{Asvesta:2022fts}
\bibinfo{author}{\bibfnamefont{K.}~\bibnamefont{Asvesta}},
  \bibinfo{author}{\bibfnamefont{L.}~\bibnamefont{Kazantzidis}},
  \bibinfo{author}{\bibfnamefont{L.}~\bibnamefont{Perivolaropoulos}},
  \bibnamefont{and} \bibinfo{author}{\bibfnamefont{C.~G.}
  \bibnamefont{Tsagas}}, \bibinfo{journal}{Mon. Not. Roy. Astron. Soc.}
  \textbf{\bibinfo{volume}{513}}, \bibinfo{pages}{2394} (\bibinfo{year}{2022}),
  \eprint{2202.00962}.

\bibitem[{\citenamefont{Cao et~al.}(2021)\citenamefont{Cao, Ryan, and
  Ratra}}]{Cao:2021ldv}
\bibinfo{author}{\bibfnamefont{S.}~\bibnamefont{Cao}},
  \bibinfo{author}{\bibfnamefont{J.}~\bibnamefont{Ryan}}, \bibnamefont{and}
  \bibinfo{author}{\bibfnamefont{B.}~\bibnamefont{Ratra}},
  \bibinfo{journal}{Mon. Not. Roy. Astron. Soc.}
  \textbf{\bibinfo{volume}{504}}, \bibinfo{pages}{300} (\bibinfo{year}{2021}),
  \eprint{2101.08817}.

\bibitem[{\citenamefont{Kessler and Scolnic}(2017)}]{Kessler:2016uwi}
\bibinfo{author}{\bibfnamefont{R.}~\bibnamefont{Kessler}} \bibnamefont{and}
  \bibinfo{author}{\bibfnamefont{D.}~\bibnamefont{Scolnic}},
  \bibinfo{journal}{Astrophys. J.} \textbf{\bibinfo{volume}{836}},
  \bibinfo{pages}{56} (\bibinfo{year}{2017}), \eprint{1610.04677}.

\bibitem[{\citenamefont{Akaike}(1974)}]{akaike1974new}
\bibinfo{author}{\bibfnamefont{H.}~\bibnamefont{Akaike}},
  \bibinfo{journal}{IEEE Transactions on Automatic Control}
  \textbf{\bibinfo{volume}{19}}, \bibinfo{pages}{716} (\bibinfo{year}{1974}).

\bibitem[{\citenamefont{Schwarz}(1978)}]{Schwarz:1978tpv}
\bibinfo{author}{\bibfnamefont{G.}~\bibnamefont{Schwarz}},
  \bibinfo{journal}{Annals Statist.} \textbf{\bibinfo{volume}{6}},
  \bibinfo{pages}{461} (\bibinfo{year}{1978}).

\bibitem[{\citenamefont{Yang et~al.}(2023)\citenamefont{Yang, Giar\`e, Pan,
  Di~Valentino, Melchiorri, and Silk}}]{Yang:2022kho}
\bibinfo{author}{\bibfnamefont{W.}~\bibnamefont{Yang}},
  \bibinfo{author}{\bibfnamefont{W.}~\bibnamefont{Giar\`e}},
  \bibinfo{author}{\bibfnamefont{S.}~\bibnamefont{Pan}},
  \bibinfo{author}{\bibfnamefont{E.}~\bibnamefont{Di~Valentino}},
  \bibinfo{author}{\bibfnamefont{A.}~\bibnamefont{Melchiorri}},
  \bibnamefont{and} \bibinfo{author}{\bibfnamefont{J.}~\bibnamefont{Silk}},
  \bibinfo{journal}{Phys. Rev. D} \textbf{\bibinfo{volume}{107}},
  \bibinfo{pages}{063509} (\bibinfo{year}{2023}), \eprint{2210.09865}.

\bibitem[{\citenamefont{Vagnozzi
  et~al.}(2021{\natexlab{a}})\citenamefont{Vagnozzi, Loeb, and
  Moresco}}]{Vagnozzi:2020dfn}
\bibinfo{author}{\bibfnamefont{S.}~\bibnamefont{Vagnozzi}},
  \bibinfo{author}{\bibfnamefont{A.}~\bibnamefont{Loeb}}, \bibnamefont{and}
  \bibinfo{author}{\bibfnamefont{M.}~\bibnamefont{Moresco}},
  \bibinfo{journal}{Astrophys. J.} \textbf{\bibinfo{volume}{908}},
  \bibinfo{pages}{84} (\bibinfo{year}{2021}{\natexlab{a}}),
  \eprint{2011.11645}.

\bibitem[{\citenamefont{Vagnozzi
  et~al.}(2021{\natexlab{b}})\citenamefont{Vagnozzi, Di~Valentino, Gariazzo,
  Melchiorri, Mena, and Silk}}]{Vagnozzi:2020rcz}
\bibinfo{author}{\bibfnamefont{S.}~\bibnamefont{Vagnozzi}},
  \bibinfo{author}{\bibfnamefont{E.}~\bibnamefont{Di~Valentino}},
  \bibinfo{author}{\bibfnamefont{S.}~\bibnamefont{Gariazzo}},
  \bibinfo{author}{\bibfnamefont{A.}~\bibnamefont{Melchiorri}},
  \bibinfo{author}{\bibfnamefont{O.}~\bibnamefont{Mena}}, \bibnamefont{and}
  \bibinfo{author}{\bibfnamefont{J.}~\bibnamefont{Silk}},
  \bibinfo{journal}{Phys. Dark Univ.} \textbf{\bibinfo{volume}{33}},
  \bibinfo{pages}{100851} (\bibinfo{year}{2021}{\natexlab{b}}),
  \eprint{2010.02230}.

\bibitem[{\citenamefont{Dhawan et~al.}(2021)\citenamefont{Dhawan, Alsing, and
  Vagnozzi}}]{Dhawan:2021mel}
\bibinfo{author}{\bibfnamefont{S.}~\bibnamefont{Dhawan}},
  \bibinfo{author}{\bibfnamefont{J.}~\bibnamefont{Alsing}}, \bibnamefont{and}
  \bibinfo{author}{\bibfnamefont{S.}~\bibnamefont{Vagnozzi}},
  \bibinfo{journal}{Mon. Not. Roy. Astron. Soc.}
  \textbf{\bibinfo{volume}{506}}, \bibinfo{pages}{L1} (\bibinfo{year}{2021}),
  \eprint{2104.02485}.

\bibitem[{\citenamefont{Abdalla et~al.}(2022)}]{Abdalla:2022yfr}
\bibinfo{author}{\bibfnamefont{E.}~\bibnamefont{Abdalla}} \bibnamefont{et~al.},
  \bibinfo{journal}{JHEAp} \textbf{\bibinfo{volume}{34}}, \bibinfo{pages}{49}
  (\bibinfo{year}{2022}), \eprint{2203.06142}.

\bibitem[{\citenamefont{Aghanim et~al.}(2020)}]{Planck:2018vyg}
\bibinfo{author}{\bibfnamefont{N.}~\bibnamefont{Aghanim}} \bibnamefont{et~al.}
  (\bibinfo{collaboration}{Planck}), \bibinfo{journal}{Astron. Astrophys.}
  \textbf{\bibinfo{volume}{641}}, \bibinfo{pages}{A6} (\bibinfo{year}{2020}),
  \bibinfo{note}{[Erratum: Astron.Astrophys. 652, C4 (2021)]},
  \eprint{1807.06209}.

\bibitem[{\citenamefont{Rosenberg et~al.}(2022)\citenamefont{Rosenberg,
  Gratton, and Efstathiou}}]{Rosenberg:2022sdy}
\bibinfo{author}{\bibfnamefont{E.}~\bibnamefont{Rosenberg}},
  \bibinfo{author}{\bibfnamefont{S.}~\bibnamefont{Gratton}}, \bibnamefont{and}
  \bibinfo{author}{\bibfnamefont{G.}~\bibnamefont{Efstathiou}},
  \bibinfo{journal}{Mon. Not. Roy. Astron. Soc.}
  \textbf{\bibinfo{volume}{517}}, \bibinfo{pages}{4620} (\bibinfo{year}{2022}),
  \eprint{2205.10869}.

\bibitem[{\citenamefont{Tristram et~al.}(2023)}]{Tristram:2023haj}
\bibinfo{author}{\bibfnamefont{M.}~\bibnamefont{Tristram}} \bibnamefont{et~al.}
  (\bibinfo{year}{2023}), \eprint{2309.10034}.

\bibitem[{\citenamefont{Brout et~al.}(2022)}]{Brout:2022vxf}
\bibinfo{author}{\bibfnamefont{D.}~\bibnamefont{Brout}} \bibnamefont{et~al.},
  \bibinfo{journal}{Astrophys. J.} \textbf{\bibinfo{volume}{938}},
  \bibinfo{pages}{110} (\bibinfo{year}{2022}), \eprint{2202.04077}.

\bibitem[{\citenamefont{Riess et~al.}(2021)\citenamefont{Riess, Casertano,
  Yuan, Bowers, Macri, Zinn, and Scolnic}}]{Riess:2021fzl}
\bibinfo{author}{\bibfnamefont{A.~G.} \bibnamefont{Riess}},
  \bibinfo{author}{\bibfnamefont{S.}~\bibnamefont{Casertano}},
  \bibinfo{author}{\bibfnamefont{W.}~\bibnamefont{Yuan}},
  \bibinfo{author}{\bibfnamefont{J.~B.} \bibnamefont{Bowers}},
  \bibinfo{author}{\bibfnamefont{L.}~\bibnamefont{Macri}},
  \bibinfo{author}{\bibfnamefont{J.~C.} \bibnamefont{Zinn}}, \bibnamefont{and}
  \bibinfo{author}{\bibfnamefont{D.}~\bibnamefont{Scolnic}},
  \bibinfo{journal}{Astrophys. J. Lett.} \textbf{\bibinfo{volume}{908}},
  \bibinfo{pages}{L6} (\bibinfo{year}{2021}), \eprint{2012.08534}.

\bibitem[{\citenamefont{Riess et~al.}(2022)}]{Riess:2022jrx}
\bibinfo{author}{\bibfnamefont{A.~G.} \bibnamefont{Riess}}
  \bibnamefont{et~al.}, \bibinfo{journal}{Astrophys. J. Lett.}
  \textbf{\bibinfo{volume}{934}}, \bibinfo{pages}{L7} (\bibinfo{year}{2022}),
  \eprint{2112.04510}.

\bibitem[{\citenamefont{Riess et~al.}(2019)\citenamefont{Riess, Casertano,
  Yuan, Macri, and Scolnic}}]{Riess_2019}
\bibinfo{author}{\bibfnamefont{A.~G.} \bibnamefont{Riess}},
  \bibinfo{author}{\bibfnamefont{S.}~\bibnamefont{Casertano}},
  \bibinfo{author}{\bibfnamefont{W.}~\bibnamefont{Yuan}},
  \bibinfo{author}{\bibfnamefont{L.~M.} \bibnamefont{Macri}}, \bibnamefont{and}
  \bibinfo{author}{\bibfnamefont{D.}~\bibnamefont{Scolnic}},
  \bibinfo{journal}{The Astrophysical Journal} \textbf{\bibinfo{volume}{876}},
  \bibinfo{pages}{85} (\bibinfo{year}{2019}).

\bibitem[{\citenamefont{Petronikolou et~al.}(2022)\citenamefont{Petronikolou,
  Basilakos, and Saridakis}}]{Petronikolou:2021shp}
\bibinfo{author}{\bibfnamefont{M.}~\bibnamefont{Petronikolou}},
  \bibinfo{author}{\bibfnamefont{S.}~\bibnamefont{Basilakos}},
  \bibnamefont{and} \bibinfo{author}{\bibfnamefont{E.~N.}
  \bibnamefont{Saridakis}}, \bibinfo{journal}{Phys. Rev. D}
  \textbf{\bibinfo{volume}{106}}, \bibinfo{pages}{124051}
  (\bibinfo{year}{2022}), \eprint{2110.01338}.

\bibitem[{\citenamefont{Aranda~Iriarte}(1998)}]{aranda_iriarte_1998}
\bibinfo{author}{\bibfnamefont{J.~I.} \bibnamefont{Aranda~Iriarte}}, in
  \emph{\bibinfo{booktitle}{XV Congreso de ecuaciones diferenciales y
  aplicaciones, V Congreso de Matem\'atica Aplicada : Vigo, 23-26 septiembre
  1997}} (\bibinfo{publisher}{Servizo de Publicacions}, \bibinfo{year}{1998}),
  vol.~\bibinfo{volume}{1}, pp. \bibinfo{pages}{157--162}, ISBN
  \bibinfo{isbn}{84-8158-093-7}.

\end{thebibliography}

\end{document}